\documentclass[traditabstract]{aa} 

\usepackage{graphicx}
\usepackage{txfonts}
\usepackage{longtable}
\usepackage{lscape}
\usepackage{hyperref}
\usepackage{soul}
\hypersetup{colorlinks=true, citecolor=blue}
\newcommand{\hi}{{\sc H\,i}}
\newcommand{\HIbf}{\mbox{H\hspace{0.155 em}{\footnotesize \bf I}}}
\newcommand{\kms}{$\,$km$\,$s$^{-1}$}
\newcommand{\mJybeam}{mJy beam$^{-1}$}
\begin{document}

   \title{ViCTORIA project: MeerKAT \HIbf\ observations of the ram pressure stripped galaxy NGC 4523\thanks{Based on observations obtained with
   MegaPrime/MegaCam, a joint project of CFHT and CEA/DAPNIA, at the Canada-French-Hawaii Telescope
   (CFHT) which is operated by the National Research Council (NRC) of Canada, the Institut National
   des Sciences de l'Univers of the Centre National de la Recherche Scientifique (CNRS) of France and
   the University of Hawaii.}
      }
   \subtitle{}
  \author{A. Boselli\inst{1,**},  
          P. Serra\inst{2},
	  F. de Gasperin\inst{3,4},
    B. Vollmer\inst{5},
    P. Amram\inst{1},
    H. W. Edler\inst{4},
 	  M. Fossati\inst{6,7},
    G. Consolandi\inst{7},
    P. C{\^o}t{\'e}\inst{8},
    J.C. Cuillandre\inst{9},
    L. Ferrarese\inst{8},
    S. Gwyn\inst{8}, 
	  J. Postma\inst{10},  
   M. Boquien\inst{11},
   J. Braine\inst{12},
   F. Combes\inst{13,14},
   G. Gavazzi\inst{6},
   G. Hensler\inst{15},
   M.A. Miville-Deschenes\inst{9},
   M. Murgia\inst{2},
   J. Roediger\inst{8},
   Y. Roehlly\inst{1},
   R. Smith\inst{16},
   H.X. Zhang\inst{17,18},
   N. Zabel\inst{19}
       }

\institute{     
                Aix Marseille Univ, CNRS, CNES, LAM, Marseille, France\thanks{Scientific associate INAF - Osservatorio Astronomico di Brera, via Brera 28, 20121 Milano, Italy}
                \email{alessandro.boselli@lam.fr}
        \and
		Osservatorio Astronomico di Cagliari, Via della Scienza 5, 09047 Selargius (CA), Italy
	\and
        INAF - Istituto di Radioastronomia, Via P. Gobetti 101, 40129 Bologna, Italy	
	\and
		Hamburger Sternwarte, University of Hamburg, Gojenbergsweg 112, 21029 Hamburg, Germany
	\and
		Université de Strasbourg, CNRS, Observatoire Astronomique de Strasbourg, UMR 7550, 67000 Strasbourg, France
    \and
        Universit\'a di Milano-Bicocca, piazza della scienza 3, 20100 Milano, Italy
	\and
		INAF - Osservatorio Astronomico di Brera, via Brera 28, 20121 Milano, Italy
	\and
        National Research Council of Canada, Herzberg Astronomy and Astrophysics, 5071 West Saanich Road, Victoria, BC, V9E 2E7, Canada
	\and
        AIM, CEA, CNRS, Universit\'e Paris-Saclay, Universit\'e Paris Diderot, Sorbonne Paris Cit\'e, Observatoire de Paris, PSL University, F-91191 Gif-sur-Yvette Cedex, France
    \and
        University of Calgary, 2500 University Drive NW, Calgary, Alberta, Canada
    \and	
	    Centro de Astronomi\'a (CITEVA), Universidad de Antofagasta, Avenida Angamos 601, Antofagasta, Chile
	\and
        Laboratoire d'Astrophysique de Bordeaux, Univ. Bordeaux, CNRS, B18N, all\'ee Geoffroy Saint-Hilaire, 33615 Pessac, France	
    \and
        College de France, 11 Pl. M. Berthelot, F-75005 Paris, France 
    \and
        LERMA, Observatoire de Paris, CNRS, PSL Research University, Sorbonne Universit\'es, UPMC Univ. Paris 06, F-75014 Paris, France
    \and
        Department of Astrophysics, University of Vienna, T\"urkenschanzstrasse 17, 1180 Vienna, Austria
    \and
        Departamento de F\'isica, Universidad T\'ecnica Federico Santa Mar\'ia, Vicu\~na Mackenna 3939, San Joaqu\'in, Santiago, Chile
    \and
        Key Laboratory for Research in Galaxies and Cosmology, Department of Astronomy, University of Science and Technology of China, Hefei 230026, China
    \and
        School of Astronomy and Space Science, University of Science and Technology of China, Hefei 230026, China
    \and
        Department of Astronomy, University of Cape Town, Private Bag X3, Rondebosch 7701, South Africa
                 }

\authorrunning{Boselli et al.}
\titlerunning{ViCTORIA: \hi\ tail in NGC 4523 }

   \date{}

 
  \abstract  
{We present the first results of a 21 cm \hi\ line pilot observation carried out with the MeerKAT radio telescope in preparation for the ViCTORIA (Virgo Cluster multi Telescope Observations in Radio of Interacting galaxies and AGN) project, an untargeted survey of the Virgo galaxy cluster. The extraordinary quality of the data in terms of sensitivity and angular resolution ($rms$ $\sim$ 0.65 mJy beam$^{-1}$ at $\sim$ 27\arcsec $\times$ 39\arcsec and 11 km S$^{-1}$ resolution) allowed us to detect 
an extended ($\sim$ 10 kpc projected length) low column density ($N(HI) \lesssim 2.5 \times 10^{20}$ cm$^{-2}$) \hi\ gas tail associated with the 
dwarf ($M_{star}$ =1.6 $\times$ 10$^9$ M$_{\odot}$) irregular galaxy NGC 4523 at the northern edge of the cluster. The morphology of the tail and of the stellar disc suggest that the galaxy is suffering a hydrodynamic interaction with the surrounding 
hot intracluster medium (ICM; ram pressure stripping). The orientation of the trailing tail, the gradient in the \hi\ gas column density at the 
interface between the cold ISM and the hot ICM, the velocity of the galaxy with respect to that of the cluster, and its position indicate that
NGC 4523 is infalling for the first time into Virgo from the north-northwest background of the cluster. Using a grid of hydrodynamic simulations we derive the impact parameters with the surrounding ICM, and estimate that the galaxy will be at pericentre ($D$ $\sim$ 500-600 kpc) in $\sim$ 1 Gyr, where ram pressure stripping will be able to remove most, if not all, of its gas. The galaxy is located on the star formation main sequence when its star formation rate is derived using H$\alpha$ narrow-band images obtained during the VESTIGE survey, suggesting that NGC 4523 is only at the beginning of its interaction with the surrounding environment. A few H{\sc ii} regions are detected in the deep H$\alpha$ narrow-band images within the \hi\ gas tail outside 
the stellar disc. Their ages, derived by comparing their H$\alpha$, FUV (ASTROSAT/UVIT), NUV (GALEX/GUViCS), and optical (NGVS)
colours with the predictions of SED fitting models, are $\lesssim$ 30 Myr, and suggest that these H{\sc ii} regions have formed within the stripped gas.
 }
   {}
   {}
   {}
   {}
   {}


   \maketitle
%

\section{Introduction}

The environment plays a major role in shaping galaxy evolution. Galaxies inhabiting rich clusters are predominantly
quiescent and relaxed systems, while those located in the field are star-forming, late-type objects (morphology segregation, Dressler 1980, 1997, Whitmore et al. 1993).
At the same time, late-type systems inhabiting dense regions are generally gas-poor and with a reduced star formation activity (e.g. Boselli \& Gavazzi 2006).
Several mechanisms able to affect the evolution of galaxies in high-density regions have been proposed in the literature to explain these differences. 
They include gravitational perturbations induced by the interaction with nearby companions (Merritt 1983), with the gravitational potential well of the cluster
dark matter halo (Byrd \& Valtonen 1990), or their combined effect after several high-speed fly-by encounters (galaxy harassment, Moore et al. 1998).
Hydrodynamic interactions can also occur between the cold interstellar medium (ISM) of galaxies and the surrounding hot ($T$ $\sim$ 10$^7$-10$^8$ K) intracluster medium (ICM) trapped within
the gravitational potential well of clusters and groups and emitting in X-rays (e.g. Sarazin 1986, Sun 2012). They include heat conduction 
between the two media at different temperature (thermal evaporation, Cowie \& Songaila 1977), mixing in the instabilities formed
at the interface between the galaxy ISM and the surrounding ICM (e.g. viscous stripping, Nulsen 1982), the stop of infall of fresh gas once
a galaxy becomes satellite of a larger system (starvation, Larson et al. 1980), and the pressure exerted by the 
ICM on galaxies moving at high velocity (500-1000 km s$^{-1}$) within it (ram pressure stripping, Gunn \& Gott 1972).
All these mechanisms are able to perturb the gas content and distribution of galaxies and as a consequence alter their production of new stars
(e.g. Boselli \& Gavazzi 2006, 2014; Cortese et al. 2021; Boselli et al. 2022). 
Over long timescales, once the gas reservoir is removed from the disc, the activity of star formation is reduced, and galaxies become quiescent
 (e.g. Peng et al. 2010; Boselli et al. 2023).

While the general picture of galaxy evolution in rich environments is now well understood, several open questions still need to be answered.
First of all, it is still unclear which of these mechanisms is dominant in different structures of the cosmic web, from filaments, loose and compact groups,
to rich clusters of galaxies. Since the efficiency of all these processes is tightly connected to the properties of the high-density regions
(gas density and temperature, galaxy number density, velocity dispersion, infall rate etc., Boselli \& Gavazzi 2006), their relative weight can also significantly change with 
cosmic time. Furthermore, the effects of these mechanisms are still not fully understood. The major difficulties are related to the fact that 
they concern a multi-phase medium, from cold atomic and molecular gas to ionised and hot plasma, and acts on a wide range of scales, from individual 
H{\sc ii} regions and giant molecular clouds (GMC, 50 pc scales) to entire clusters of galaxies (Mpc scales). For the same reasons, hydrodynamic simulations 
of these perturbing mechanisms are also challenging (e.g. Roediger 2009). One particular aspect which deserves further investigation is the fate 
of the stripped gas. The most recent observations seem to indicate that the gas is principally stripped from the galaxy discs as cold atomic hydrogen
and once mixed with the surrounding ICM gets first ionised and then hot (e.g. Boselli et al. 2016a, 2021, 2022; Poggianti et al. 2018; Campitiello et al. 2021; 
Sun et al. 2021; Bartolini et al. 2022), thus contributing to the pollution of the diffuse ICM (e.g. Longobardi et al. 2020a, 2020b). Part of the molecular gas can also be removed during the interaction (Fumagalli et al. 2009; Boselli et al. 2014b; Zabel et al. 2019, 2022).
The stripped gas can collapse into GMCs to form new stars (Jachym et al. 2014, 2017, 2019; Moretti et al. 2020a, 2020b), 
although this process is not ubiquitous (e.g. Gavazzi et al. 2001; Boselli et al. 2016a; Pedrini et al. 2022).

The local Universe is a perfect target for environmental studies. First of all, it gives us the end point of the evolutionary path leading to local galaxies,
and different samples selected according to different criteria can be used to drive strong statistical results. It also allows us to study 
at high physical resolution different samples at various wavelengths, giving us a unique view of the effects of the perturbations 
on the different galaxy components (gas in its different phases, dust, stars, cosmic rays, etc.). Finally, dwarf galaxies, the most 
fragile objects to external perturbations and to internal energy release because of their shallow gravitational potential well, are only here accessible at all frequencies.
For these reasons the Virgo cluster, the closest cluster of galaxies, has been the target of several untargeted multifrequency surveys. These include
GALEX UV observations (GALEX Ultraviolet Virgo Cluster Survey, GUViCS, Boselli et al. 2011), deep optical imaging (Next Generation Virgo cluster Survey, NGVS, Ferrarese et al. 2012), mid- (Wide-field Infrared Survey Explorer, WISE, Wright et al. 2010) and far-IR 
imaging (Herschel Virgo Cluster Survey, HeViCS, Davies et al. 2010). A deep untargeted H$\alpha$ narrow-band imaging of the cluster, especially designed to study the 
effects of the perturbations on the star formation activity of galaxies and to detect extended low surface brightness ionised gas features formed
during the interactions of galaxies with their surrounding environment, has been almost completed (Virgo Environmental Survey Tracing Ionised Gas Emission, VESTIGE, Boselli et al. 2018a).
Given their untargeted nature, these surveys are ideally designed to identify galaxies undergoing a perturbation and the very 
nature of the perturbing mechanism on a statistically complete sample, thus minimising any possible bias related to selection effects 
present in targeted observations. 

The Virgo cluster has been covered also in the radio domain by the NRAO VLA Sky Survey (NVSS, Condon et al. 1998)
at 1.4 GHz, and in the \hi\ line by the HI Parkes All Sky Survey (HIPASS, Barnes et al. 2001) and Arecibo Legacy Fast ALFA  survey (ALFALFA, Giovanelli et al. 2005) surveys. The data collected by these projects 
have been crucial for identifying spiral galaxies with an enhanced radio continuum emission in the cluster environment with respect to the field
(Gavazzi \& Boselli 1999) and for deriving the radial decrease of the \hi\ gas content of galaxies as a function of the distance from the cluster 
centre (Gavazzi et al. 2013). This last measurement is critical for comparison with the predictions of hydrodynamic cosmological simulations.
The untargeted nature of the ALFALFA survey also allowed us to discover \hi\ in extended tails of gas associated with perturbed galaxies (Haynes et al. 2007)
or free floating within the cluster (Haynes 2007; Kent et al. 2007, 2009; Kent 2010). While being sensitive to diffuse gas emission, the ALFALFA data are still relatively shallow and have a
low angular resolution (3.2$\arcmin$). They are thus not optimal for resolved studies. High-quality interferometric data are available 
only for $\sim$ 50 galaxies in the \hi\ 21 cm (VIVA, Chung et al. 2009) and $^{12}$CO(2-1), $^{13}$CO(2-1), and C$^{18}$O(2-1) lines (VERTICO, Brown et al. 2021).
For this reason we are undertaking an untargeted survey of the Virgo cluster in the radio domain using different ground-based facilities:
MeerKAT in the continuum at 1.4 GHz and in the \hi\ line at 21 cm, and Low Frequency Array (LOFAR) at 144 MHz (Edler et al. 2023). These surveys will allow us to study with unprecedented 
sensitivity and angular resolution the radio emission properties of cluster galaxies, adding thus a new 
major milestone in the study of galaxy evolution in rich environments.

Our radio survey of the Virgo cluster (Virgo Cluster multi-Telescope Observations in Radio of Interacting galaxies and AGN: ViCTORIA) will be presented in a future communication (de Gasperin et al. in prep.). Here we present the first results obtained from a pilot observation of $\sim10$ deg$^2$ of Virgo with MeerKAT. With only 42 minutes of integration per pointing, the excellent quality of the MeerKAT data allowed us to discover an extended tail of \hi\ gas associated with the star-forming galaxy NGC 4523, which is undergoing ram-pressure stripping as discussed in this publication. 
The paper is structured as follows: in Sect. 2 we present the MeerKAT observations and the data reduction, in Sect. 3 the multifrequency data 
gathered during other surveys, including unpublished narrow-band H$\alpha$ (VESTIGE) and FUV (ASTROSAT UltraViolet Imaging Telescope (UVIT)) imaging. In Sect. 4 we compare the 
\hi\ data to those available in the literature, and in Sect. 5 we combine them with those available at other frequencies for a multifrequency analysis. 
Discussion and conclusions are given in Sec. 6 and 7, respectively. Throughout this work we assume the galaxy to be at the distance of the cluster (16.5 Mpc, Gavazzi et al. 1999; Mei et al. 2007; Blakeslee et al. 2009; Cantiello et al. 2018).
 
\begin{table}
\caption{Properties of the galaxy NGC 4523 (VCC 1524)}
\label{gal}
{
\[
\begin{tabular}{ccc}
\hline
\noalign{\smallskip}
\hline
Variable                        & Value                                                 & Ref.          \\      
\hline
$Type$                          & SAB(s)m                                               & 1       \\
$cz_{Hel}$                      & 271 km s$^{-1}$                                       & TW       \\
$M_\mathrm{star}^a$             & 1.6$\times$10$^{9}$ M$_{\odot}$                       & 2       \\ 
$P.A.^b$                        & 19$^{\circ}$$\pm$2$^{\circ}$                          & TW      \\
$incl.^b$                       & 32$^{\circ}$$\pm$4$^{\circ}$                          & TW      \\
$M(HI)$                         & 1.46$\times$10$^{9}$ M$_{\odot}$                      & TW       \\ 
$HI-def$                        & 0.39                                                  & TW       \\ 
$R_\mathrm{e}(g)$               & 3.33 kpc                                              & 3       \\              
$Distance$                      & 16.5 Mpc                                              & 4,5,6,7 \\
$Proj.~ distance~from~M87$      & 0.83 Mpc, 0.85 $r/r_\mathrm{200}^c$                   & TW       \\
log$f(H\alpha+[NII])^d$         & -12.22$\pm$0.02  erg s$^{-1}$ cm$^{-2}$               & 2       \\
$SFR^a$                         & 0.08  M$_{\odot}$yr$^{-1}$                            & 2       \\
\noalign{\smallskip}
\hline
\end{tabular}
\]
References: 1) NED; 2) Boselli et al. (2023); 3) Ferrarese, private comm.; 4) Mei et al. (2007); 5) Gavazzi et al. (1999); 6) Blakeslee et al. (2009); 
7) Cantiello et al. (2018); TW) this work.\\
Notes:  All quantities are scaled to a distance of 16.5 Mpc, the mean distance of the cluster, for a fair comparison with other VESTIGE works. The 
uncertainty on the distance can be assumed to be $\sim$ 1 Mpc, the virial radius of the Virgo cluster. \\
a) $M_\mathrm{star}$ and $SFR$ are derived assuming a Chabrier (2003) IMF and the Calzetti et al. (2010) calibration.
b) derived from the kinematics of the H{\sc i} gas.
c) assuming $r_\mathrm{200}(Virgo)$ = 0.974 Mpc (Boselli et al. 2022).
d) corrected for Galactic attenuation. }
\end{table}

\section{MeerKAT observations and data reduction}

ViCTORIA pilot observations of the Virgo cluster were carried out using MeerKAT (Camilo et al. 2018; Mauch et al. 2020)
 on September 10 and 12, 2022. These observations have been programmed to test the feasibility of a large untargeted radio survey at 20 cm covering the a large fraction of the Virgo cluster, planned to be carried out during 2023 ($\sim60$ deg$^2$, with a total of 125 h allocated; PI F. de Gasperin). The observations are part of the ViCTORIA project. The project includes data from MeerKAT and from the low- and high-band systems of LOFAR (Elder et al. 2023), and it will greatly increase the sensitivity and resolution of the wide-field coverage of Virgo in radio continuum between 42-1700\,MHz and in the 21\,cm line. Polarization data from MeerKAT observations will also be obtained. 

The present pilot observations concern 10 partially overlapping fields covering a L-shaped pattern centred on M87 and stretching towards both the north and east. The data were taken with the MeerKAT 32k mode, covering the frequency range 856 - 1712 MHz with 32,768 channels of width 26.123 kHz. We split the 10 fields in 2 groups of 5. Each group is observed in a single $\sim 4.5$ hour-long MeerKAT observing session, during which we cycle 9 times through the 5 fields integrating for 280 sec on each field at each visit. We observe a phase calibrator for 2 min in between 5-field cycles (i.e., every $\sim23$ min) and observe the bandpass and polarisation calibrators for 10 min each at the end of the observing session. The total integration time is 42 min for each of our Virgo fields, which thanks to the excellent  MeerKAT \it uv \rm coverage and to our observing strategy ensures that all the relevant angular scales are fully covered for each field.

For the purpose of this paper, we reduce the MeerKAT pilot data using the Containerized Automated Radio Astronomy Calibration (CARACal) pipeline (J{\'o}zsa et~al. 2020). We adopt an identical data reduction strategy as in Serra et al. (2023), and refer to that paper for a detailed description. Here we summarise the main data reduction steps and highlight any differences compared to that paper. Given that our focus is on \hi\ science, we only process a subset of the MeerKAT data (HH and VV correlations only), focusing on the frequency range 1395.48 - 1437.22 MHz with $2\times$ binning in frequency. This results in 800 channels with a width of 52.246 kHz, which corresponds to $\sim 11.0$ \kms\ for \hi\ at redshift $z = 0$.

Following automated flagging of radio frequency interference (RFI) we perform standard cross-calibration. This includes solving for antenna-based, time-independent delays and bandpass, and for antenna-based, frequency-independent, time-dependent complex gains (one solution every 23 min). We apply these calibration terms to the target fields, and flag RFI in the resulting visibilities. We then perform standard frequency-independent, phase-only self-calibration on each field independently, adopting a 2 min solution interval. The imaging-calibration iterations are automated, with a continuum source-finding strategy fine-tuned to obtain a model of M87 of sufficient quality for our goal of obtaining good \hi\ cubes. Science-quality radio continuum images are obtained with a separate data reduction using a wider frequency band, and will be described in a separate paper (Edler et al. in prep.). For each field, we subtract the radio continuum emission in two steps: \it i) \rm we Fourier transform and subtract the continuum clean components from the self-calibrated visibilities; and \it ii) \rm we fit 3$^\mathrm{rd}$-order polynomials to the visibility spectra, and subtract the result from the data. The latter step is performed together with Doppler correction to a common barycentric frequency grid. We image the resulting continuum-subtracted visibilities with \it i) \rm Brigg's \it robust \rm $=-0.5$ and $6''$ \it uv \rm tapering, and \it ii) \rm Brigg's \it robust \rm $=0$ and $15''$ \it uv \rm tapering. This results in \hi\ cubes with a resolution of $12'' \times 21''$ ($PA =170$ deg) and $27'' \times 39''$ ($PA =178$ deg), respectively, and typical noise level $\sim0.65$ \mJybeam\ per 11-\kms-wide channel for each field and at both resolutions --- within $\sim10\%$ of the expected value. The 3 fields closest to M87 have higher noise. Finally, we mosaic the \hi\ cubes. Due to the limited overlap between the fields included in this pilot observations, the noise in the mosaic does not improve significantly over that of the single fields. We use the Source Finding Application (\texttt{SoFiA}, Serra et al. 2015; Westmeier et al. 2021) to detect and parameterise \hi\ sources in the mosaic. The uncertainty on the derived parameters is a combination of statistical uncertainties and a $\sim10\%$ uncertainty on the flux scale.

One of our \hi\ detections is NGC~4523, the target of this work,  located at a projected angular distance of 2.87$^{\circ}$ from M87 in the north direction. The noise of the HI mosaic cubes at the position of this galaxy is 0.75 mJy beam$^{-1}$ at both resolutions of $12''\times21''$ and $27''\times39''$. This is slightly worse than the noise level mentioned above for the single fields because none of the observed fields is centred on this galaxy, combined with the limited overlap between adjacent pointings in our MeerKAT pilot data. The corresponding $3 \sigma$ column density sensitivity is $1.6\times10^{20}$ cm$^{-2}$ at high resolution and $3.9\times10^{19}$ cm$^{-2}$ at low resolution assuming a line width of 25 km s$^{-1}$. During the following analysis we will use both the low- and the high-resolution data cubes, the former to study the extended low column density features such as the extended \hi\ gas tail, the latter to study the gas kinematics in the inner regions minimising beam smearing effects.



\section{Multifrequency data}

The \hi\ MeerKAT data presented in this work are combined with those available at other frequencies to study the perturbing mechanism and its effect on
NGC 4523. These include narrow-band H$\alpha$ imaging, sensitive to the distribution of the young stellar population (age $\lesssim$ 10 Myr, 
Kennicutt 1998, Boselli et al. 2009) and to the presence of diffuse ionised gas possibly removed during the interaction of the galaxy with the surrounding environment
(e.g. Gavazzi et al. 2001, 2018; Yoshida et al. 2002, 2012; Yagi et al. 2007, 2010, 2017; Sun et al. 2007, 2010; 
Fossati et al. 2012, 2018; Boselli et al. 2016, 2018b; Poggianti et al. 2017), UV data, sensitive to the distribution of
young stars (age $\sim$ 100 Myr, Kennicutt 1998, Boselli et al. 2009), and optical imaging tracing the distribution of the bulk of the stellar population.

\subsection{Optical imaging}

High quality optical images of NGC 4523 in the $u,g,i,z$ filters are  available thanks to NGVS, a broad-band imaging survey of the entire Virgo
cluster carried out at the Canada France Hawaii Telescope (CFHT) with MegaCam (Ferrarese et al. 2012). These data have been gathered using a specific
observing strategy and reduced with the Elixir LSB pipeline (Ferrarese et al. 2012) especially tailored to detect possible extended, low surface brightness structures 
(shells, tidal tails) possibly produced during the interaction of galaxies with their surrounding environment (Duc et al. 2015). Thanks to this 
optimised data analysis, NGVS is sensitive to extended features of surface brightness $\mu(g)$ $\simeq$ 29 mag arcsec$^{-2}$. 
The optical data will be analysed in Sec. 5.1.

\subsection{VESTIGE narrow-band H$\alpha$ imaging}

High quality narrow-band H$\alpha$ imaging data are available thanks to VESTIGE, a deep untargeted survey of the Virgo cluster
carried out with MegaCam at the CFHT (Boselli et al. 2018a). NGC 4523 was observed in the narrow-band H$\alpha$ filter 
MP9603 ($\lambda_c$ = 6591 \AA; $\Delta\lambda$ = 106 \AA) with an integration time of 2~h, and 12 minutes in the broad-band $r$-filter
to secure the subtraction of the stellar continuum (Boselli et al. 2019). The camera has a pixel scale of 0.187 arcsec pixel$^{-1}$.
The data were reduced using Elixir LSB (Ferrarese et al. 2012), a specific procedure especially tailored to detect low surface brightness and extended features such as those produced during the interaction of galaxies with the surrounding environment.
The data were reduced as described in Boselli et al. (2018a), securing an accurate astrometric and photometric calibration of the data (few \%\ uncertainty).
The survey has a sensitivity of $f(H\alpha)$ $\simeq$ 4 $\times$ 10$^{-17}$ 
erg s$^{-1}$ cm$^{-2}$ (5$\sigma$) for point sources and $\Sigma(H\alpha)$ $\simeq$ 2 $\times$ 10$^{-18}$ erg s$^{-1}$ cm$^{-2}$ arcsec$^{-2}$
(1$\sigma$ after smoothing the data to $\sim$ 3\arcsec\ resolution) for extended sources, and has been carried out under excellent seeing conditions 
($FWHM$=0.76\arcsec\ $\pm$ 0.07\arcsec). It is thus perfectly suited to detect star-forming regions of luminosity 
$L(H\alpha)$ $\simeq$ 10$^{36.5}$ erg s$^{-1}$.\footnote{These luminosities correspond to star formation rates of \it SFR \rm $\geq$ 2 $\times$ 10$^{-5}$ M$_{\odot}$ yr$^{-1}$ assuming a standard calibration. We recall, however, that at these regimes the conversion of H$\alpha$ luminosities into star formation rates is not always possible because several conditions (stationarity of the star formation activity, stochastic sampling of the IMF, etc..., see Boselli et al. 2023) are not necessary satisfied.}.
These extremely low H$\alpha$ luminosities are comparable to those emitted by a single early-B star, among the young stars the one with  
the lowest mass and temperature able to ionise the gas (Sternberg et al. 2003).
The H$\alpha$ data will be analysed in Sec. 5.3. 

\subsection{ASTROSAT/UVIT and GALEX UV imaging}

We also compare the \hi\ data to UV data obtained during the ASTROSAT/UVIT legacy survey of the Virgo cluster (PI. A. Boselli) and the
GUViCS (A GALEX Ultraviolet Virgo Cluster Survey) survey (Boselli et al. 2011). ASTROSAT/UVIT (Agrawal et al. 2006; Tandon et al. 2020) 
imaging data of NGC 4523 were obtained thanks to a pointed observation of the galaxy in the FUV filter BaF2 ($\lambda_c$ = 1541 \AA; $\Delta\lambda$ = 380 \AA),
within the field of view of the instrument ($\simeq$ 28$\arcmin$) with an angular resolution of $\simeq$ 1.5$\arcsec$.
The observations were carried out with an integration time of 3635 s, reaching a typical surface brightness of $\mu(FUV)$ $\simeq$ 25.9 AB mag arcsec$^{-2}$.
The data were reduced as described in Tandon et al. (2020) using a zero point of $z_p$ = 17.771 mag. The astrometry of the field was checked against the accurate
NGVS imaging data (Ferrarese et al. 2012).
The galaxy has been also observed during the GUViCS survey with GALEX in the NUV band ($\lambda_c$ = 2316 \AA, $\Delta\lambda$ = 1060 \AA), with an integration time of 1629 s, able to reach
a surface brightness limit of $\simeq$ 27 AB mag arcsec$^{-2}$. The angular resolution of the NUV band image is of $\simeq$ 5$\arcsec$. The FUV UVIT and NUV GALEX images of the galaxy will be analysed in Sec. 5.3.

\section{Derived parameters}


The MeerKAT data are used to extract the \hi\ parameters of the galaxy: the integrated H{\sc i} flux $SHI = 23.4\pm0.2$ Jy km s$^{-1}$, the flux-weighted recessional velocity $VHI = 271\pm 6$ km s$^{-1}$ (see Sec. \ref{sec:gaskin}), and the line width of the integrated spectrum measured at 50\%\ of the peak $WHI_{50} = 121\pm6$ km s$^{-1}$.
These numbers are in line with those derived
in the literature (see Table \ref{letteratura}). 

\begin{figure*}
\centering
\includegraphics[width=0.48\textwidth]{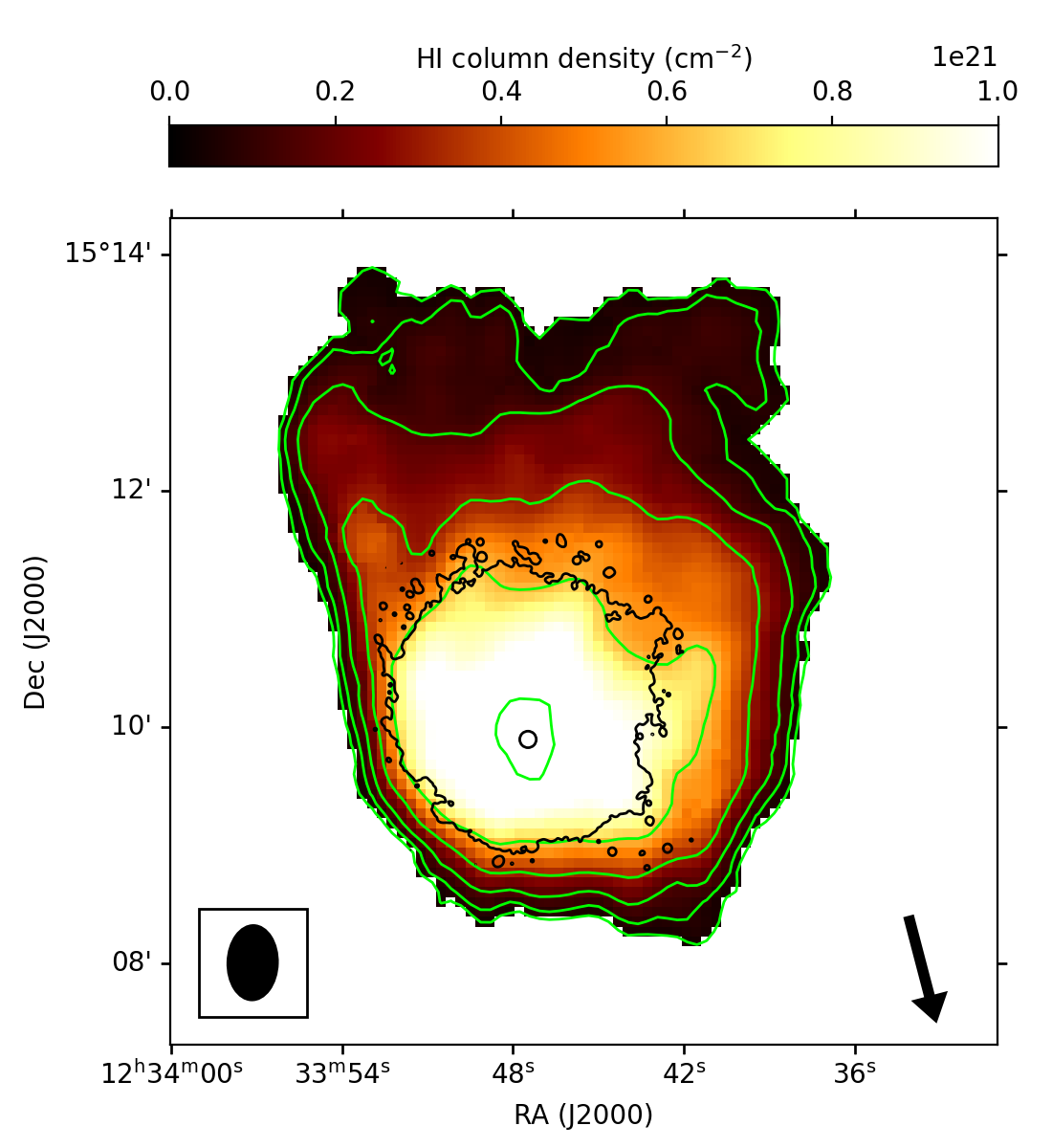}
\includegraphics[width=0.48\textwidth]{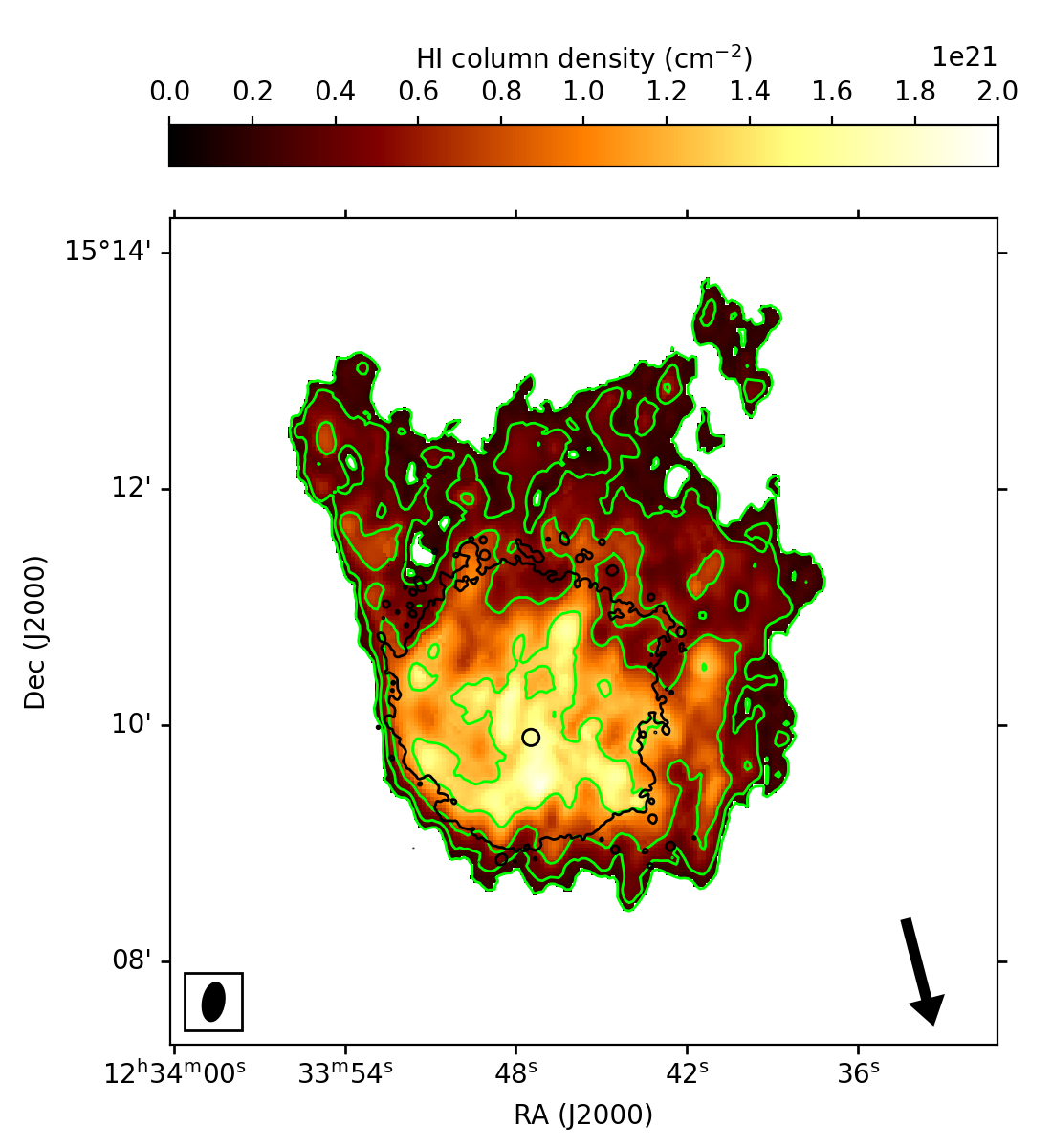}\\
\caption{Low-resolution (27\arcsec $\times$ 39\arcsec; left) and high-resolution (12\arcsec $\times$ 21\arcsec; right) \hi\ gas distribution of NGC 4523. Green contours are at column densities of $N(HI) = 2^n \times 3.9 \times 10^{19}$ cm$^{-2}$ with $n$=0,1,..5. 
The black contour shows the $i$-band isophote at 24.5 mag arcsec$^{-2}$.
The open dot shows the kinematic centre of the galaxy, the ellipse indicates the size of the beam, the arrow the direction of the cluster centre (M87), located at 0.83 Mpc projected distance.
}
\label{VCC1524_HI_lr}%
\end{figure*}

The $SHI$ flux derived in this work can be converted in a \hi\ gas mass
of $M_{HI}$ = 1.50 $\times$ 10$^9$ M$_{\odot}$ using eq. 9 of Haynes \& Giovanelli (1984), and used to estimate a \hi\--deficiency parameter $HI-def$ = 0.39 using the recent calibration of Cattorini et al. (2023). This calibration, which is based on the optical diameter, is optimal for the present purpose since it is the only one which extends to the dwarf systems regime. This \hi\--deficiency is slightly larger than the typical dispersion of the scaling relation used to define this parameter in galaxies living in low-density environments ($\sigma$ $\sim$ 0.3, Cattorini et al. 2023). 
\rm 
Of the measured \hi\ content, 67\%\ (9.8 $\times$ 10$^8$ M$_{\odot}$) is located on the galaxy, 33\%\ (4.8 $\times$ 10$^8$ M$_{\odot}$) in the tail outside the stellar disc at the 24.5 mag arcsec$^{-2}$ $i$-band isophote, as shown in Fig. \ref{VCC1524_HI_lr}.

\begin{table}
\caption{HI measurements of NGC 4523 in the literature }
\label{letteratura}
{
\[
\begin{tabular}{lllcc}
\hline
\noalign{\smallskip}
\hline
$VHI$         & $WHI_{50}$ & $SHI$            & Telescope    & Ref.   \\ 
km s$^{-1}$   & km s$^{-1}$& Jy km s$^{-1}$   &            &      \\  
\hline
+271$\pm$6   & 121$\pm6$ & 23.4$\pm$0.2 & MeerKAT   & T.W.  \\
+258          & 124$\pm$3     & 24.30$\pm$0.08 & Arecibo & 1    \\
-             & -             & 24.42$\pm$0.1  & Arecibo & 2    \\
+262          & 137           & 25.9           & Arecibo & 3    \\
+260$\pm$2    & 120           & 20.54          & Effelsberg & 4 \\
+259$\pm$2    & 120$\pm$6     & 19.54$\pm$0.20 & -          & 5 \\
\noalign{\smallskip}
\hline
\end{tabular}
\]
References: T.W. : this work; 1) Haynes et al. (2018); 2) Hoffman et al. (2019), 3) Hoffman et al. (1987); 4) Huchtmeier et al. (2000); 5) Bottinelli et al. (1990).}
\end{table}







\section{Analysis}

\begin{figure}
\centering
\includegraphics[width=0.49\textwidth]{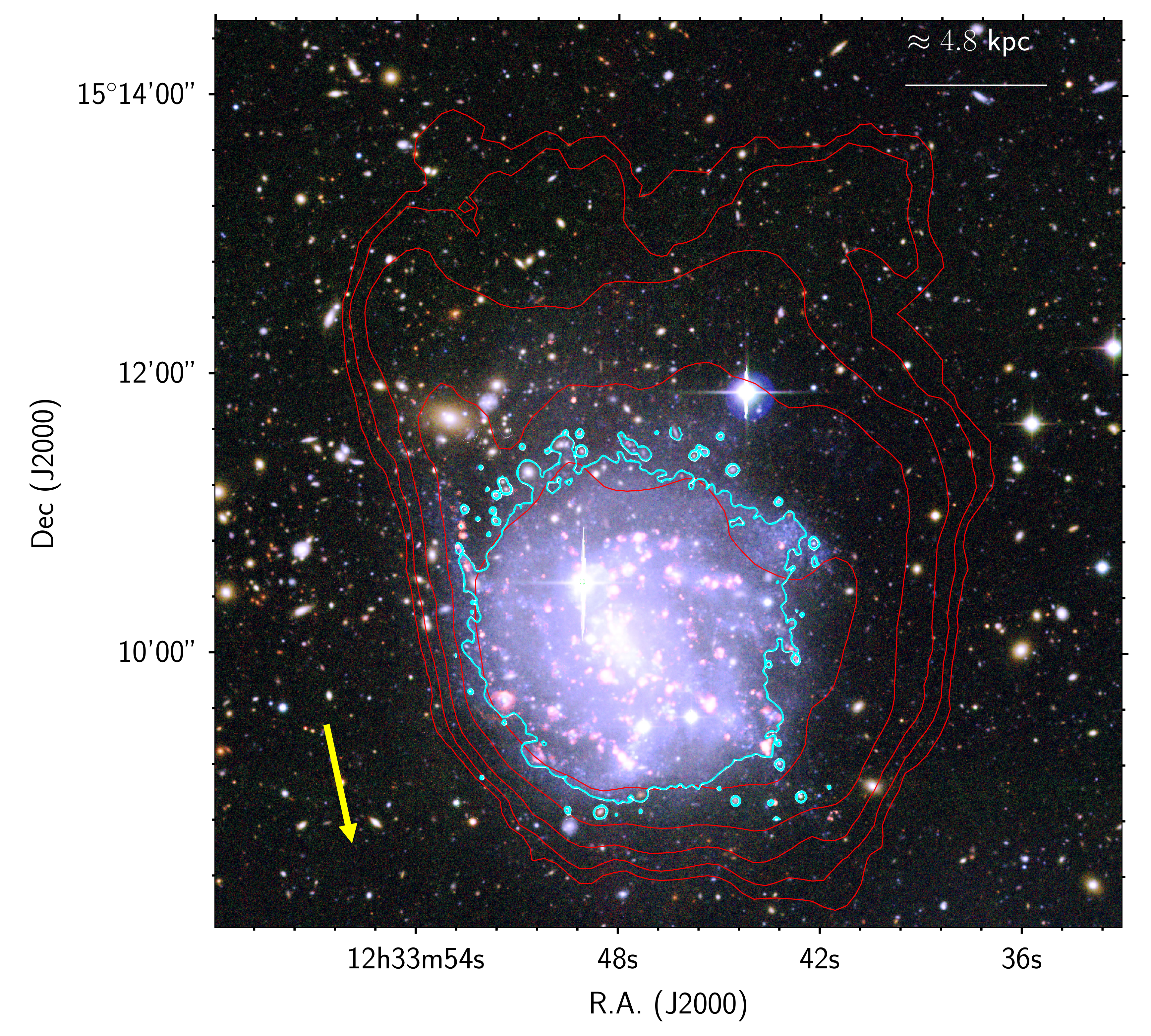}
\caption{Pseudo-colour image of the galaxy NGC 4523 obtained combining the NGVS (Ferrarese et al. 2012) and VESTIGE
(Boselli et al. 2018a) optical $u$ and $g$ in the blue channel, the $r$ and 
the narrow-band H$\alpha$ in the green, and the $i$ and the continuum-subtracted H$\alpha$ in the red.
Red contours are shown at column densities of $N(HI) = 2^n \times 3.9 \times 10^{19}$ cm$^{-2}$ with $n$=0,1,..5. 
The cyan contours show the $i$-band isophote at 24.5 mag arcsec$^{-2}$. The yellow arrow indicates the position of the cluster centre (M87), located at 0.83 Mpc projected distance. 
}
\label{VCC1524_UGRHINET_HI}%
\end{figure}

\subsection{Gas distribution and origin of the extended \hi\ tail}

Figure \ref{VCC1524_HI_lr} shows the \hi\ gas distribution (moment-0 map) of NGC 4523. The same gas distribution is compared to that of the stars in Fig. \ref{VCC1524_UGRHINET_HI}. At the angular resolution of 27\arcsec$\times$39\arcsec\ the data reach a $3\sigma$ column density of $N(HI)$ = 3.9 $\times$ 10$^{19}$ cm$^{-2}$ for a line width of 25 km s$^{-1}$. The \hi\ is distributed asymmetrically, with an extended tail of low-column-density gas ($N(HI) \lesssim 2.5 \times 10^{20}$ cm$^{-2}$) stretching out to 10 kpc (projected distance) from the edge of the stellar disc measured at the $i$-band 24.5 mag arcsec$^{-2}$ in the north direction. We stress that this 10 kpc length is a lower limit to the extension of the tail because the galaxy has an important velocity component along the line-of-sight (LoS) ($\sim$ 800 km s$^{-1}$) with respect to the cluster ($\sim$ 1040 km s$^{-1}$, Kashibadze et al. 2020). 

The possible presence of similar features in deep optical images is crucial for the identification of the dominant perturbing mechanism. As shown in Fig. \ref{VCC1524_HI_lr}, the deep NGVS frames indicate that stars are also distributed asymmetrically, as often observed in galaxies of this morphological type (SAB(s)m). The $i$-band 24.5 mag arcsec$^{-2}$ isophote has a circular shape, while it gets elliptical ($b/a$ = 0.73, where $a$ and $b$ are the major and minor diameters) and elongated in the north-south direction ($P.A.$ $\sim$ 15$^{\circ}$) at the surface brightness limit of 26.5 mag arcsec$^{-2}$. The northern side is more elongated (147.6$\arcsec$) than the southern one (120.8$\arcsec$) when measured from the galaxy centre. This asymmetric stellar distribution at a low surface brightness limit in the same direction as the HI gas tail suggest that the galaxy might have been perturbed by a gravitational interaction.


There are several arguments suggesting that tidal interactions cannot be the main or only cause of the observed asymmetric distribution of the \hi\ out to large radius, and that ram pressure must play an important role. First of all, the deep optical images do not show any evident structure such as streams, tidal tails, or shells --- which might have formed during a gravitational perturbation (Duc et al. 2015) --- down to a surface brightness limit of $\mu(g)$ $\simeq$ 29 mag arcsec$^{-2}$ across the full extent of the \hi\ tail. Such features cannot be totally ruled out if their surface brightness is below the detection limit of the NGVS and VESTIGE data, as some simulations suggest ($\sim$ 33 mag arcsec$^{-2}$ Mancillas et al. 2019), but observations of a large sample of galaxies gathered with MegaCam at the CFHT rather indicate that tidal features are rare at a surface brightness limit below 27.5 mag arcsec$^{-2}$ (Sola et al. 2022). Furthermore, the presence of the \hi\ tail combined with the relatively large content of atomic gas on the galaxy disc and the lack of ionised gas in the tail, which requires $\gtrsim$ 100 Myr to get mixed with the surrounding hot IGM and change of phase (e.g. Boselli et al. 2022), suggest that the perturbing process is still ongoing, as also suggested by our simulations. It is thus unlikely that the tidal streams, if formed, have had enough time to disappear (optical features are relatively long lived: 0.7-4 Gyr according to Mancillas et al. 2019). Finally, the galaxy is located at the northern periphery of the cluster and is falling into it for the first time. Here the number of massive galaxies which might have recently perturbed NGC 4523 is very limited: IC 800 (VCC 1532), a late-type system of comparable stellar mass, is located at a projected distance $d_{proj}$ $\sim$ 55 kpc to the north, but has a relative LoS velocity with respect to NGC 4523 of $>$ 2000 km s$^{-1}$. 
Assuming as relative distance between the two objects their projected distance (lower limit) we can measure the gravitational acceleration exerted by this perturber on the stellar disc of NGC 4523 and compare it with that keeping the matter linked to its gravitational potential well as indicated in Henriksen \& Byrd 1996 and conclude that an efficient gravitational perturbation with this object is unlikely. This is also the case for the galaxies IC 3522 (VCC 1585) east of NGC 4523 ($d_{proj}$ $\sim$ 75 kpc, with a difference in the LoS velocity of only $\Delta(v)$ $\sim$ 400 km s$^{-1}$ but with a low stellar mass of $M_{star}$ $\simeq$ 8.5 $\times$ 10$^7$ M$_{\odot}$), IC 797 (VCC 1393) west of NGC 4523 ($d_{proj}$ $\sim$ 135 kpc, $\Delta(v)$ $\sim$ 1800 km s$^{-1}$), NGC 4540 to the north-east ($d_{proj}$ $\sim$ 135 kpc, $\Delta(v)$ $\sim$ 1000 km s$^{-1}$), or the two massive galaxies to the south (NGC 4501, $d_{proj}$ $\sim$ 250 kpc, $\Delta(v)$ $\sim$ 2000 km s$^{-1}$, NGC 4548, $d_{proj}$ $\sim$ 230 kpc, $\Delta(v)$ $\sim$ 200 km s$^{-1}$). We recall, however, that in a cluster environment where the velocity dispersion of galaxies is of the order of 1000 km s$^{-1}$ the relative position of the perturbed object and of its perturber can drastically change ($\sim$ 1 kpc Myr$^{-1}$), making the identification of any possible perturber very challenging. 
A further possibility is that the tail is a remnant of gas accreted by NGC 4523 after a minor merging event with a gas-rich system. This picture, however, seems ruled out by the regular velocity field of the gaseous component (see Sec. 5.2) which would keep a perturbed shape if the merging event is recent with respect to the galaxy revolution time ($\lesssim$ 500 Myr). We also remind that merging events in clusters such as Virgo are very unlikely given the high velocity dispersion of galaxies (Makino \& Hut 1997, Boselli \& Gavazzi 2006).

All these arguments suggest that the \hi\ gas tail has formed during a hydrodynamic interaction with the surrounding hot ICM (namely, ram pressure stripping). The diffuse stellar emission associated to the tail (see Fig. \ref{VCC1524_UGRHINET_HI}) has a blue colour indicating that it is composed of stars of age $\lesssim$ 300 Myr and might thus have been formed in the stripped gas (see Sec. 7.3). 

\begin{figure}
\centering
\includegraphics[width=0.49\textwidth]{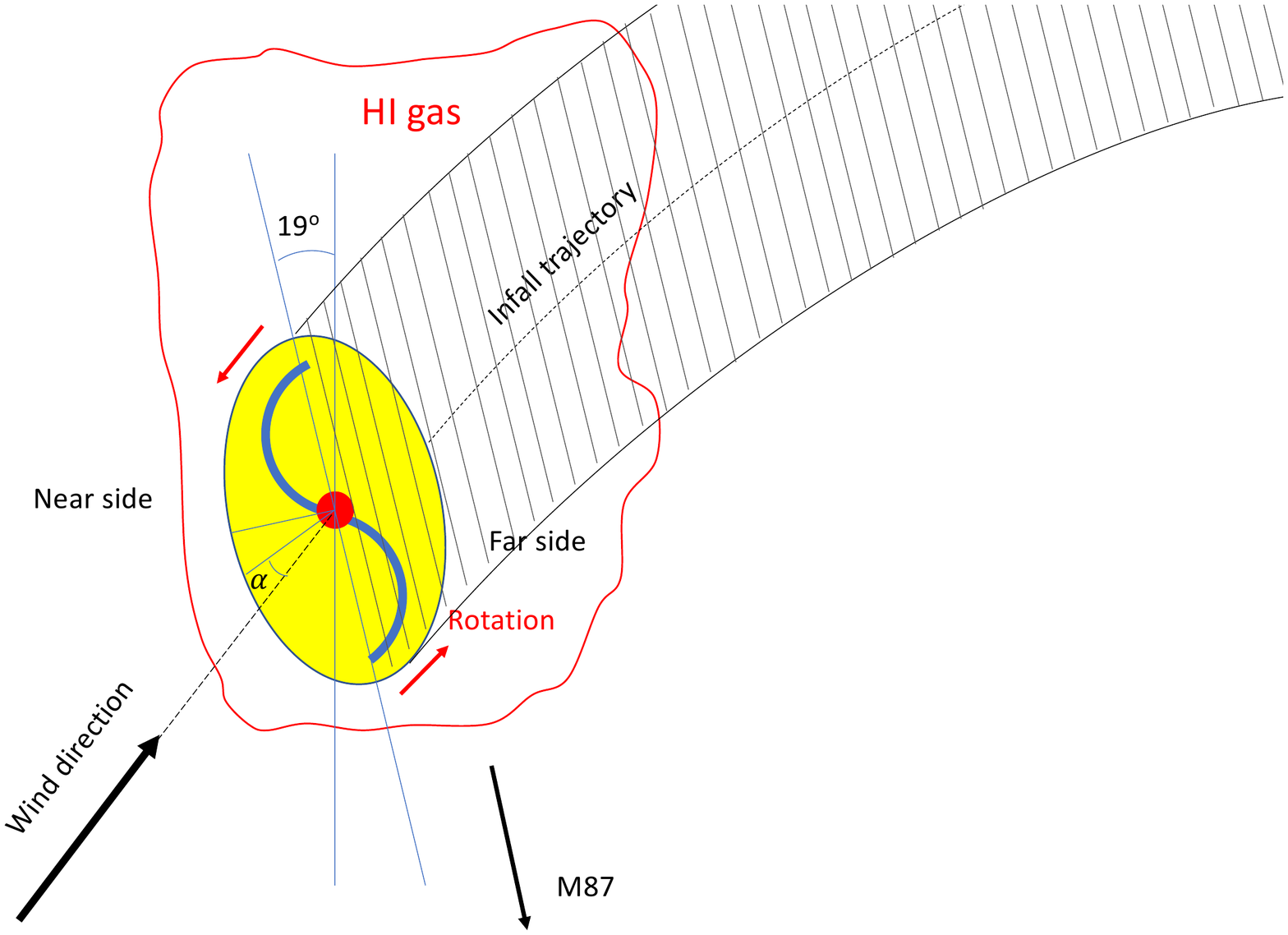}
\caption{Cartoon representing the orbit of the galaxy and its orientation within the cluster. The yellow ellipse represents the stellar disc, the red contour the H{\sc i} distribution, the black shaded region the trajectory of the galaxy during its first infall into the cluster. The position of the cluster centre (M87) and the direction of the wind are indicated by the black arrows. $\alpha$ is the angle between the wind direction and the plane of the stellar disc.
}
\label{cartoon}%
\end{figure}

The projected orientation of the tail, approximately in the opposite direction of the cluster centre, suggests that the galaxy is at its first infall into the cluster. More in detail, the fact that the \hi\ gas column density has a steep gradient at the south-east edge of the stellar disc (Figs. \ref{VCC1524_HI_lr} and \ref{VCC1524_UGRHINET_HI}), opposite to the north-west maximum extension of the tail, suggests that the infall vector has a south-east component on the plane of the sky. Combined with the fact that the galaxy is blue-shifted relative to the cluster (its recessional velocity is $v_{sys}$ = 271 s$^{-1}$ while the mean velocity of the main body of Virgo, cluster A, is $v_{cluster A}$ = 1040 km s$^{-1}$; Kashibadze et al. 2020) this leads us to believe that NGC~4523 is infalling into Virgo from its back, north-western side, as depicted in Fig. \ref{cartoon}. In the phase-space diagram, the galaxy is located in between the first infall and intermediate regions defined by the simulations of Rhee et al. (2017) (see Fig. 12 in Boselli et al. 2023).
At the projected distance from the cluster centre of 0.83 Mpc the electron density of the hot ICM is $n_e$ $\sim$ 2$\times$10$^{-4}$ cm$^{-3}$ (Boselli et al. 2022). Assuming that the galaxy has a radius of $R_{gal}$ $\simeq$ 8 kpc and a rotational velocity $v_{flat}$ $\simeq$ 100 km s$^{-1}$, we can use the relation (Fujita \& Nagashima 1999):

\begin{equation}
{\rho_{ICM}V^2_{\perp} > \frac{v_{flat}^2\Sigma_{HI}}{R_{gal}}}
\end{equation}

\noindent
to estimate the velocity of the galaxy with respect to the ICM necessary to strip atomic hydrogen with a typical column density of $\Sigma_{HI}$ $\simeq$ 5 M$_{\odot}$ pc$^{-2}$. This gives $V_{\perp}$ $\simeq$ 1000 km s$^{-1}$, and suggests that the velocity of the galaxy on the plane of the sky is $V_{sky}$ $\sim$ 600 km s$^{-1}$.

 The shape and the orientation of the spiral arms indicate that the galaxy is rotating counterclockwise on the plane of the sky by assuming trailing arms. Given that the northern side is blue-shifted and the southern side red-shifted, this indicates that the eastern side of the disc is the one closer to the observer (see Fig. \ref{cartoon}). Therefore, as \hi\ is pushed away from the disc towards north-west by ram pressure, it continues rotating counterclockwise until, on the eastern side of the galaxy, its motion is directed against the ram-pressure wind. This explains the orientation of the low column density \hi\ tail to the north as well as the compressed \hi\ contours on the north-eastern side of the tail. It should also lead to a piling up of \hi\ on the northern side of the galaxy. In Sec. 5.2 we will see what consequences this has on the \hi\ kinematics.

\subsection{Gas kinematics}
\label{sec:gaskin}

\begin{figure}
\centering
\includegraphics[width=0.35\textwidth]{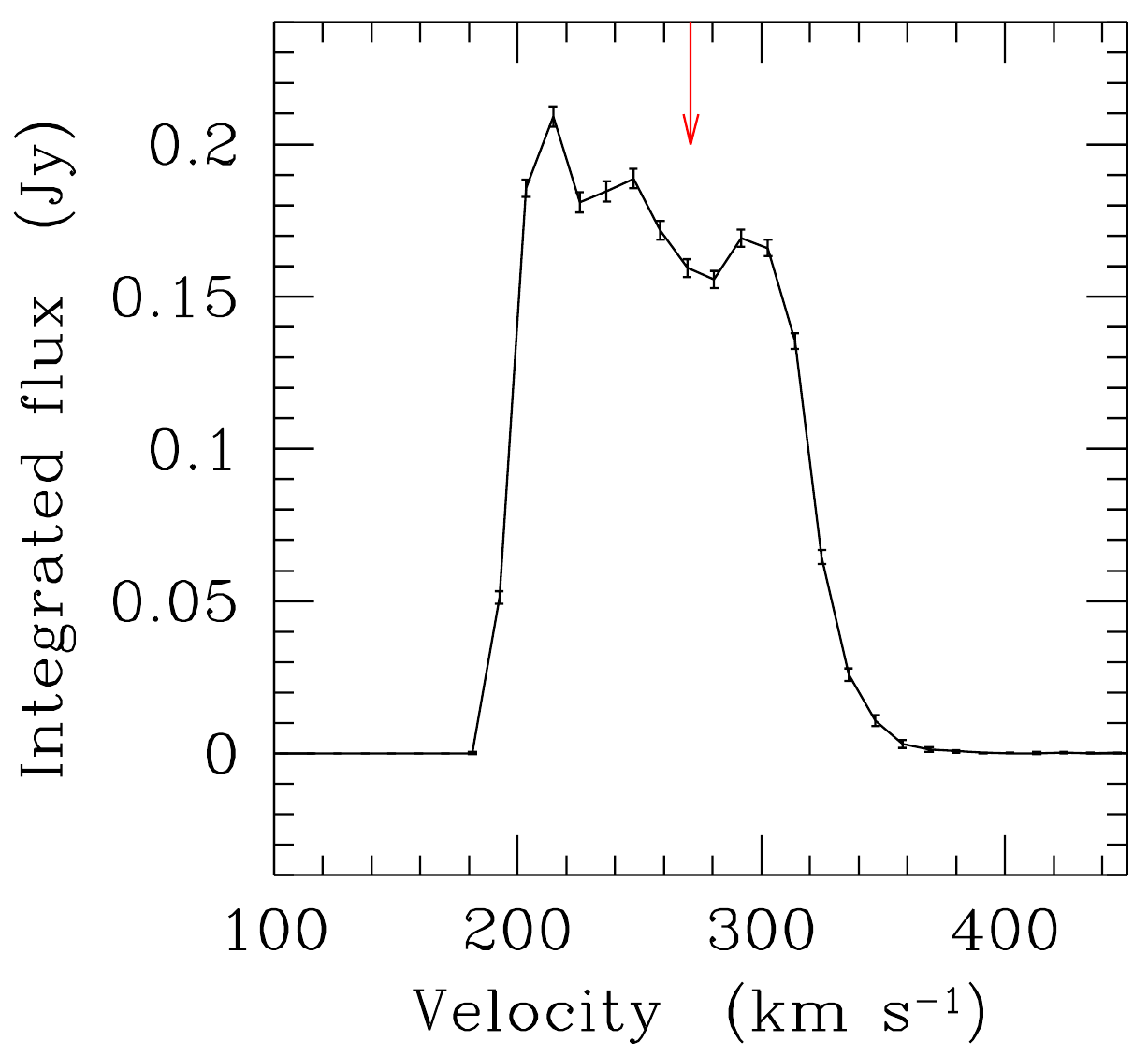}
\caption{Integrated \hi\ line profile of NGC 4523.
The red arrow indicates the systemic heliocentric velocity
(271 \kms).}
\label{profilo}%
\end{figure}

\begin{figure*}
\centering
\includegraphics[width=0.48\textwidth]{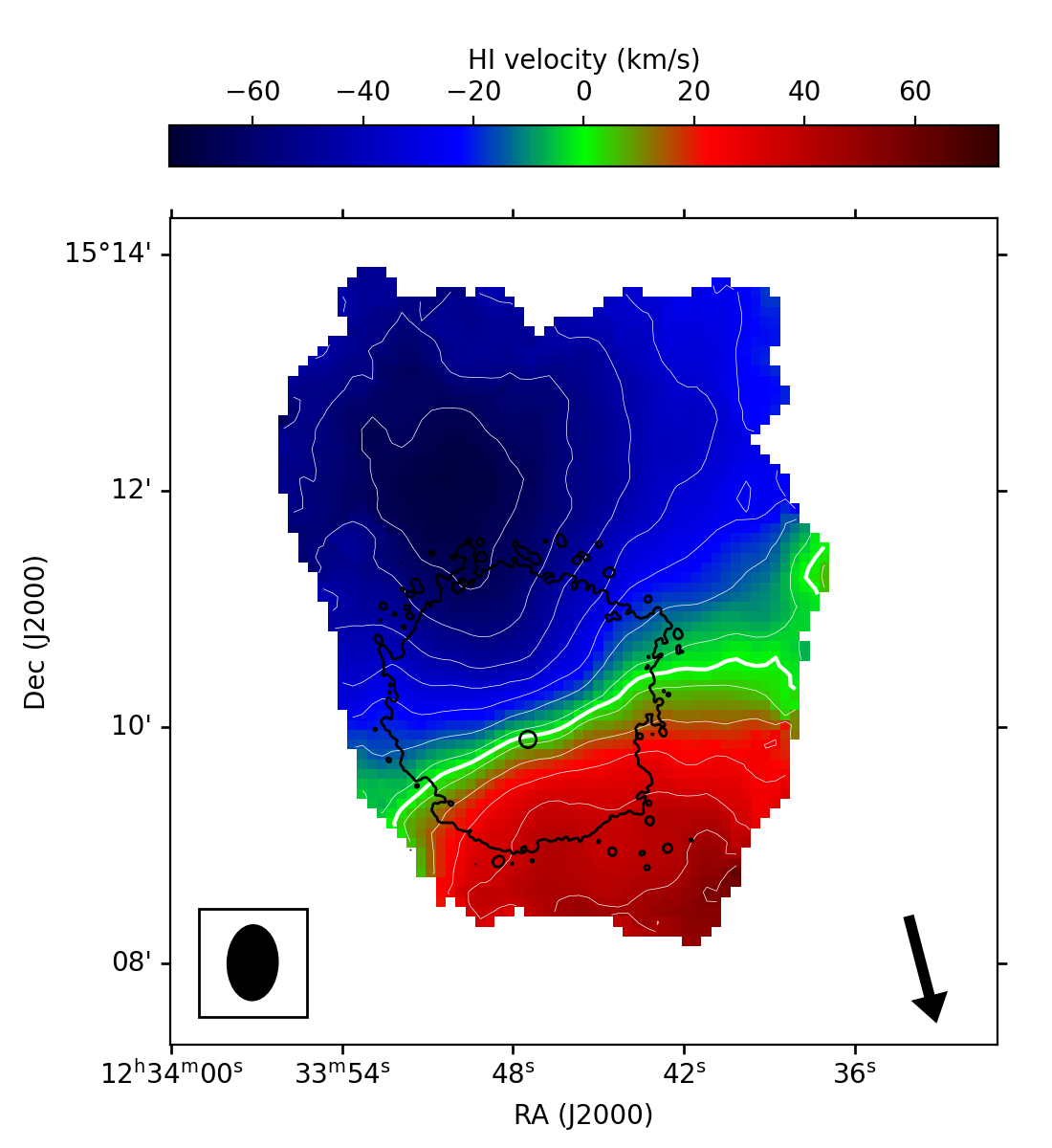}
\includegraphics[width=0.48\textwidth]{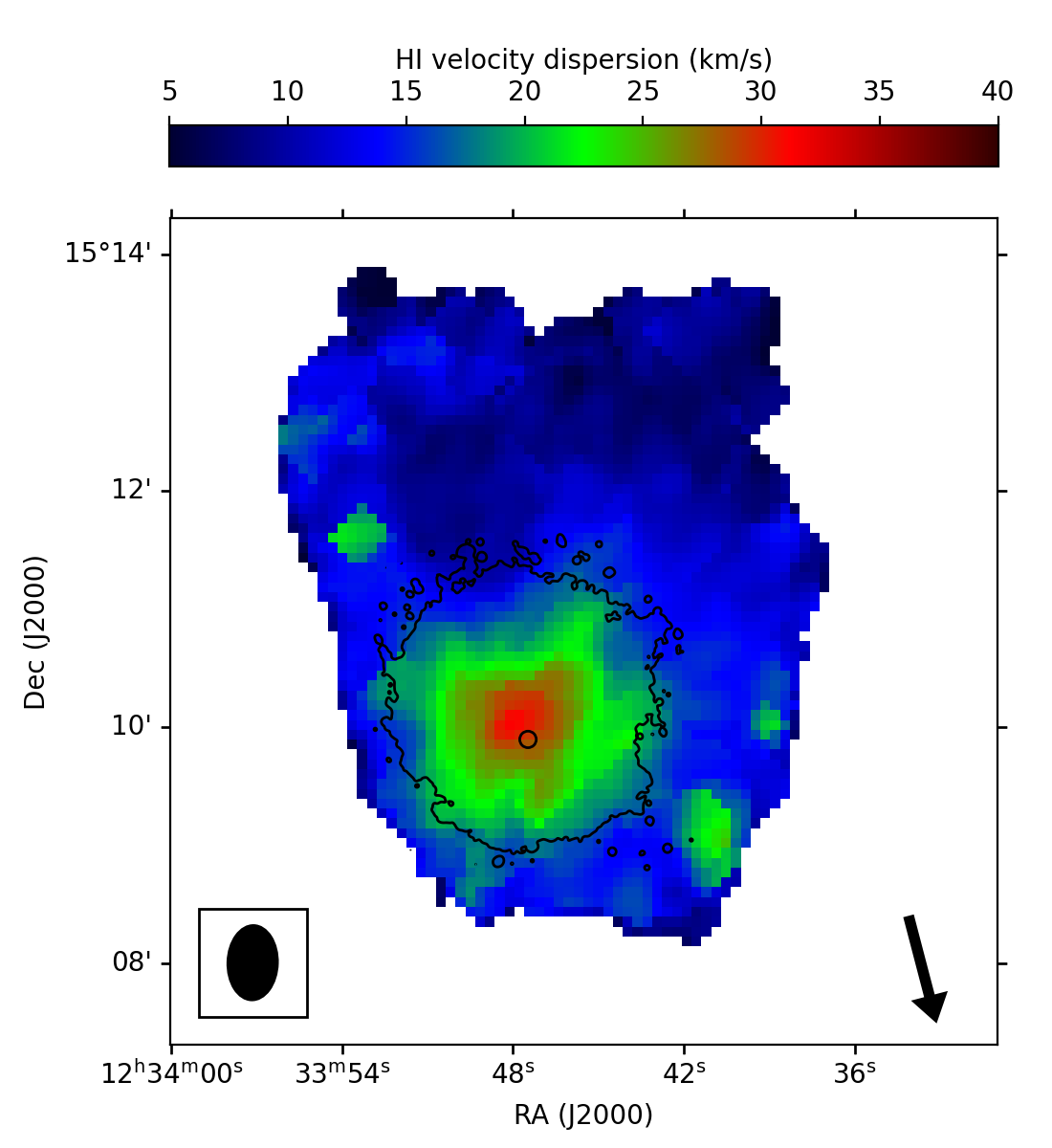}\\
\includegraphics[width=0.48\textwidth]{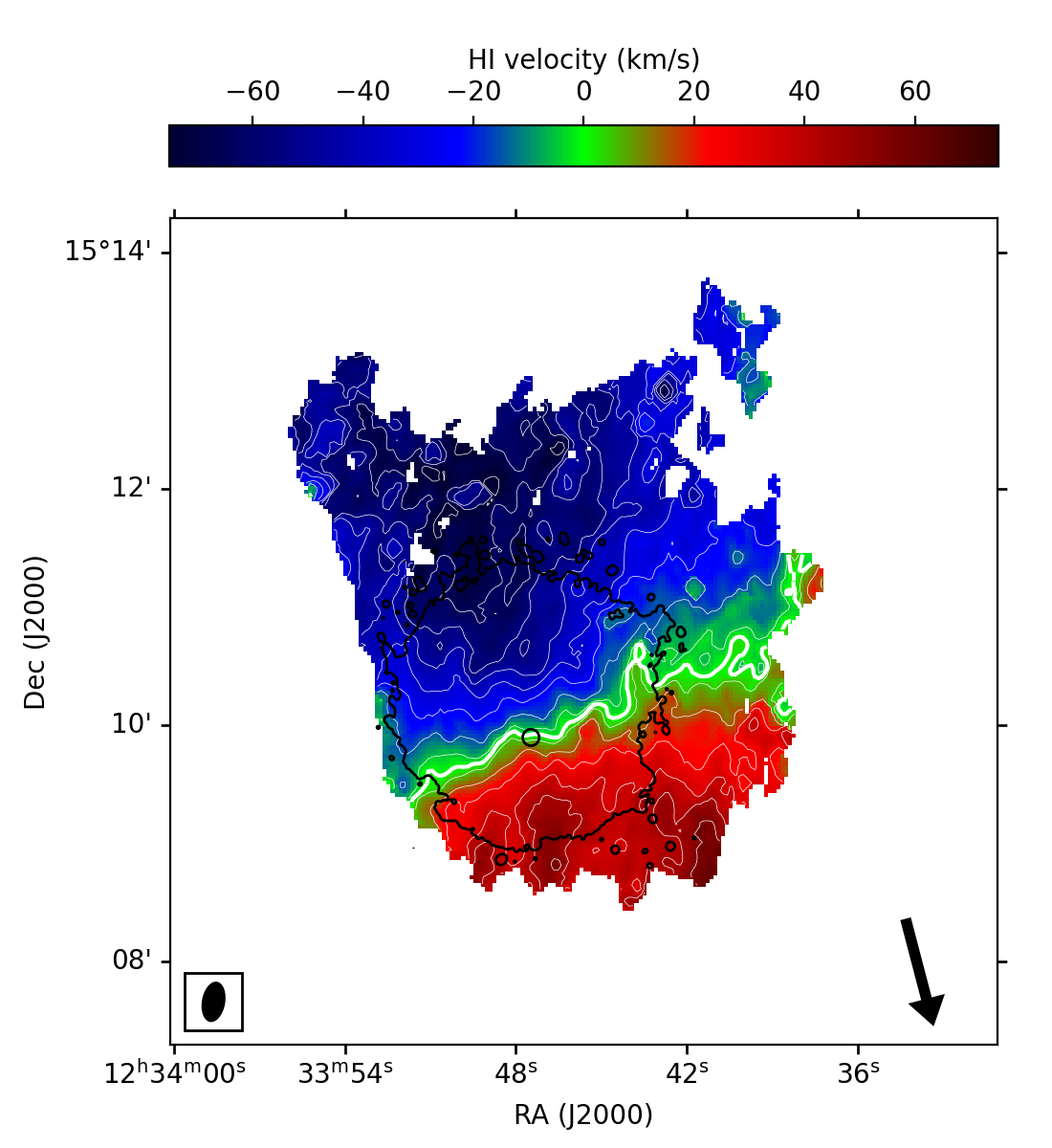}
\includegraphics[width=0.48\textwidth]{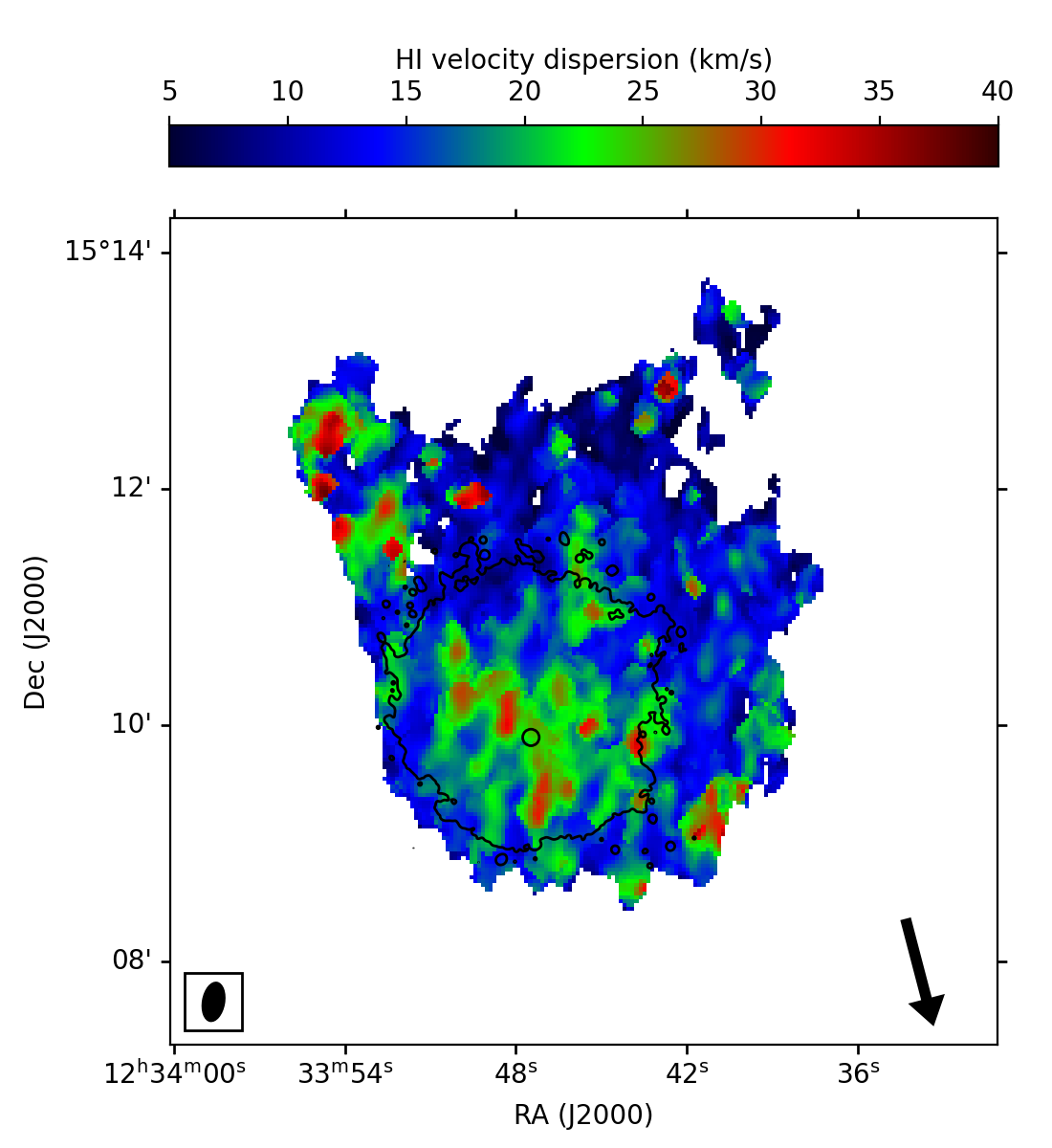}\\
\caption{Low-resolution (27\arcsec $\times$ 39\arcsec; upper panels) and high-resolution (12\arcsec $\times$ 21\arcsec; lower panels) \hi\ gas velocity field (moment-1 , left) and velocity dispersion (moment 2, right) of NGC 4523. The black contour shows the stellar distribution as traced by the $i$-band 24.5 mag arcsec$^{-2}$ isophote. The open dot show the kinematic centre of the galaxy, the black ellipse indicates the beam size and the arrow shows the direction to the cluster centre (M87). The thick white contour represents the systemic velocity of 271 \kms, and the thinner white contours are drawn at intervals of 10 \kms.
}
\label{Velmap_hr}%
\end{figure*}


\begin{figure*}
\centering
\includegraphics[width=0.48\textwidth]{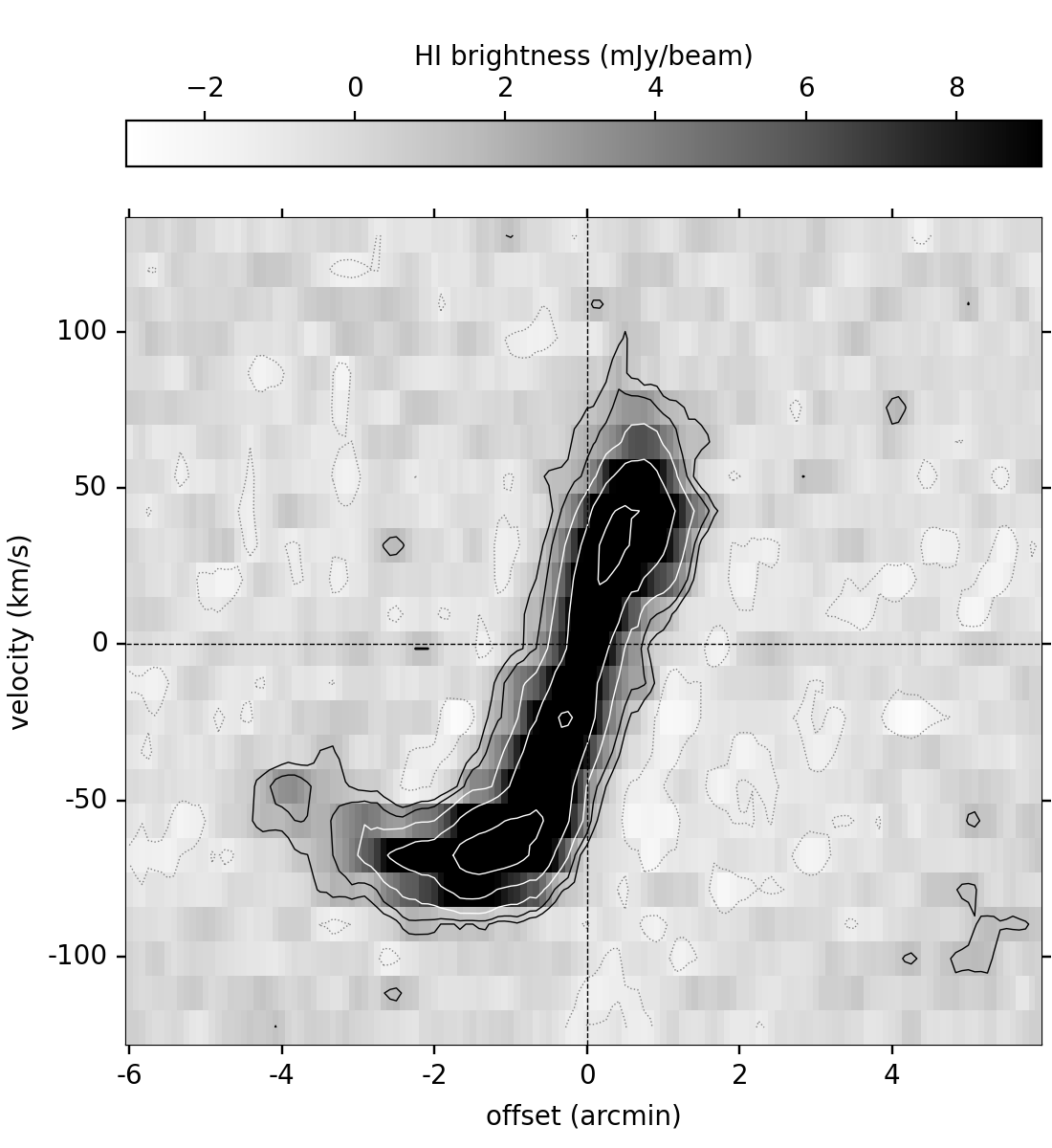}
\includegraphics[width=0.48\textwidth]{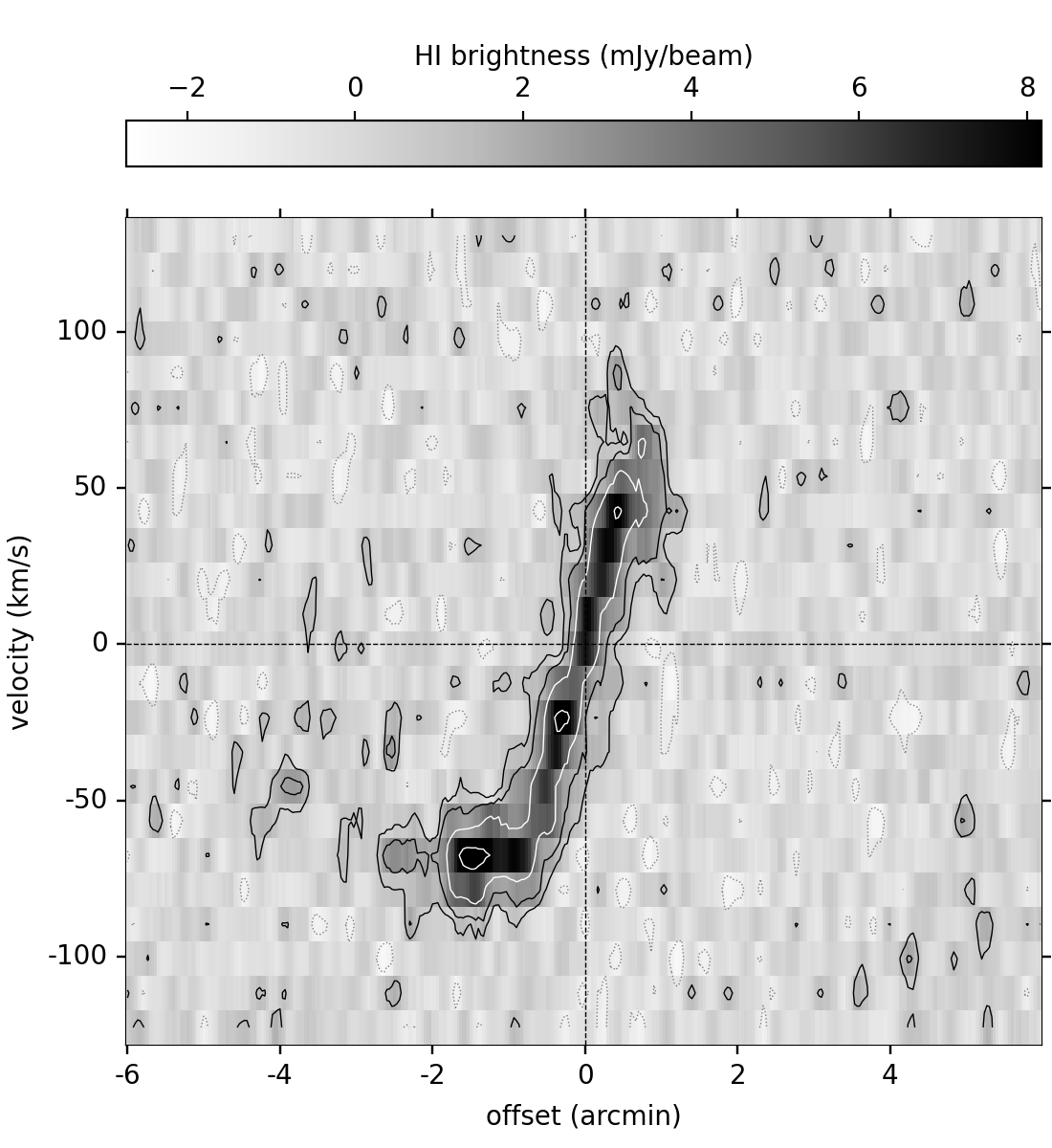}
\caption{Low- (left) and high-resolution (right) \hi\ position-velocity diagram of NGC 4523 derived on the major axis ($P.A.$ = 19.4$^{\circ}$). North is left, South is right.}
\label{pv}%
\end{figure*}

The integrated \hi\ profile of NGC 4523 is fairly asymmetric (Fig. \ref{profilo}). It has significantly more flux below than above systemic, and it is steeper at low velocity near the peak of the emission ($\sim$ 200 km s$^{-1}$) and shallower at higher velocities ($\sim$ 340 km s$^{-1}$). 
The \hi\ velocity field (Fig. \ref{Velmap_hr}, left panels) is also asymmetric due to the presence of the low column density tail to the north. Note that the northern side of the galaxy is blue-shifted compared to systemic. Therefore, the piling-up of gas on the northern side of the galaxy discussed in the previous section explains the steep rise of the integrated spectrum at low velocities (Fig. \ref{profilo}).

The position-velocity diagram drawn along the kinematical major axis ($PA$ = 19$^{\circ}$, Fig \ref{pv}) shows that the gas stripped from the disc and now located in the northern tail decelerates to velocities closer to the systemic velocity of the galaxy; i.e., the \hi\ LoS velocity increases towards systemic with increasing distance from the centre along the tail. 
This is consistent with a red-shifting ram-pressure wind caused by the galaxy's blue-shift relative to the ICM. Because of the orientation of the galaxy relative to the wind, the stripping process occurs mainly edge-on. The gas removed at the south-eastern edge of the disc remains gravitationally bound to the galaxy, while that at the south-western edge gets stripped, producing the extended tail along the LoS in the integrated profile.

To derive the main kinematical parameters of the galaxy and quantify the impact of the external perturbation on the \hi\ kinematics we fitted the high-resolution \hi\ velocity field of Fig. \ref{Velmap_hr} following the 2D fitting method described in Epinat et al. (2008). This method is based on the Levenberg-Marquardt non-linear least-square algorithm and has been upgraded using the MocKinG software\footnote{\url{https://gitlab.lam.fr/bepinat/MocKinG}}. We fitted the \hi\ velocity field, assuming an analytic axisymmetric velocity field purely based on circular motion,  with a solid-body rise in the inner regions and a flat part in the outer regions. The parameters of this modelled velocity field (i.e., the radius where the velocity field changes from solid-body to flat, and the circular velocity of the flat part) are optimised during the fit in minimising the residual velocity field (the difference between the observed and the model velocity fields). All other model parameters included in the fit (central position and velocity, inclination, position angle) are assumed to be independent of radius.
Given that we fit the velocity field by weighting pixels based on their \hi\ flux intensity, the best-fitting parameter values are dominated by the inner galaxy regions, where the effects of the environmental interaction under investigation are minimal. We assessed the robustness of the fit by weighting the pixels using different methods, and found that the output parameters are remarkably stable. 

\begin{figure*}
\centering
\includegraphics[width=0.49\textwidth]{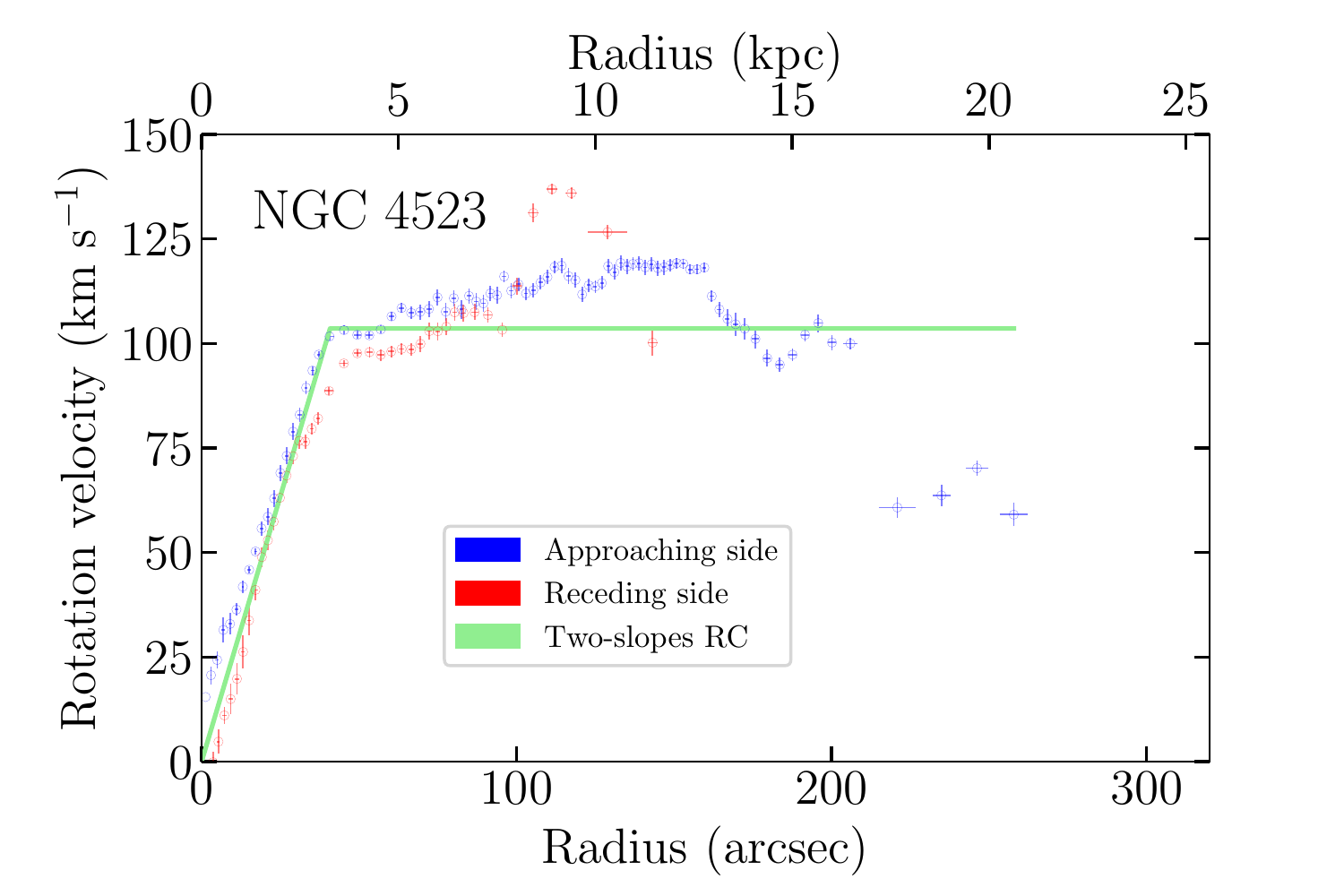}
\includegraphics[width=0.49\textwidth]{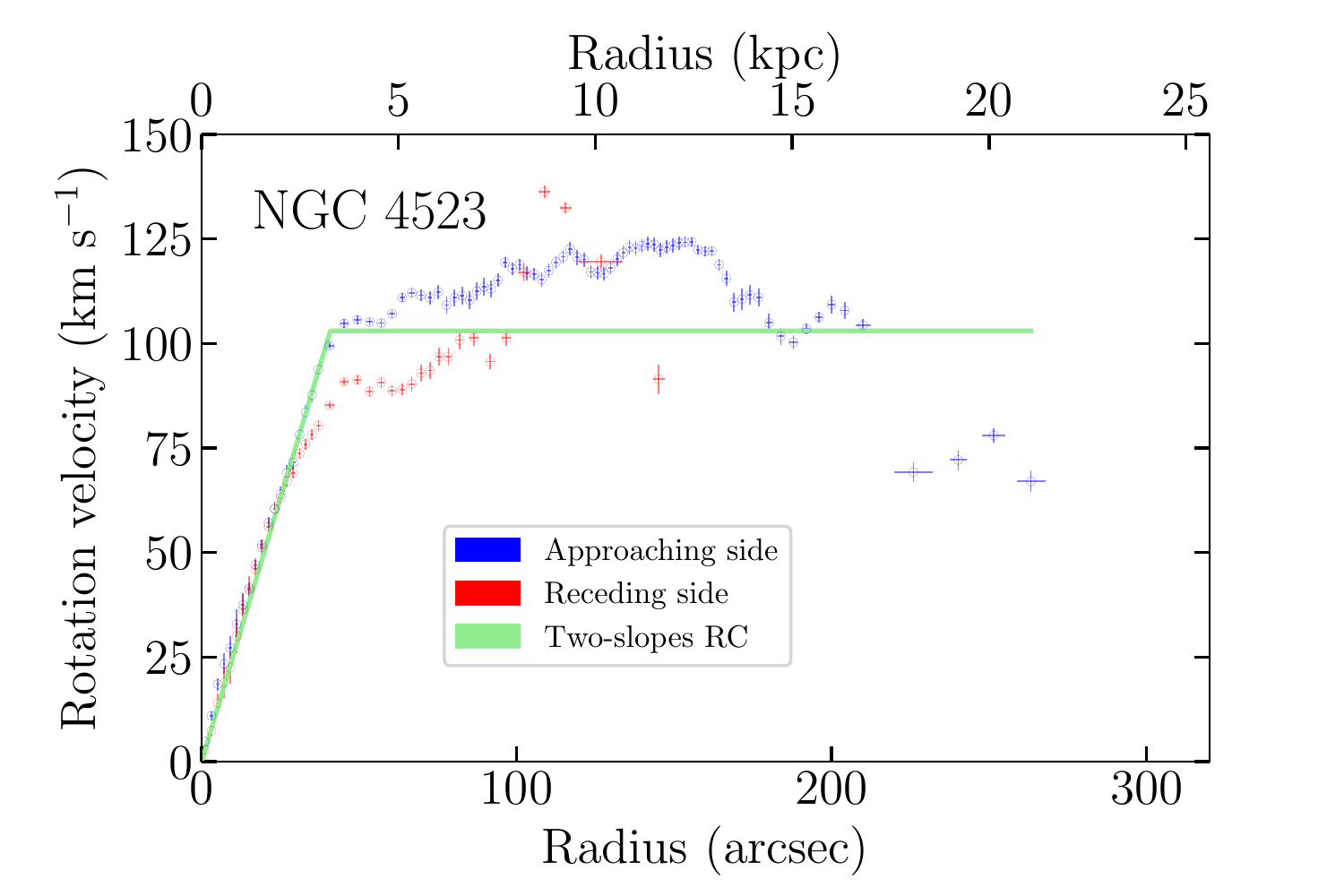}\\
\caption{Rotation curve of NGC 4523 measured from the high-resolution \hi\ velocity field within the modelling approach described in Sec. \ref{sec:gaskin}. Blue and red squares show the approaching and receding sides of the rotation curve, respectively. The green solid line indicates the best-fitting analytic rotation curve (solid body in the inner regions, and flat in the outer parts). Left panel: rotation curve derived leaving all the fitting parameters free. Right panel: rotation curve derived by shifting the position of the kinematic centre by $4.4''$ compared to that in the left panel in order to minimise the difference between the approaching and receding side in the inner regions. }
\label{RC}%
\end{figure*}


Having the observed velocity field and the parameter determined from the fit of the velocity field, we can derive an observed rotation curve for both sides of the galaxy. For this purpose, we defined an angular sector of inclusion of 67.5$^\circ$ around the major axis (on the galaxy plane) in order to minimise possible contamination from radial motions\footnote{Because of projection effects, velocities along and close to the minor axis can be affected by radial motions (expansion and contraction along the radius), while velocities along and close to the major axis are dominated by galactic rotation.}. Within this sector, we calculated the mean velocity along circular arcs (in the galaxy plane) whose geometry is defined by our best-fitting model, correcting the velocity values with the cosine of the angular distance from the major axis on the galaxy plane. The left panel of Fig. \ref{RC} shows that the resulting rotation curve exhibits a systematic difference between the approaching and receding side in the inner regions ($R$ $\lesssim$ 2 kpc). This discrepancy can be removed by shifting the kinematic centre of the rotating disc by $4.4''$ (350 pc) (see Fig. \ref{RC}, right panel). Such a small shift of the dynamical centre is consistent with the idea that the external perturbation only marginally affects the innermost regions of the disc, where the gravitational potential well is deep. In order to reach the best matching between both sides of the solid body shape of the rotation curve, we shift the coordinates of the centre to the position RA = 12:33:47.47 and Dec = 15:09:53.8, at $\sim$ 14 arcsec to the north-west of the photometric centre (RA(J2000) = 12:33:47.95 and Dec = 15:10:05.7; NGVS catalogue, Ferrarese, private com.). We repeated the fit of the velocity field and obtained $v_\mathrm{sys}$ = 271 km s$^{-1}$, $incl$ = 32.2$^{\circ}$, $PA$ = 19.4$^{\circ}$, and $v_\mathrm{flat}=103$ \kms. This is our final model. Its parameters are given in Table \ref{gal} and adopted throughout this paper.
For comparison, the inclination derived from the baryonic Tully-Fisher relation of Ponomareva et al. (2018) is $incl$ = 33$^{\circ}$ $\pm$ 5$^{\circ}$ derived assuming a baryonic mass of $M_{bar}$ = 4.5 $\times$ 10$^9$ M$_{\odot}$ ($M_{bar}$ = $M_{star}$ + 1.4 $\times$ $M_{HI}$ + 1.4 $\times$ $M_{H_2}$, where $M_{H_2}$ = 0.38 $\times$ $M_{HI}$) and the observed rotational velocity of $WHI_{50}$ = 121 km s$^{-1}$.

An accurate inspection of the rotation curve derived using the above best-fitting model parameters (Fig. \ref{RC}, right panel) clearly shows that the receding side of the rotating disc begins to be perturbed at $R$ $\simeq$ 2.5 kpc from the galaxy centre. Here the rotational velocity begins to be lower than expected for a solid body rotation, it is fairly flat in between 3.5 $\lesssim$ $R$/kpc $\lesssim$ 6, and then steeply increases up to $\simeq$ 9.5 kpc, dropping again at larger radius. The approaching side follows a solid body rotation with a steep gradient for $R$ $\lesssim$ 3 kpc, then has a flatter but fairly constant slope up to $\simeq$ 13 kpc. A mild increase of the rotation curve up to these radii is not surprising in a low-mass galaxy such as NGC 4523. What is, however, unusual is the abrupt change of slope observed at $R$ $\simeq$ 3 kpc and the presence of two well defined components inside and outside this radius. These kind of asymmetries are expected in galaxies undergoing a ram pressure stripping event (Kronberger et al. 2008a). Finally, the approaching side rotational velocity decreases more rapidly than Keplerian for $R$ $\gtrsim$ 13 kpc. All these peculiarities in the rotation curve might be related to the gas compression on the receding side and gas removal from the galaxy disc on the approaching side.

\begin{figure}
\centering
\includegraphics[width=0.49\textwidth]{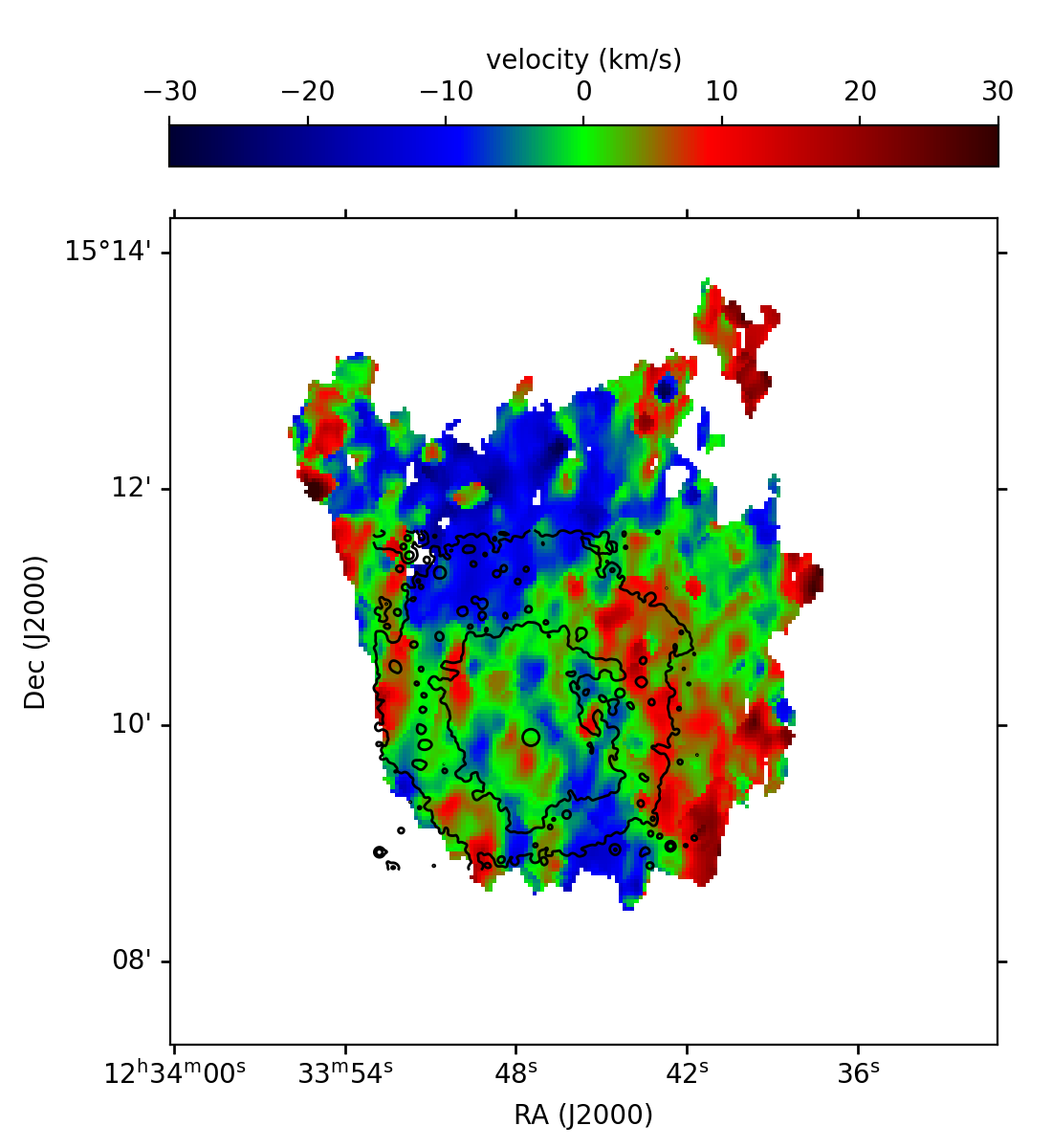}
\caption{High-resolution residual velocity field considering a two-slope rotation curve. The black contours show the $i$-band 23.5 and 25.0 mag arcsec$^{-2}$ isophotes, the black open dot the kinematic centre.
}
\label{residual}%
\end{figure}

The difference between our data and the best model can be associated to the external perturbation. We thus subtract the best-fitting model velocity field from the observed velocity field, and show the result in Fig. \ref{residual}. The figure shows that within a large circular inner region approximately delimited by the $i$-band 23.5 mag arcsec$^{-2}$ isophote, both sides of the disc show residual velocities close to zero \kms. This suggests that the inner velocity field is not strongly perturbed by the interaction with the Virgo environment. In a region delimited between the 25 and 23.5 mag arcsec$^{-2}$ isophotes, the residual velocities are abruptly positive on both sides of the galaxy along the minor axis. This difference with respect to the model is stronger on the west side, where the tail is more extended. Outside the $i$-band 25 mag arcsec$^{-2}$ isophote we observe the following. First, within a wide double cone of $\sim45\degr$ around the minor axis in the west direction, the residual velocities are again close to zero km s$^{-1}$. Because we are around the minor axis of the galaxy, this shows that radial motions (if any) are not significant. Second, within the complementary double cone of $\sim45\degr$ around the major axis, the residual velocities are negative. This is evident on the approaching side (north), where the \hi\ tail is located, but it is observable in a narrow region on the receding side (south), too. Finally, in the region in between these two double cones the residual velocities are typically positive. 


Figure \ref{residual} gives some interesting information on the \hi\ kinematics on small physical scales, too. It shows that the peaks in the residual do not correlate with those observed in the velocity dispersion map shown in Fig. \ref{Velmap_hr} (right panel). This implies that the \hi\ kinematics is not dominated by the turbulence within the star forming regions, but by local motions due to the external pressure that the galaxy is suffering (large scale shearing).


To summarise, the \hi\ gas distribution, the velocity field, the integrated \hi\ line profile, and the position-velocity diagrams can be explained if the galaxy is entering into the cluster from the back, north-west side and interacting with the surrounding ICM through a ram-pressure wind opposite to the infall and non perpendicular to the stellar disc (Fig. \ref{cartoon}). The gas on the southern-west side of the disc, which has a LoS velocity of $\sim$ 340 km s$^{-1}$, is removed from the disc mainly along the LoS (and is thus red-shifted by ram pressure), producing the red-shifted shallow shoulder of the integrated \hi\ spectrum and the compressed \hi\ contours visible in Fig. \ref{VCC1524_UGRHINET_HI}. The gas in the northern side, which is rotating against the ICM, is more efficiently removed, producing the northern diffuse tail. The LoS velocity of \hi\ in the tail is pushed back towards systemic, likely producing the steep shoulder of the integrated \hi\ spectrum. In the outer regions this gas is still rotating but on a different plane than the one of the stars.

\subsection{Gas vs. star formation}

\begin{figure}
\centering
\includegraphics[width=0.45\textwidth]{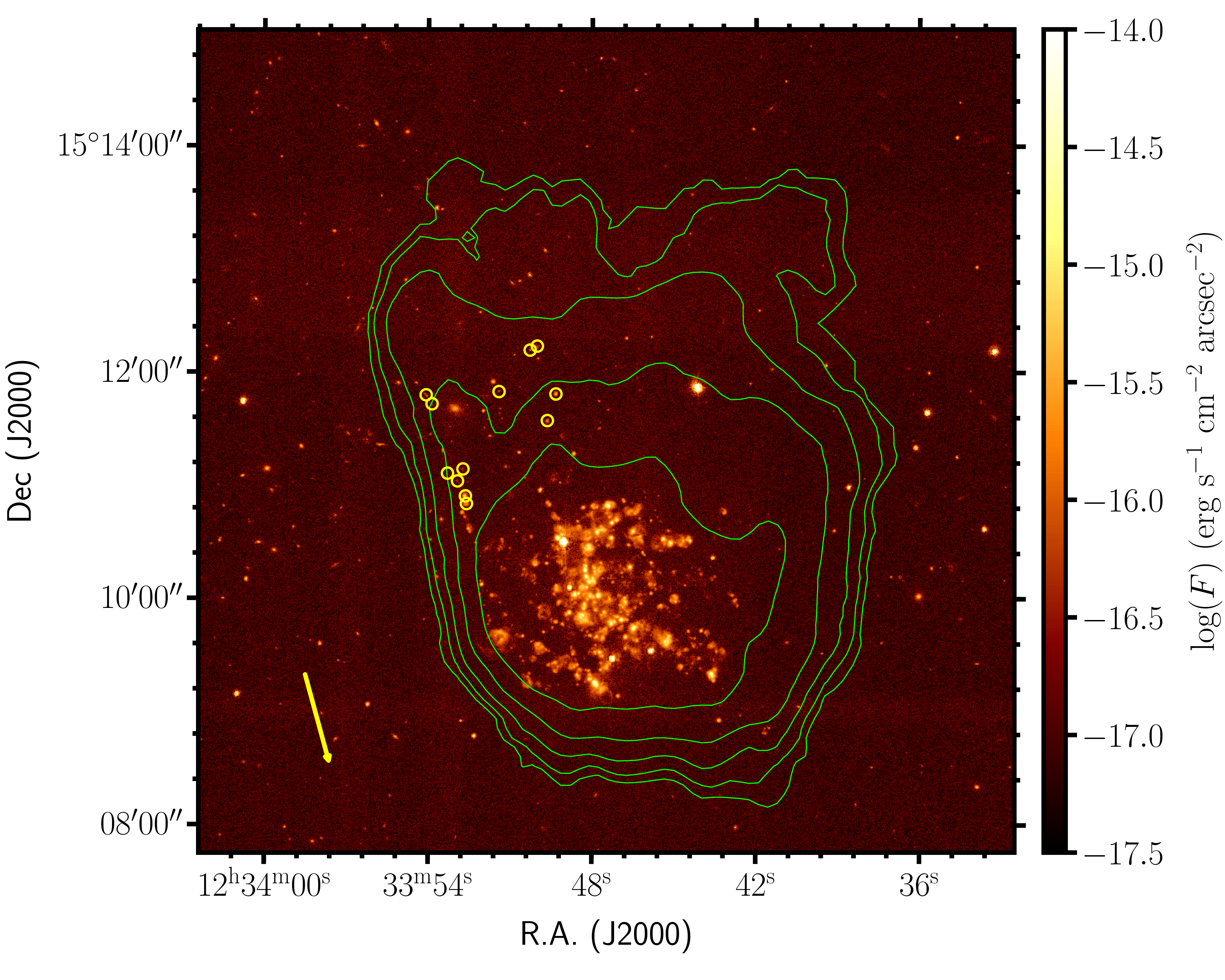}
\caption{\hi\ gas distribution (green contours) overlayed on the continuum-subtracted H$\alpha$ image of NGC 4523
derived using the VESTIGE narrow-band imaging data. Green contours are shown at column densities of $N(HI) = 2^n \times 3.9 \times 10^{19}$
cm$^{-2}$ with $n$=0,1,..5. The colour scale on the right gives the H$\alpha$ surface brightness in units of erg s$^{-1}$ cm$^{-2}$ arcsec$^{-2}$. 
The H{\sc ii} regions located outside the stellar disc and analysed in Sec. 6 are identified with yellow circles. The yellow arrow indicates the direction of the cluster centre (M87), located at a projected distance of 0.83 Mpc.
}
\label{HaHI}%
\end{figure}

\begin{figure*}
\centering
\includegraphics[width=0.49\textwidth]{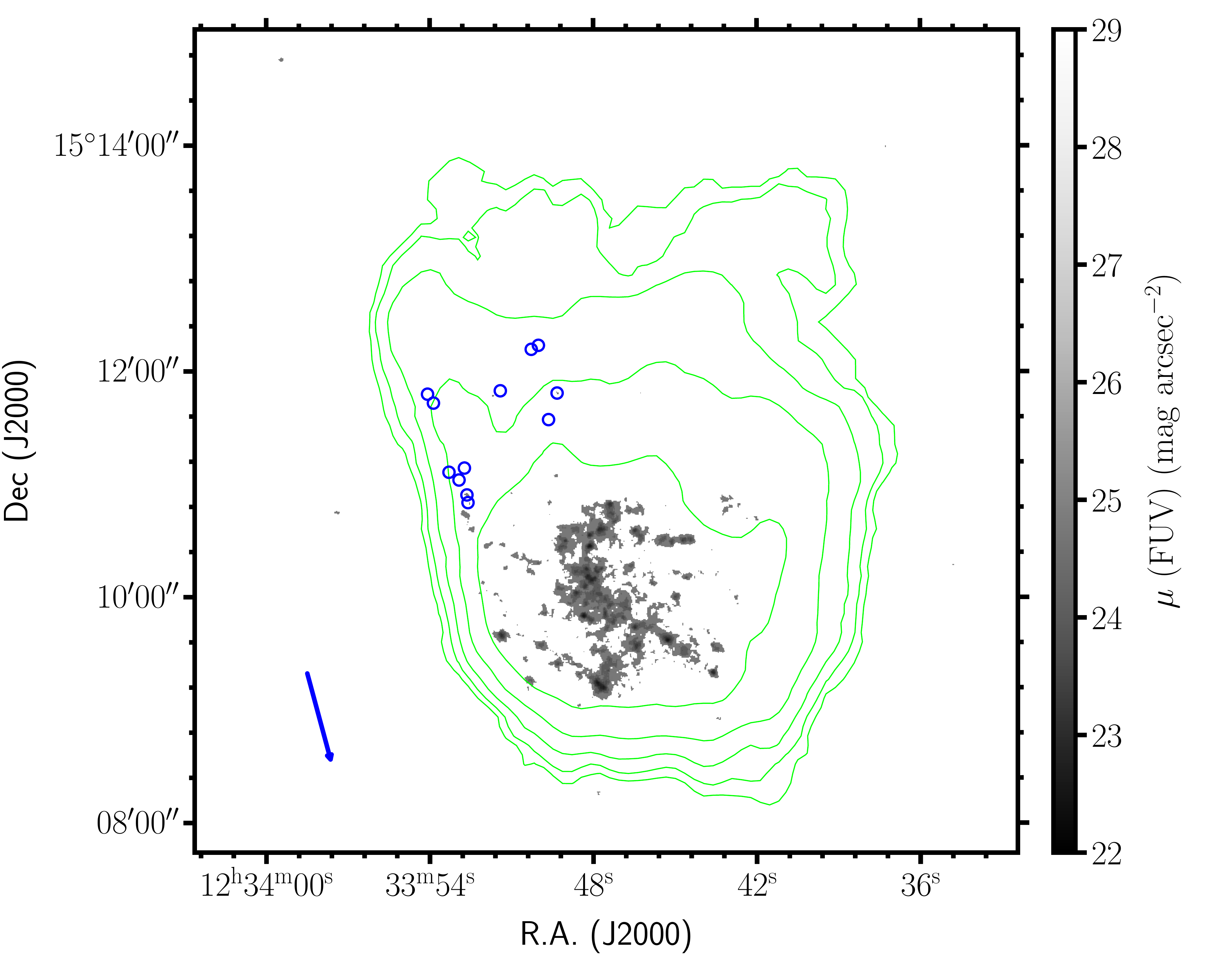}
\includegraphics[width=0.49\textwidth]{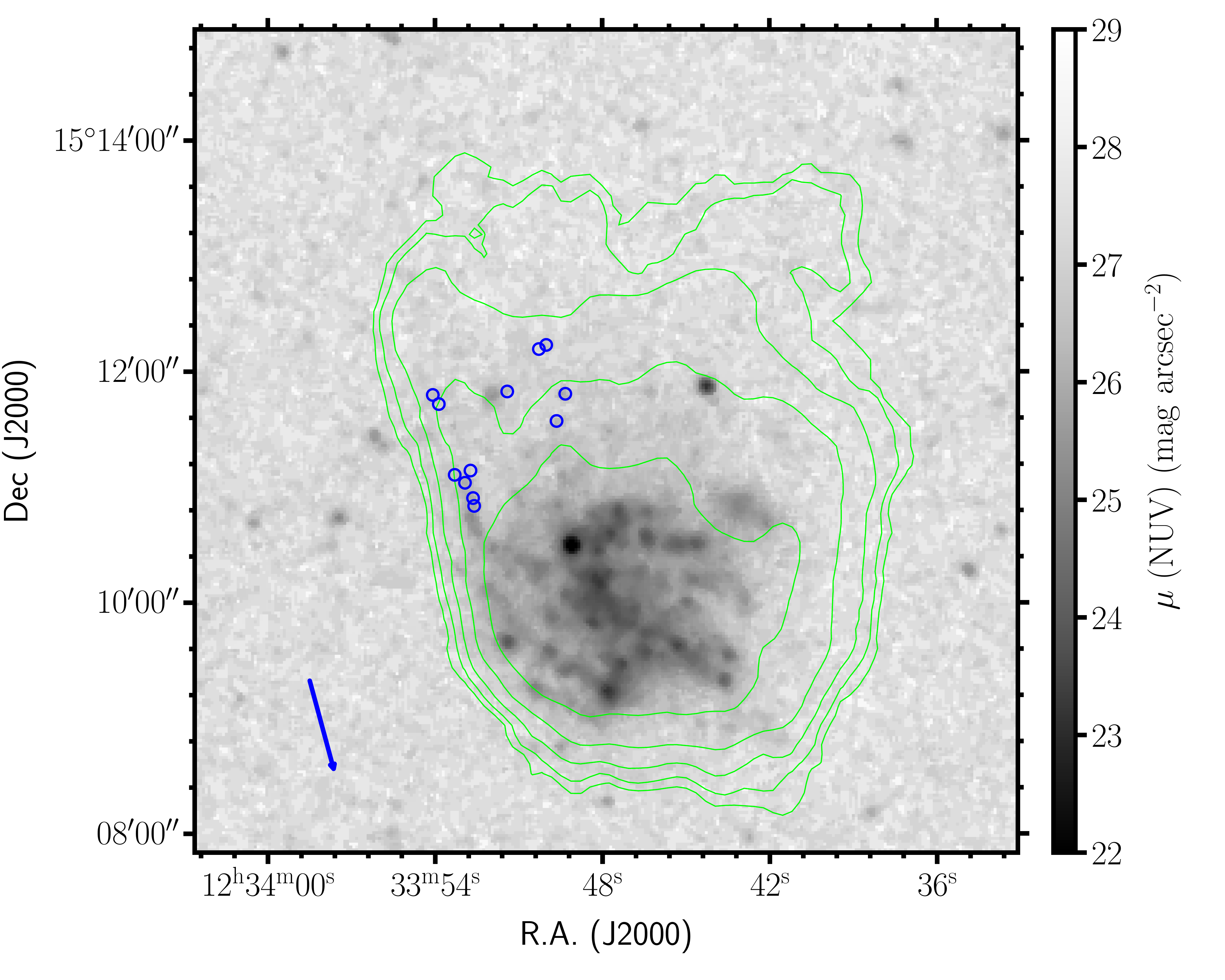}\\
\caption{\hi\ gas distribution (green contours) overlayed on the FUV ASTROSAT/UVIT image in the BaF2 filter 
($\lambda_c$ = 1541 \AA; $\Delta\lambda$ = 380 \AA) (left panel) and on the GALEX image of NGC 4523 in the NUV filter ($\lambda_c$ = 2316 \AA\, $\Delta\lambda$ = 1060\AA; right panel). 
Green contours are for \hi\ gas column densities of $N(HI) = 2^n \times 3.9 \times 10^{19}$
cm$^{-2}$ with $n$=0,1,..5. The H{\sc ii} regions located outside the stellar disc and analysed in Sec. 6 are identified with blue circles. The blue arrow indicates the direction of the cluster centre (M87), located at 0.83 Mpc projected distance.
}
\label{UVHI}%
\end{figure*}

Figure \ref{HaHI} compares the distribution of the ionised gas traced by the Balmer H$\alpha$ line with that of the cold atomic hydrogen. Despite its sensitivity, the 
H$\alpha$ narrow-band image does not show any diffuse ionised gas emission associated with the \hi\ gas tail in the northern side of the disc. A few compact H{\sc ii} regions, however, are detected at the north-east edge of the \hi\ gas tail outside the stellar disc defined by the $i$-band surface brightness isophote of $\mu(i)$ = 24.5 mag arcsec$^{-2}$. They have been identified as those objects with an excess in emission in the narrow-band H$\alpha$ image not associated to background galaxies in the broad-band $r$-image. These regions are unresolved in the low
resolution \hi\ gas map, and only a few of them are detected in the shallow FUV ASTROSAT/UVIT image (Fig. \ref{UVHI}). The GALEX NUV band image, which is sensitive to the emission of 
a somewhat older stellar population (100-500 Myr), shows some diffuse emission not associated with any recent star-forming region emitting in H$\alpha$.
In particular, there is an extended region at the north-west edge of the stellar disc (R.A.(2000) = 12:33:43.0, Dec = 15:10:49) without any associated H$\alpha$ emission.


\section{Simulations \label{sec:simulations}}

We simulated the galaxy using a 3D N-body code with two components: a non-collisional component that simulates the stellar bulge/disc and
the dark halo, and a collisional component that simulates the ISM. 
The non-collisional component consists of $81920$ particles that simulate the galactic halo, bulge, and disc. The
characteristics of the different galactic components are approximately adapted to the observed properties.
We adopted a model where the ISM is simulated as a collisional component, i.e., as discrete particles that possess a
mass and a radius and can have inelastic collisions (sticky particles). The $20000$ particles of the collisional
component represent gas cloud complexes that evolve in the gravitational potential of the galaxy. During the disc evolution,
the particles can have inelastic collisions, the outcome of which (coalescence, mass exchange, or fragmentation) is simplified
following Wiegel (1994). This results in an effective gas viscosity in the disc.
The model galaxy is somewhat smaller and has a somewhat steeper rotation curve than NGC~4523.

As the galaxy moves through the ICM, its gas clouds are accelerated by ram pressure. Within the galaxy’s inertial system, its clouds are exposed to a wind caused by the galaxy's motion through the ICM.
The effect of ram pressure on the clouds is simulated by an additional force on the clouds
in the wind direction. Only clouds that are not protected by other clouds against the wind are affected.
Since the gas cannot develop instabilities, the influence of turbulence on the stripped gas is not included in
the model. The mixing of the intracluster medium into the ISM is very crudely approximated by a finite penetration length of
the intracluster medium into the ISM, i.e., up to this penetration length the clouds undergo an additional acceleration
caused by ram pressure. A scheme for star formation was implemented where stars are formed during cloud collisions and
then evolve as non-collisional particles (see Vollmer et al. 2012). These newly formed star particles carry their
time of formation. Model H$\alpha$ maps were produced with all star particles whose ages are smaller than  $10$~Myr.
The UV emission of a star particle in the ASTROSAT/UVIT FUV band is modeled by the FUV flux from single stellar
population models from STARBURST99 (Leitherer et al. 1999). The total FUV distribution is then the extinction-free distribution
of the FUV emission of the newly created star particles.

Since NGC~4523 is located far away from the cluster centre ($\sim \sqrt{2} \times 0.83$~Mpc$\sim 1.2$~Mpc), we assume that the
galaxy's velocity and the encountered ICM density do not significantly vary during the last $360$~Myr. With a velocity
of $\sim 1000$~km\,s$^{-1}$ this corresponds to a distance of $380$~kpc. We also tested slowly increasing temporal ram-pressure
profiles. Since these simulations gave similar results, we only discuss the constant ram-pressure simulations.
All gas clouds beyond the galaxy's effective radius are assumed to have a constant gas surface density of $10$~M$_{\odot}$pc$^{-2}$.
The surface density of the gas clouds increases towards the galaxy centre (e.g., Vollmer et al. 2008b).
The surface density of atomic hydrogen was calculated via the gas density (Sect.~5.1 of Vollmer et al. 2008b).
For the different simulations we assumed a galaxy velocity of $1000$~km\,s$^{-1}$ and varied the ICM density $n_{\rm ICM}$ and the angle
between the ram pressure wind and the the galaxy's disc plane $\alpha$. The parameters of the different simulations are presented
in Table~\ref{tab:sim}.

\begin{table}
      \caption{Simulations}
         \label{tab:sim}
      \[
       \begin{tabular}{lccccc}
        \hline
        name & $v_{\rm gal}^{\rm{(a)}}$ & $n_{\rm ICM}^{\rm{(b)}}$ & $\alpha^{\rm{(c)}}$ & timestep$^{\rm{(d)}}$ & $az^{\rm{(e)}}$ \\
         & (km\,s$^{-1}$) & cm$^{-3}$ & degrees & Myr & degrees\\
        \hline
        A & 1000 & 10$^{-4}$ & 30 & 360 & 0 \\
        B & 1000 & 2 $\times$ 10$^{-4}$ & 30 & 220 & 0 \\
        C & 1000 & 10$^{-4}$ & 45 & 400 & 5 \\
        D & 1000 & 2 $\times$ 10$^{-4}$ & 45 & 330 & 5 \\
        E & 1000 & 10$^{-4}$ & 10 & 360 & 20 \\
        F & 1000 & 2 $\times$ 10$^{-4}$ & 10 & 220 & 20 \\
        \hline
        \end{tabular}
       \]
        \begin{list}{}{}
        \item[$^{\rm{(a)}}$] galaxy velocity within the ICM
        \item[$^{\rm{(b)}}$] ICM density
        \item[$^{\rm{(c)}}$] angle between the ram pressure wind and the disc plane
        \item[$^{\rm{(d)}}$] time since the beginning of the external pressure
        \item[$^{\rm{(e)}}$] azimuthal viewing angle
      \end{list}
\end{table}

\begin{figure*}
\centering
\includegraphics[width=0.49\textwidth]{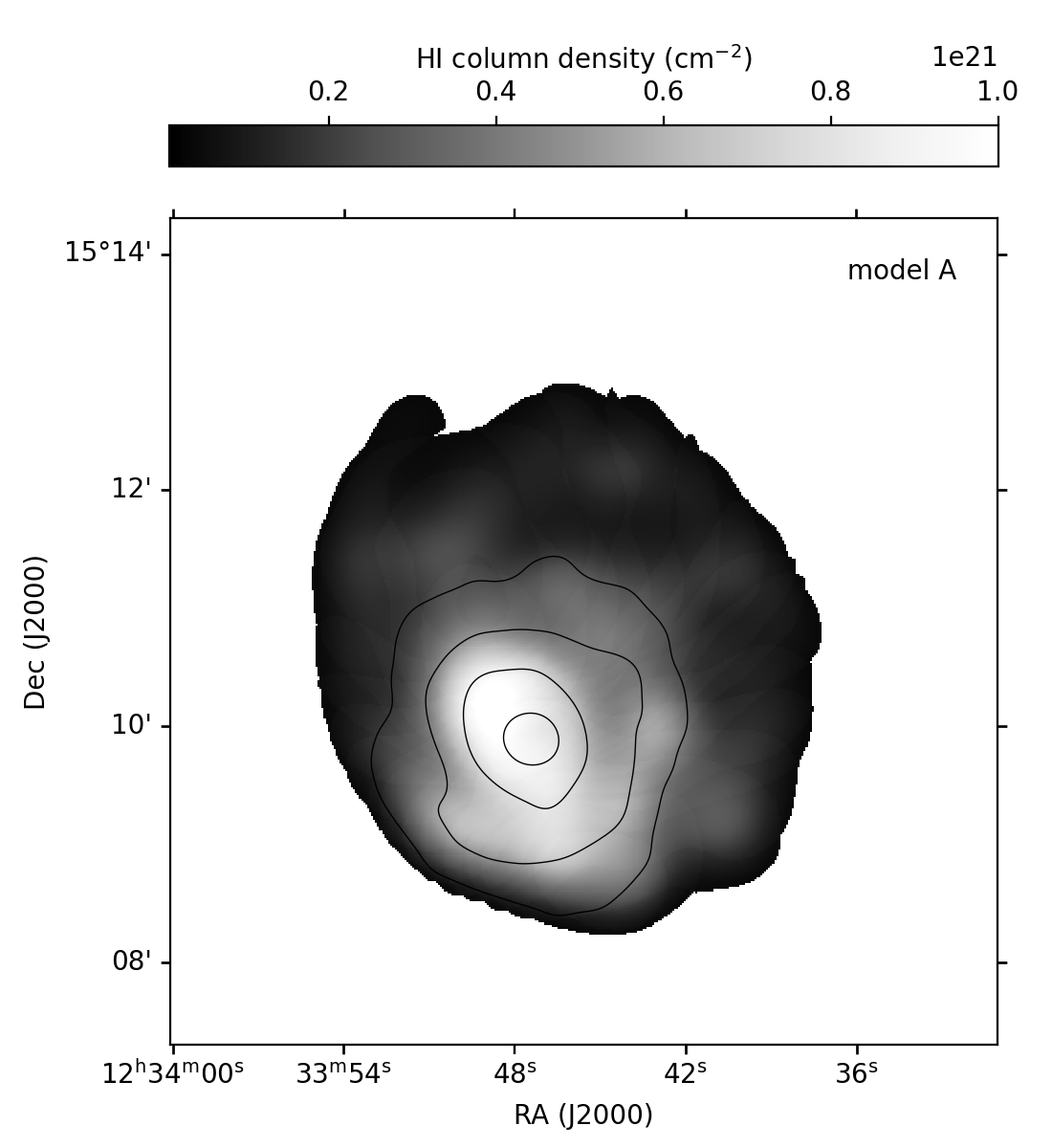}
\includegraphics[width=0.49\textwidth]{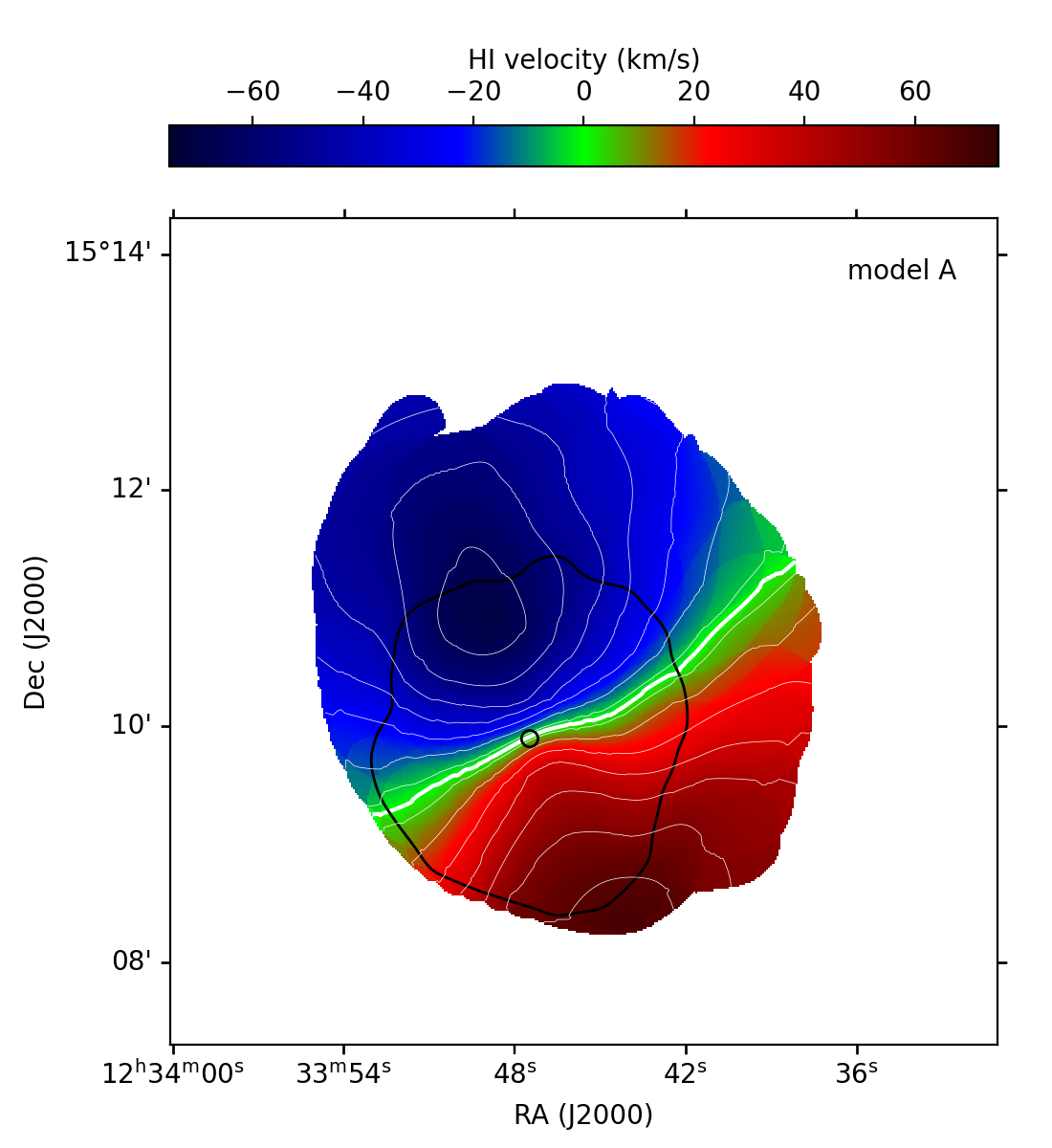}\\
\includegraphics[width=0.49\textwidth]{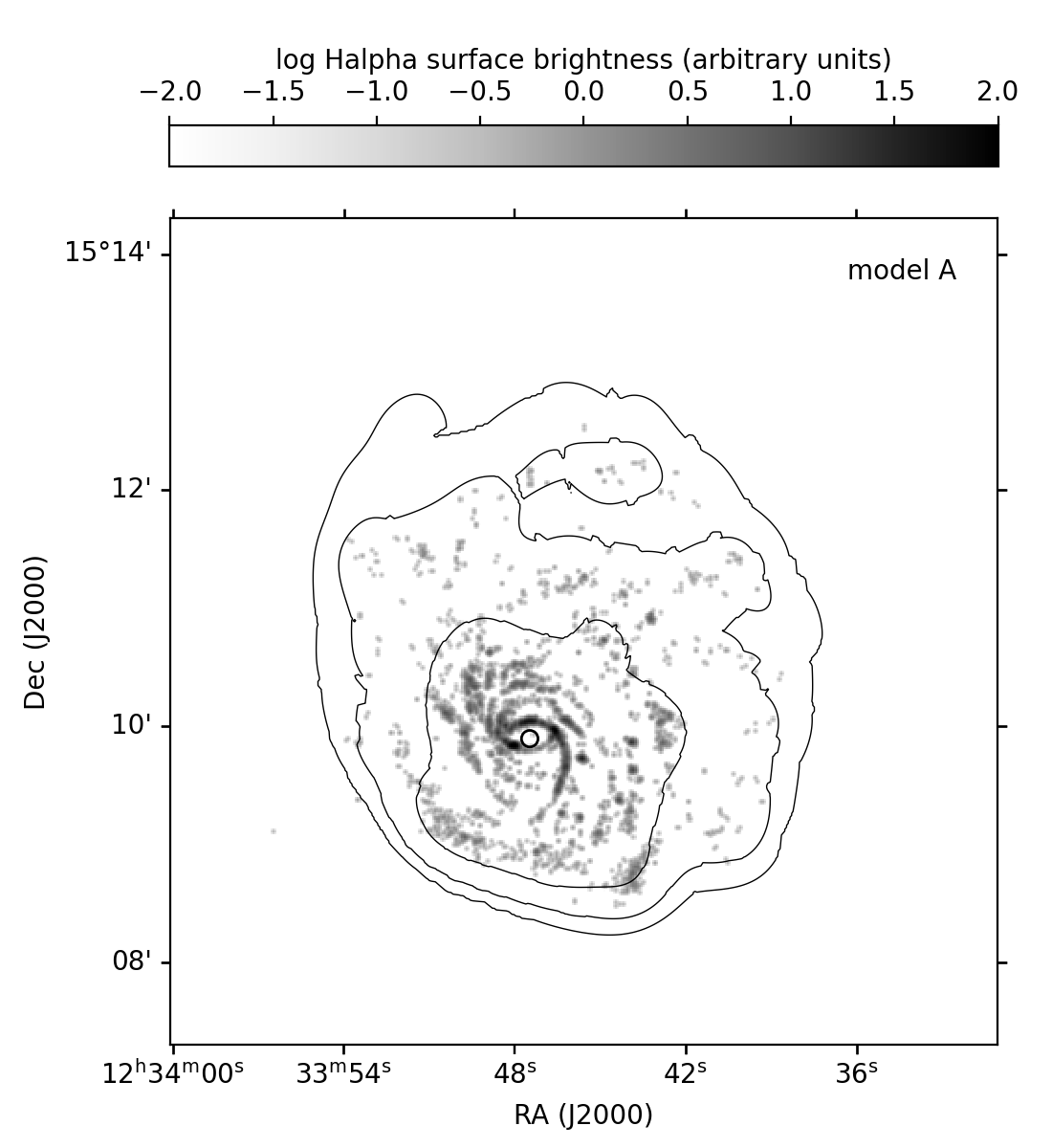}
\includegraphics[width=0.49\textwidth]{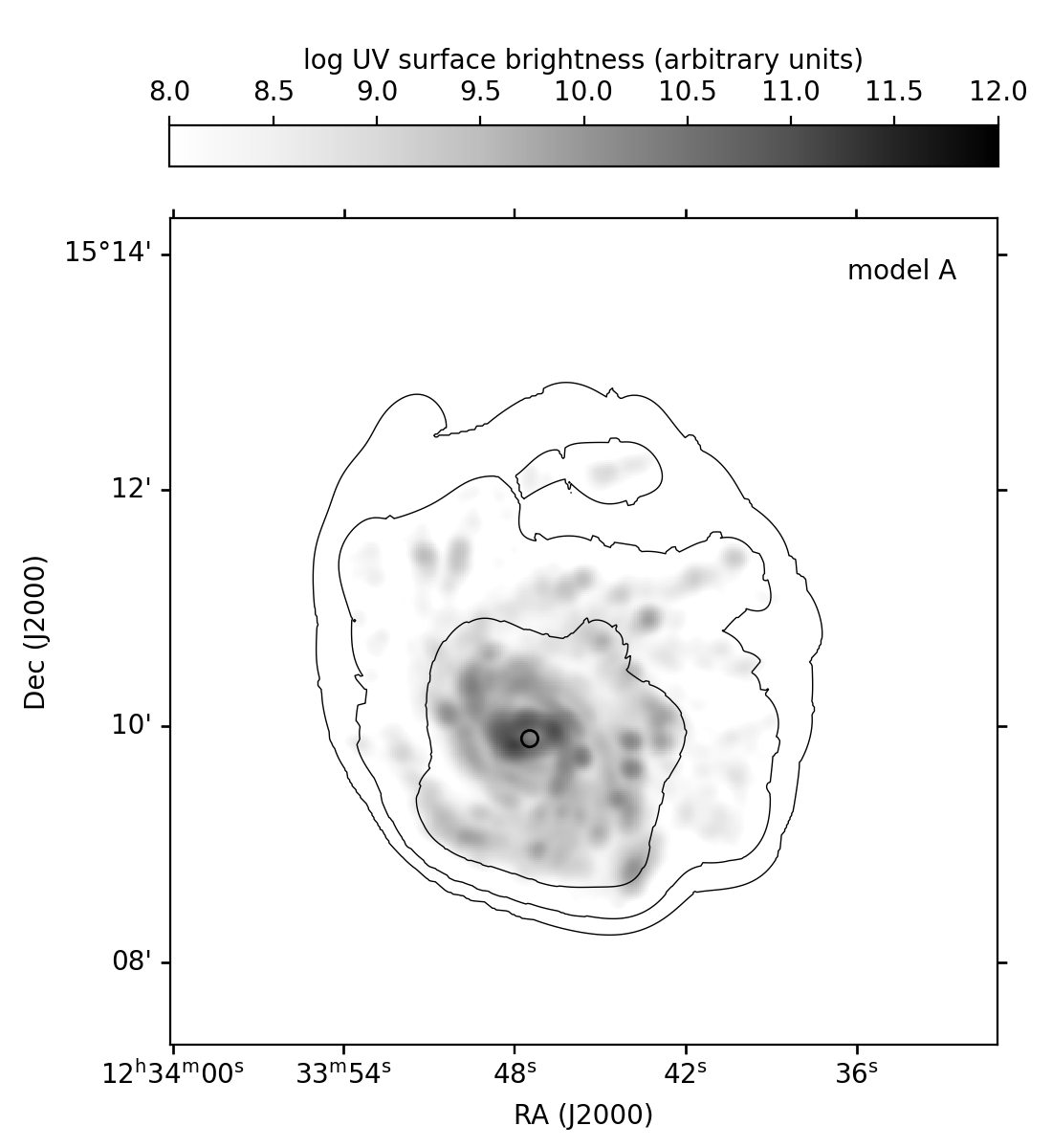}\\
\caption{Upper panels: model~A; left: model H{\sc i} moment~0 map, the contour levels show the model stellar distribution at 
10 $\times$ 10$^{0.5 \times n}$ ($n$=0,1,2,..) M$_{\odot}$ pc$^{-2}$;
right: model H{\sc i} moment~1 map, the white contour levels are from $-70$ to $70$~km\,s$^{-1}$ in steps of $10$~km\,s$^{-1}$, 
the black contour the model stellar distribution at 10 M$_{\odot}$ pc$^{-2}$. The empty dot shows the kinematic centre.
Lower panels: left: model H{\sc i} contours on model H$\alpha$ distribution; right:
model H{\sc i} contours on model FUV distribution. The H{\sc i} contour 
levels on the H$\alpha$ and the FUV images are $N(HI)$ = 3.9$\times$10$^{19}$
$\times$ 10$^{0.5 \times n}$ ($n$=0,1,2...) cm$^{-2}$. }
\label{modelA}
\end{figure*}
The timesteps of interest for the $\alpha=45^{\circ}$ simulations are $40$ to $50$~Myr later than those of the
$\alpha=10^{\circ}$ and $\alpha=30^{\circ}$ simulations: in the case of a more face-on stripping ($\alpha=45^{\circ}$) more time is needed
to push the ISM towards the north and west.
For the projection angles we use an inclination of $i=40^{\circ}$ and a position angle of $PA=20^{\circ}$.
The eastern half of the galactic disc is the near side.
Moreover, the azimuthal viewing angle $az$ (e.g., Vollmer et al. 2012) was chosen to best reproduce the observations.
The galaxy's normalised velocity vectors are $(-0.06,-0.09,-0.99)$ for the $\alpha=45^{\circ}$ stripping,
$(-0.31,-0.11,-0.95)$ for the $\alpha=30^{\circ}$ stripping, and $(-0.44,-0.52,-0.73)$ for the $\alpha=10^{\circ}$ stripping (the parameters of the orbit of model A are given in the Appendix).
Therefore, the galaxy's dominant velocity component is the radial ($z$) component in all simulations.
For $\alpha=45^{\circ}$ and $\alpha=10^{\circ}$ the galaxy moves approximately to the southeast, whereas it moves more to the east
for $\alpha=30^{\circ}$. The associated velocities on the plane of the sky are $\sim 100$, $\sim 330$, and $\sim 680$~km\,s$^{-1}$ for $\alpha$ = 45$^{\circ}$, 30$^{\circ}$, and 10$^{\circ}$, respectively.

All six models show the same characteristics, which are similar to those of NGC~4523 (see Fig.~\ref{modelA} for the low resolution view of model A; the high resolution view of the same model as well both low- and high resolution images of all other models are shown in Appendix A): (i) the east--west and north--west
H{\sc i} asymmetry, (ii) a relatively dense outer gas arm at the southeast edge of the H{\sc i} disc pointing towards the northeast direction, (iii) an overdense gas arm to the northeast,
(iv) an asymmetric velocity field in the north-south direction with increasing velocities to the south and an almost
constant velocity to the north with a slight increase to large radii, and (v)
an asymmetric velocity field in the east-west direction with higher velocity than expected by symmetry in the northwestern quadrant.
The observed characteristics, which are not reproduced by the models, are the following: (1) the model southeastern overdense arm
is shifted to the west with respect to the H{\sc i} observations, (2) the model northeastern gas arm bends to the northwest whereas the observed northeastern arm bends to the northeast, and (3) the observed northwestern low surface density H{\sc i} extension is not
reproduced by the model. 

Point (v) is too pronounced in model B. Point (iv), especially the increase of the radial velocity to the north at large radii,
is less present in model~C. Models~A, D, E, and F give almost equivalent results (Fig.~\ref{modelA}, and 
figures in the Appendix).
The shape of the outer H{\sc i} contours of model~E are closest to that of the observations.
The H{\sc i} distribution of model~D is somewhat more extended to the east compared to the observed H{\sc i} distribution and that of
models~A, E, and F. The western extent of the observed H{\sc i} distribution is best reproduced by models~A and F.
The velocity field along the galaxy's major axis is also somewhat different between the model:
whereas the northern sides of models~A and E better reproduce the observations,
the southern side is slightly better reproduced by model~D. The southern lowest radial velocities of models~E and F are significantly
lower than the corresponding observed velocities. In addition, in models E and F the eastern half of the zero-velocity contour
is horizontal or slightly bent to the north, contrary to observations.

Only models~A and F show H{\sc ii} regions to the northeast of the stellar disc as it is observed (lower left panel of
Fig.~\ref{modelA}). The southeastern star-forming arm observed in H$\alpha$ and UV is reproduced by models~A, D, and F
(lower right panels of Fig.~\ref{modelA}, and the figures in the Appendix).

We conclude that models~A, D, E, and F reproduce the observed characteristics of NGC~4523 in a satisfactory way.
Model~A ($\alpha=30^{\circ}$) is somewhat preferred because it simultaneously reproduces the main characteristics of the observed H{\sc i} distribution and
kinematics and the observed distribution of H{\sc ii} regions outside the optical disc.

If the galaxy's velocity within the ICM is $1000$~km\,s$^{-1}$, the encountered ICM density is $1$--$2 \times 10^{-4}$~cm$^{-3}$.
The model ICM density is consistent with the value derived from X-ray observations (Schindler et al. 1999).
The galaxy moves to the southeast with a velocity of $400 \pm 300$~km\,s$^{-1}$.
For the position of NGC~4523 with respect to M~87 we assume $(-188, 810, 834)$~kpc.
Based on these numbers and the orbit modelling of Vollmer et al. (2001), we estimate
NGC~4523's closest approach to the cluster centre (M~87) to be $D \sim 500$--$600$~kpc in about $1$~Gyr.
The galaxy's velocity at closest approach will be $\sim 1300$~km\,s$^{-1}$. The galaxy will thus undergo a
maximum ram pressure that will be five to ten times higher than the current ram pressure. Since the current
stripping radius is about $8$~kpc, it is probable that NGC~4523 will lose most, if not all, of its ISM within the
next $\sim 1.5$~Gyr.

\section{Discussion}

\subsection{The MeerKAT survey of the Virgo cluster}

Pilot observations of the Virgo cluster undertaken with the MeerKAT radio telescope are confirming, once again, the power of deep \hi\ 21 cm observations 
in identifying galaxies undergoing a perturbation with their surrounding environment. With only 42 minutes of exposure per field, MeerKAT is able to detect 
\hi\ tails of stripped gas at column densities as low as $3.9 \times 10^{19}$ cm$^{-2}$ (0.3 M$_{\odot}$ pc$^{-2}$; 3$\sigma$ assuming a 25 km s$^{-1}$ line width) at a spectral resolution of 11 km s$^{-1}$ and an angular 
resolution of $\sim$ 30\arcsec, sufficient to resolve most of the galaxies at the distance of the cluster (16.5 Mpc, Gavazzi et al. 1999, Mei et al. 2007).
We thus expect that, once completed, the full survey will detect several star-forming systems and resolve most of them. Given the tight relation
between the atomic gas content and the star formation activity of galaxies (e.g. Boselli et al. 2001), valid also within the Virgo cluster (e.g. Gavazzi et al. 2013, 
Boselli et al. 2014), we expect to detect $\sim$ 200 of the 384 galaxies detected in H$\alpha$ by the VESTIGE survey (Boselli et al. 2023). Not all H$\alpha$ emitting sources will be detected because of
the limited coverage of the cluster (60 deg$^2$ for the proposed MeerKAT observations vs. 104 deg$^2$ for the VESTIGE survey) and sensitivity, able to detect 
only objects with \hi\ gas masses $M(HI)$ $\gtrsim$ 2 $\times$ 10$^6$ M$_{\odot}$ (3$\sigma$ over a 100 km s$^{-1}$ linewidth).
Despite this limitation, the MeerKAT observations will increase by at least a factor of four the number of \hi\ resolved galaxies observed by the VIVA survey
in Virgo (Chung et al. 2009), extending the parameter space to dwarf irregular and early-type gas-rich systems.

\subsection{Ram pressure at the periphery of the cluster}

The extraordinary data gathered for NGC 4523 are a further confirmation that ram pressure is an efficient mechanism responsible for the stripping of 
the atomic gas content of galaxies of stellar mass $M_{star}$ $\simeq$ 10$^9$ M$_{\odot}$ entering a cluster of mass $M_{200}$ $\simeq$ 10$^{14}$ M$_{\odot}$ (Boselli et al. 2022).
The fact that the galaxy is located at 2.87$^{\circ}$ from M87, the centre of the cluster, thus at $R$ $\sim$ 0.85 $\times$ $r_{200}$ (see Table \ref{gal}), suggests once again that 
in similar environments ram pressure stripping is efficient at least up to the virial radius of the cluster (Boselli et al. 2022). As mentioned in Sec. 5,
the orientation of the \hi\ tail in the direction opposite to the cluster centre suggests that the galaxy is at its first infall into the cluster. Since
the amount of \hi\ already removed by the interaction with the ICM, if any, is still minor as deduced from the \hi\--deficiency parameter, this implies that ram pressure stripping is a rapid process, as already deduced by targeted observations of other representative objects 
(Vollmer et al. 2004; Boselli et al. 2006; Pappalardo et al. 2010; Abramson et al. 2011; Fossati et al. 2018),
of large statistical samples (Crowl \& Kenney 2008; Boselli et al. 2008a,b, 2014, 2016b; Vollmer 2009; Gavazzi et al. 2013; Vulcani et al. 2020), or of tuned 
hydrodynamic simulations (Vollmer et al. 2006a, 2006b, 2008a, 2008b, 2012, 2018, 2021; Boselli et al. 2021). A recent stripping process during its first infall at the periphery of the cluster is also consistent with the lack of a ionised gas tail. Indeed, the stripped cold gas requires a few hundreds million years to mix with the hot ICM and change of phase, as indicated by the hydrodynamic simulations of IC 3476 (Boselli et al. 2021). 

\begin{figure}
\centering
\includegraphics[width=0.5\textwidth]{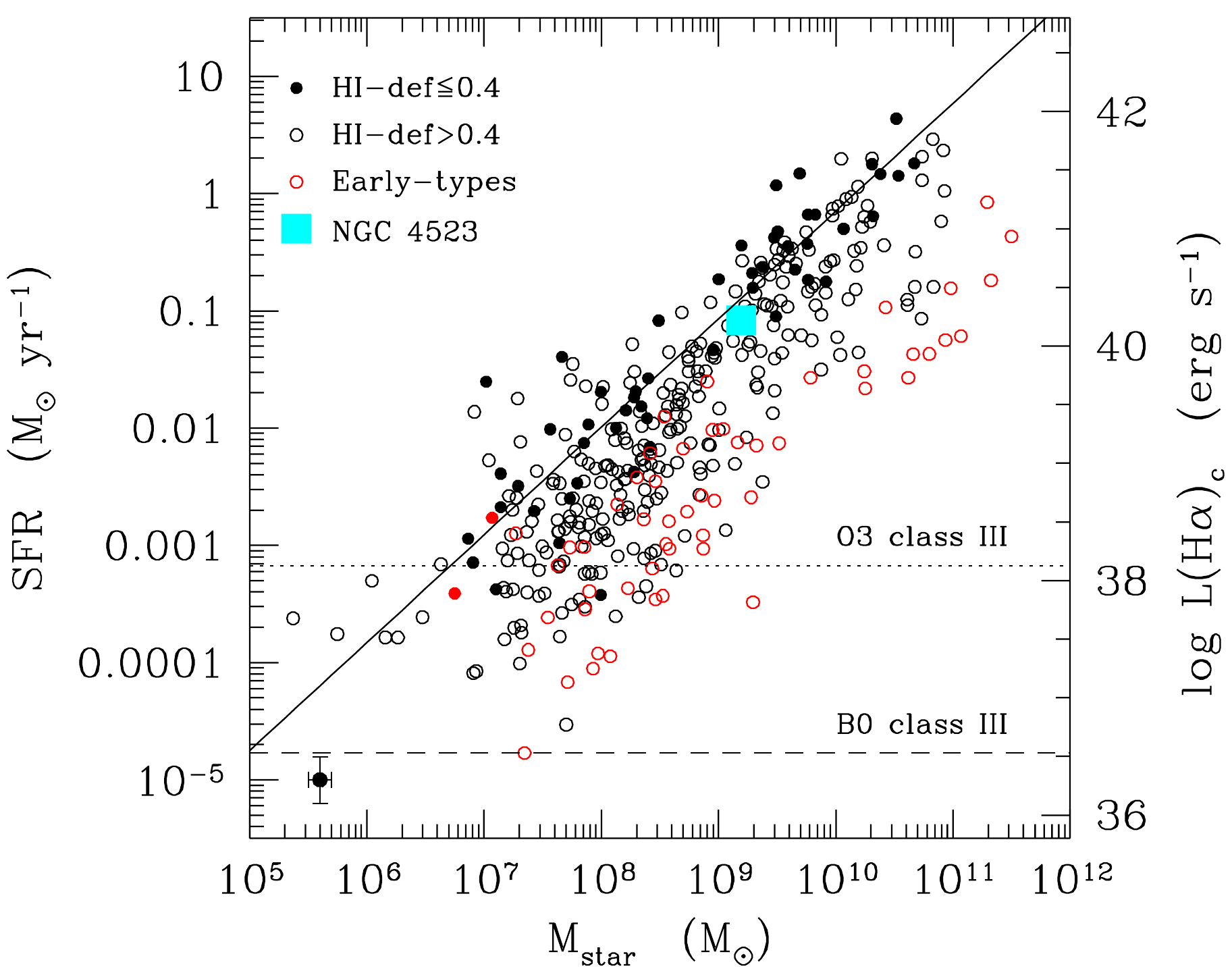}
\caption{Location of NGC 4523 on the main sequence relation (filled blue square) defined by star-forming galaxies in the Virgo cluster (adapted from Boselli et al. 2023).
Late-type galaxies ($\geq$ Sa) are indicated by black symbols, 
early-type galaxies by red symbols. Filled dots are for \hi\ gas rich objects ($HI-def$ $\leq$ 0.4), while \hi\-deficient objects ($HI-def$ $>$ 0.4) by empty circles. Star formation rates 
have been derived assuming stationary conditions. The Y-axis on the right side gives the corresponding H$\alpha$ luminosities corrected for dust attenuation.
The black solid line shows the best fit obtained for gas-rich star-forming systems. The horizontal dotted and dashed lines indicate the corresponding 
SFR derived using the number of ionising photons produced by an O3 class III and a B0 class III star as derived using the model atmospheres of Sternberg et al. (2003).
The typical uncertainty in the data is shown at the lower left corner.}
\label{main}%
\end{figure}

Simulations suggest that under specific conditions ram pressure stripping can trigger the star formation activity of the perturbed galaxies (Fujita \& Nagashima 1999; 
Bekki \& Couch 2003; Steinhauser et al. 2012; Bekki 2014; Henderson \& Bekki 2016; Steyrleithner et al. 2020;  Lee et al. 2020; Troncoso-Iribarren et al. 2020).
Figure \ref{main} shows the position of NGC 4523 on the main sequence relation defined by all the star-forming objects detected by the VESTIGE survey (Boselli et al. 2023). The total star formation rate of NGC 4523 and those of the other VESTIGE galaxies are derived assuming a Chabrier IMF after correcting the data for [NII] contamination and dust attenuation as described in Boselli et al. (2023).
The galaxy is located on the main sequence relation of gas-rich systems ($HI-def$ $\leq$ 0.4), here considered as unperturbed objects. 
In NGC 4523, as in most of the ram pressure stripped galaxies identified within the cluster, the overall star formation activity is not increased (Boselli et al. 2022, 2023). Given that the ram pressure exerted on the gas disc is close to edge-on, and that this geometrical configuration is the most favourable for enhancing the activity of star formation (Kronberger et al. 2008b; Boselli et al. 2022), it is conceivable that the increased activity observed in the GASP sample (Vulcani et al. 2018) or in other ram pressure stripped galaxies in Coma and A1367 (Molnar et al. 2022, Boselli et al. 2022) might be due to selection effects (optically selected tails might favour galaxies undergoing a more violent or more advanced gas stripping phase than \hi\ selected objects), or to the more hostile environment typical of these more massive clusters ($M_{200}$ $\simeq$ 10$^{15}$ M$_{\odot}$ in Coma vs. 
$M_{200}$ $\simeq$ 10$^{14}$ M$_{\odot}$ in Virgo). However, episodes of increased activity have been observed also in Virgo (e.g. IC 3476, Boselli et al. 2021) and in other local low-mass clusters (JO206, Ramatsoku et al. 2019).
Their short duration and the moderate increase in the star formation activity makes a negligible effect on a complete star-forming population (Boselli et al. 2022, 2023). If compared to the predictions of the 2D models of gas stripping in cluster galaxies presented in Boselli et al. (2006, 2014, 2023), the position of NGC 4523 on the main sequence
relation and its \hi\ gas deficiency ($HI-def$ = 0.39) suggest that the galaxy has recently started its interaction with the surrounding ICM ($\lesssim$ 100 Myr, see Fig. 9 of Boselli et al. 2023; notice that the model plotted in this figure is for a galaxy with a rotational velocity of 70 km s$^{-1}$, similar to the one
of NGC 4523). The star formation activity is marginally reduced probably because most of the stripped gas is located in the outer disc where star formation does not necessarily take place. This evolutionary picture is consistent with the simulations presented in Sec. 6. 

\begin{table}
\caption{Properties of the H{\sc ii} regions outside the stellar disc}
\label{pallini}
{
\[
\begin{tabular}{ccccc}
\hline
\noalign{\smallskip}
\hline
Region  & log$L(H\alpha)$   & log$SFR$              & log$M_{star}$ & Age   \\ 
Units   & erg s$^{-1}$      & M$_{\odot}$ yr$^{-1}$ & M$_{\odot}$   & Myr   \\  
\hline
 1  & 37.80$\pm$0.01     & -3.50$\pm$0.01	 &  3.94$\pm$0.23 &   8.84$\pm$1.76	\\
 2  & 37.59$\pm$0.02     & -3.71$\pm$0.02	 &  3.92$\pm$0.18 &  10.92$\pm$1.50	\\
 3  & 37.43$\pm$0.02     & -3.87$\pm$0.02	 &  4.42$\pm$0.11 &  17.06$\pm$1.26	\\
 4  & 37.23$\pm$0.04     & -4.07$\pm$0.04	 &   -            &    -   		\\
 5  & 36.79$\pm$0.10     & -4.51$\pm$0.10	 &   -            &  11.34$\pm$25.80	\\
 6  & 36.82$\pm$0.09     & -4.48$\pm$0.09	 &   -            &    -     		\\
 7  & 36.51$\pm$0.19     & -4.79$\pm$0.19	 &   -            &    -   		\\
 8  & 36.48$\pm$0.20     & -4.82$\pm$0.20	 &   -            &    -   		\\
 9  & 36.64$\pm$0.14     & -4.66$\pm$0.14	 &   -            &  62.93$\pm$87.34	\\
10  & 37.08$\pm$0.05     & -4.22$\pm$0.05	  &   -            &    -    		\\
11  & 36.94$\pm$0.07     & -4.36$\pm$0.07	  &   -            &    -    		\\
12  & 36.71$\pm$0.12     & -4.59$\pm$0.12	  &   -            &    -    		\\
\noalign{\smallskip}
\hline
\end{tabular}
\]
Notes:H$\alpha$ luminosity are derived after correcting the VESTIGE H$\alpha$ fluxes for [NII] contamination (N[II]$\lambda$6583\AA/H$\alpha$=0.1) and dust attenuation ($A(H\alpha)$ = 0.02 mag) as in Boselli et al. (2018b). Star formation rates are derived from the H$\alpha$ luminosity using the calibration of Calzetti et al. (2010). Given the low star formation regime, these numbers should be taken as indicative (see Boselli et al. 2023 for details). Both $SFR$ and $M_{star}$ have been derived assuming a Chabrier IMF.}
\end{table}

\subsection{Star formation in the tail}

Twelve star-forming regions have been detected in the continuum-subtracted H$\alpha$ narrow-band image outside the $i$-band surface brightness 
isophote of $\mu(i)$ = 24.5 mag arcsec$^{-2}$. They are all within the \hi\ gas tail of NGC 4523. 
These regions are located only on the north-east side of the tail (approaching side) where ram pressure stripping is expected to be the most efficient
given the combined effect of rotation and the galaxy journey within the cluster, infalling from the north-east and back side. It is possible that these H{\sc ii} regions
have been formed within the turbulent structures at the interface of the two media (ISM and ICM), as indeed suggested by our simulations. We can age-date these H{\sc ii} regions following the prescription given in Boselli et al. (2018b) by comparing their observed (H$\alpha$-FUV) and (FUV-NUV) age-sensitive colours to the typical synthetic colours of H{\sc ii} regions as described in Boselli et al. (2018b).
%
%
%
We have calculated the colours of these regions after extracting their fluxes as described in Fossati et al. (2018). 
As in Boselli et al. (2018b) we run the CIGALE code (Boquien et al. 2019) to derive star formation rates, stellar masses and ages of each individual region using the flux densities measured in the VESTIGE H$\alpha$, UVIT BaF2 FUV, GALEX NUV, CFHT $ugriz$, and Spitzer IRAC1 and IRAC2 bands, but derived these physical parameters only for a few of them given the very low detection rate in the optical and near-IR bands (see Table \ref{pallini}).

\begin{figure}
   \centering
   \includegraphics[width=9cm]{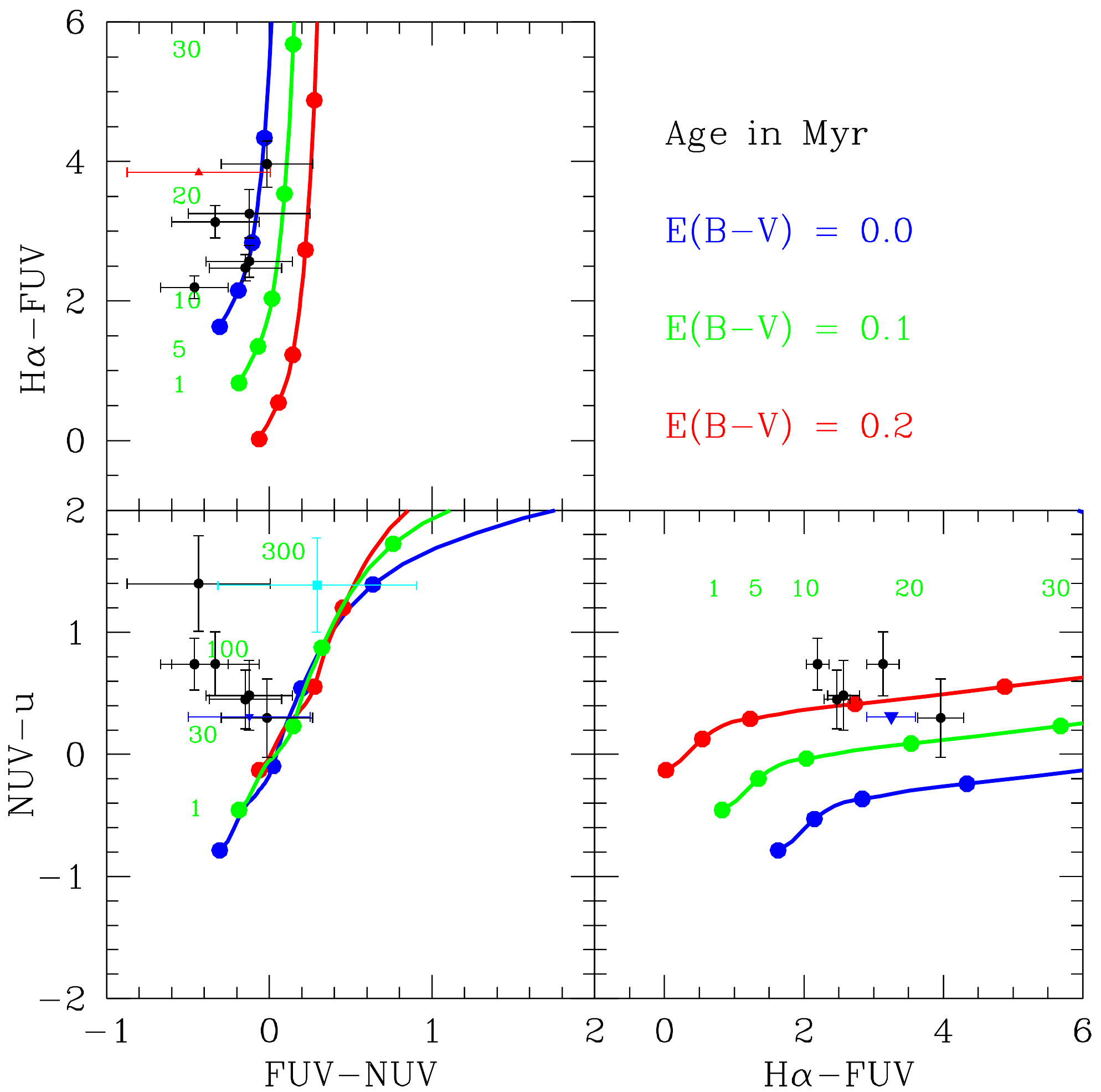}
   \caption{Age-sensitive colour-colour diagrams. Upper left: H$\alpha$-$FUV$ vs. $FUV-NUV$, lower left: $NUV-u$ vs. $FUV-NUV$,
   lower right: $NUV-u$ vs. H$\alpha$-$FUV$. 
   Black filled dots are for regions detected in all bands, red filled triangles
   are lower limits in H$\alpha$-$FUV$, blue filled triangles upper limits in $NUV-u$, 
   while the cyan filled square the mean value for the diffuse extended tail. 
   The blue, green, and red filled dots
   and solid lines indicate the expected colours at different ages (in Myr) for the assumed star formation history for an $E(B-V)_Y$ = 0.0, 0.3, and 0.6, 
   respectively. The green numbers are the ages in Myr for $E(B-V)$=0.1.
 }
   \label{FUVNUVHa}%
   \end{figure}

As expected, the typical colour of these regions suggests that they are recent since they are composed of stars with a typical age of $\lesssim$ 30 Myr, Fig. \ref{FUVNUVHa}.
These ages are comparable to those derived for the H{\sc ii} regions formed within the tails of other perturbed galaxies in Virgo, such as 
the harassed gas of NGC 4254 (Boselli et al. 2018b) or in the ram pressure stripped gas of IC 3418 (Fumagalli et al. 2011). For comparison, if the galaxy is moving on the plane of the sky at $\sim$ 1000 km s$^{-1}$, a velocity comparable to the velocity dispersion of cluster A (the main body of the Virgo cluster, $\sigma$ = 799 km s$^{-1}$, Boselli et al. (2014a)) or to that of the late-type systems within the whole Virgo cluster ($\sigma$ = 1150 km s$^{-1}$, Boselli \& Gavazzi (2006)), it would take 10 Myr to travel 10 kpc, the mean distance of these H{\sc ii} regions from the outer stellar disc. Similarly, in 30 Myr the galaxy would make $\sim$ 10\%\ of its revolution.

Worth noticing is also the fact that
star formation within the tails of ram pressure stripped gas is not ubiquitous. The VESTIGE survey, which has covered the whole cluster with exquisite 
sensitivity and angular resolution, has identified tails without any star-forming region (e.g. NGC 4569, Boselli et al. 2016a), others with a few 
(NGC 4388, Yoshida et al. 2002; NGC 4330, Fossati et al. 2018; IC 3476, Boselli et al. 2021). There are also examples of galaxies with several H{\sc ii} regions outside their stellar discs
but without any apparent associated diffuse gas tail (e.g. IC 3418, Hester et al. 2010; Fumagalli et al. 2011; Jachym et al. 2013; Kenney et al. 2014; 
Hota et al. 2021), at least at the sensitivity of the available observations. 
The process of star formation and the conditions under which this occurs in the tails of stripped material is still poorly known. Star formation requires gas cooling in the molecular phase to take place. The molecular gas is hardly removed during a ram pressure stripping event since it is principally located within giant molecular clouds with a limited cross section with respect to the external pressure (Boselli et al. 2022). Molecular gas in clumpy regions has been observed in the tails of several ram pressure stripped galaxies (e.g. Vollmer et al. 2005b, 2008b; Jachym et al. 2014, 2017, 2019; Moretti et al. 2018, 2020b). It is thus conceivable that the cold atomic gas located in the tails, under some still unclear conditions, collapses to form new stars. This might happen in the turbulent regions formed by the instabilities at the edges of the stripped cold ISM and the surrounding hot ICM. The mixing of the different gas phases can induce cooling of the ICM into dense clouds (Tonnesen \& Bryan 2021), while magnetic fields can keep the cold gas confined within filaments, thus affecting the cooling process (Tonnesen \& Stone 2014; Ruszkowski et al. 2014; M\"uller et al. 2021). Furthermore, it is also unknown which is the efficiency with which the molecular gas component is converted into stars within ram pressure stripped tails (e.g. Boissier et al. 2012; Villanueva et al. 2022). 

The completion of the MeerKAT survey of the Virgo cluster will certainly boost our understanding of this important physical process. Despite the presence of the bright radio galaxy M87 (Virgo A) in the field, the quality of the data is excellent in terms of sensitivity and angular resolution, and will be increased by a factor of $\sim$ 2 once the survey will be completed thanks to an increased overlap of pointings. ViCTORIA will provide us 
with critical information on the atomic gas content, the main gas component stripped during the interaction since loosely bound to the gravitational potential well of the 
galaxy. Combined with that of the other gas phases (molecular, VERTICO survey, Brown et al. 2021; ionised, VESTIGE survey, Boselli et al. 2018a; hot, eROSITA, Predhel et al. 2021) 
\hi\ data will be critical to understand under which conditions the stripped gas can cool and condense to form stars or heat up changing of phase and becoming 
first ionised, then hot gas. This complete set of multifrequency data on the different gas phases will thus be crucial for constraining tuned hydrodynamic simulations of interacting galaxies now able to resolve giant molecular clouds and star-forming regions within the tails. ViCTORIA will also provide us with a unique set of radio continuum data at different frequencies, where the emission is sensitive to the energy loss of relativistic electrons in weak magnetic fields, two other major ingredients in the study of galaxy evolution in rich environments.

\section{Conclusion}

During pilot observations of the Virgo cluster carried out with the MeerKAT radio telescope we discovered an extended (10 kpc projected distance) low column density atomic gas tail ($N(HI) \lesssim 2.5 \times 10^{20}$ cm$^{-2}$) in a dwarf galaxy 
(NGC 4523, $M_{star}$ = 1.6 $\times$ 10$^9$ M$_{\odot}$) at the northern periphery of the cluster ($R$ $\simeq$ 0.85 $\times$ $r_{200}$).
The multifrequency analysis of the data consistently suggests that
NGC 4523 is suffering a ram pressure stripping event which has already removed part of the atomic gas. The overall 
star formation activity of the galaxy, estimated using new H$\alpha$ narrow-band imaging gathered during the VESTIGE survey, is not significantly
reduced if compared to that of similar unperturbed objects. A few compact H{\sc ii} regions, whose age has been estimated comparing the prediction
of SED models to their observed H$\alpha$, ASTROSAT/UVIT FUV, GALEX NUV, and NGVS $u$ colours ($\lesssim$ 30 Myr),
have been formed within the tail of stripped gas. NGC 4523 is a further example indicating that ram pressure stripping is efficient in intermediate 
mass clusters such as Virgo ($M_{200}$ $\simeq$ 10$^{14}$ M$_{\odot}$) up to the virial radius, confirming once again the main role of
this hydrodynamic process in shaping galaxy evolution in rich environments. 

The results presented in this work are a further demonstration of the power of untargeted surveys of nearby clusters in identifying
galaxies undergoing a perturbation. Once completed, the MeerKAT survey of the Virgo cluster will provide us with a unique view of the
effects of the environment on the cold atomic gas of infalling galaxies. Combined with similar information gathered thanks to the VESTIGE
survey for the ionised gas component, the \hi\ data will be crucial to understand the effects induced by the perturbations on the process of 
star formation. They will be necessary also to understand the fate of the stripped gas, the formation of possible compact sources outside the
stellar discs of perturbed galaxies, and ultimately the pollution of the ICM.

\begin{acknowledgements}

We thank the anonymous referee for constructive comments which helped improving the quality of the manuscript. The MeerKAT telescope is operated by the South African Radio Astronomy Observatory, which is a facility
of the National Research Foundation, an agency of the Department of Science and Innovation.
We are grateful to the whole CFHT team who assisted us in the preparation and in the execution of the observations and in the calibration and data reduction: 
Todd Burdullis, Daniel Devost, Bill Mahoney, Nadine Manset, Andreea Petric, Simon Prunet, Kanoa Withington.
We acknowledge financial support from ``Programme National de Cosmologie and Galaxies" (PNCG) funded by CNRS/INSU-IN2P3-INP, CEA and CNES, France,
and from ``Projet International de Coop\'eration Scientifique" (PICS) with Canada funded by the CNRS, France.
This research has made use of the NASA/IPAC Extragalactic Database (NED) 
which is operated by the Jet Propulsion Laboratory, California Institute of 
Technology, under contract with the National Aeronautics and Space Administration
and of the GOLDMine database (http://goldmine.mib.infn.it/) (Gavazzi et al. 2003).
Part of the data published here have been reduced using the CARACal pipeline, partially supported by ERC Starting grant number 679627 “FORNAX”, MAECI Grant Number ZA18GR02, DST-NRF Grant Number 113121 as part of the ISARP Joint Research Scheme, and BMBF project 05A17PC2 for D-MeerKAT. Information about CARACal can be obtained online under the URL: https://caracal.readthedocs.io.
MB gratefully acknowledges support by the ANID BASAL project FB210003 and from the FONDECYT regular grant 1211000.
N.Z. is supported through the South African Research Chairs Initiative of the
Department of Science and Technology and National Research Foundation.

\end{acknowledgements}

\begin{appendix}

\section{Outputs of the simulations}

\begin{figure*}
\centering
\includegraphics[width=0.49\textwidth]{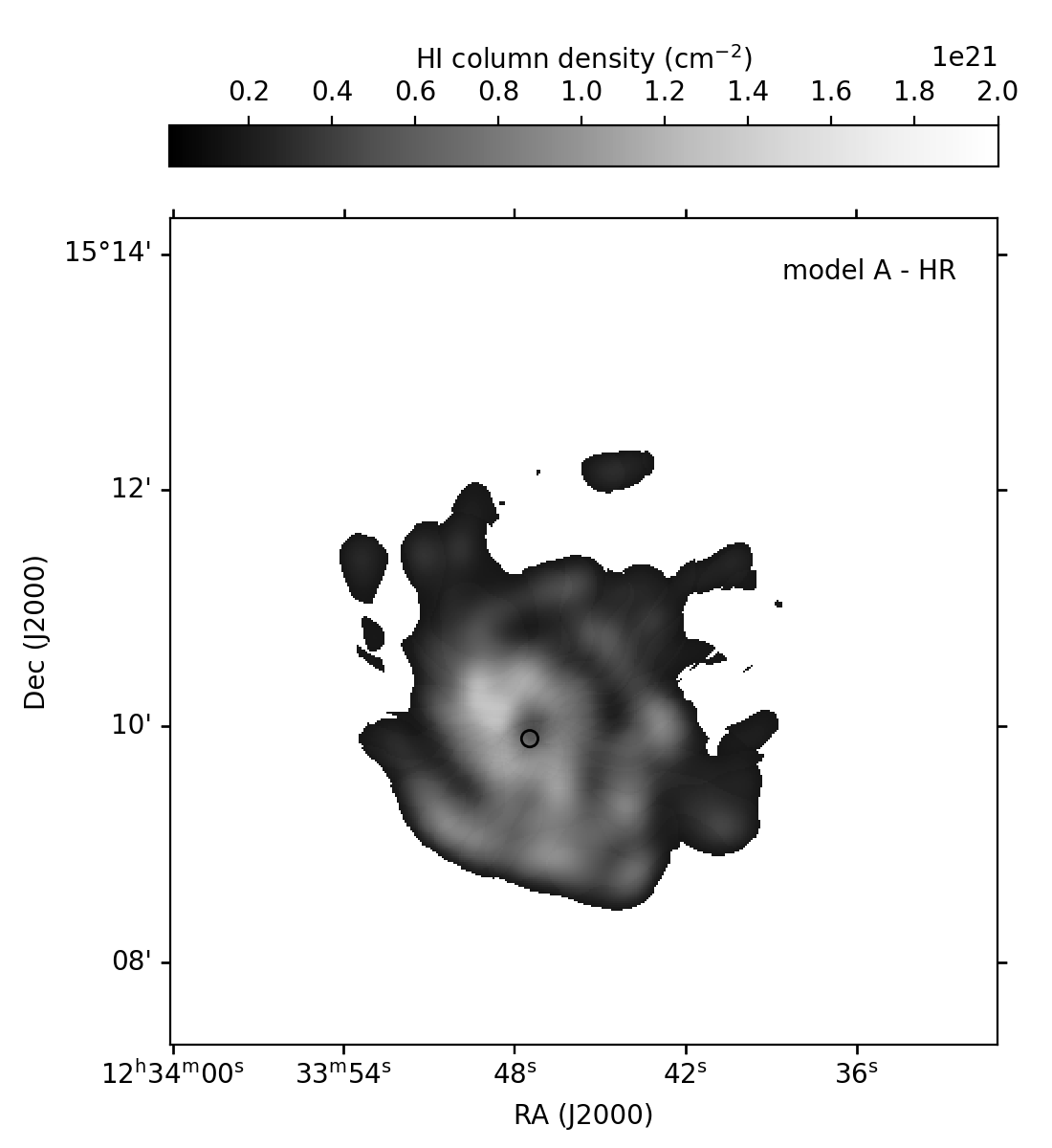}
\includegraphics[width=0.49\textwidth]{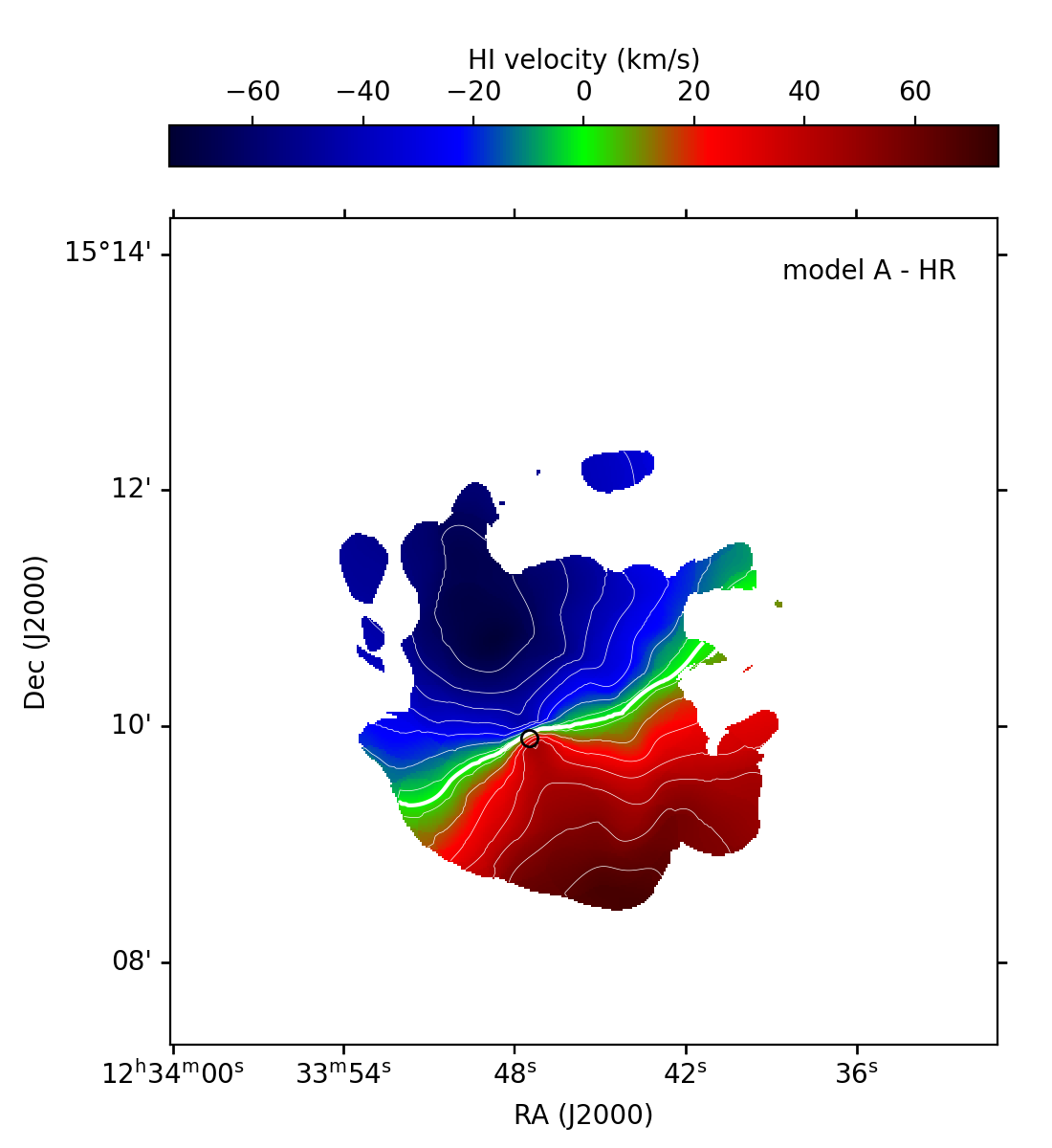}\\
\caption{Upper panels: model~A; left: high resolution model H{\sc i} moment~0 map, the contour levels show the model stellar distribution at 
10 $\times$ 10$^{0.5 \times n}$ ($n$=0,1,2,..) M$_{\odot}$ pc$^{-2}$;
right: high resolution model H{\sc i} moment~1 map, the white contour levels are from $-70$ to $70$~km\,s$^{-1}$ in steps of $10$~km\,s$^{-1}$, the black contour the model stellar distribution at 10 M$_{\odot}$ pc$^{-2}$. The empty dot shows the kinematic centre.}
\label{modelAHR}
\end{figure*}

\begin{figure*}
\centering
\includegraphics[width=0.49\textwidth]{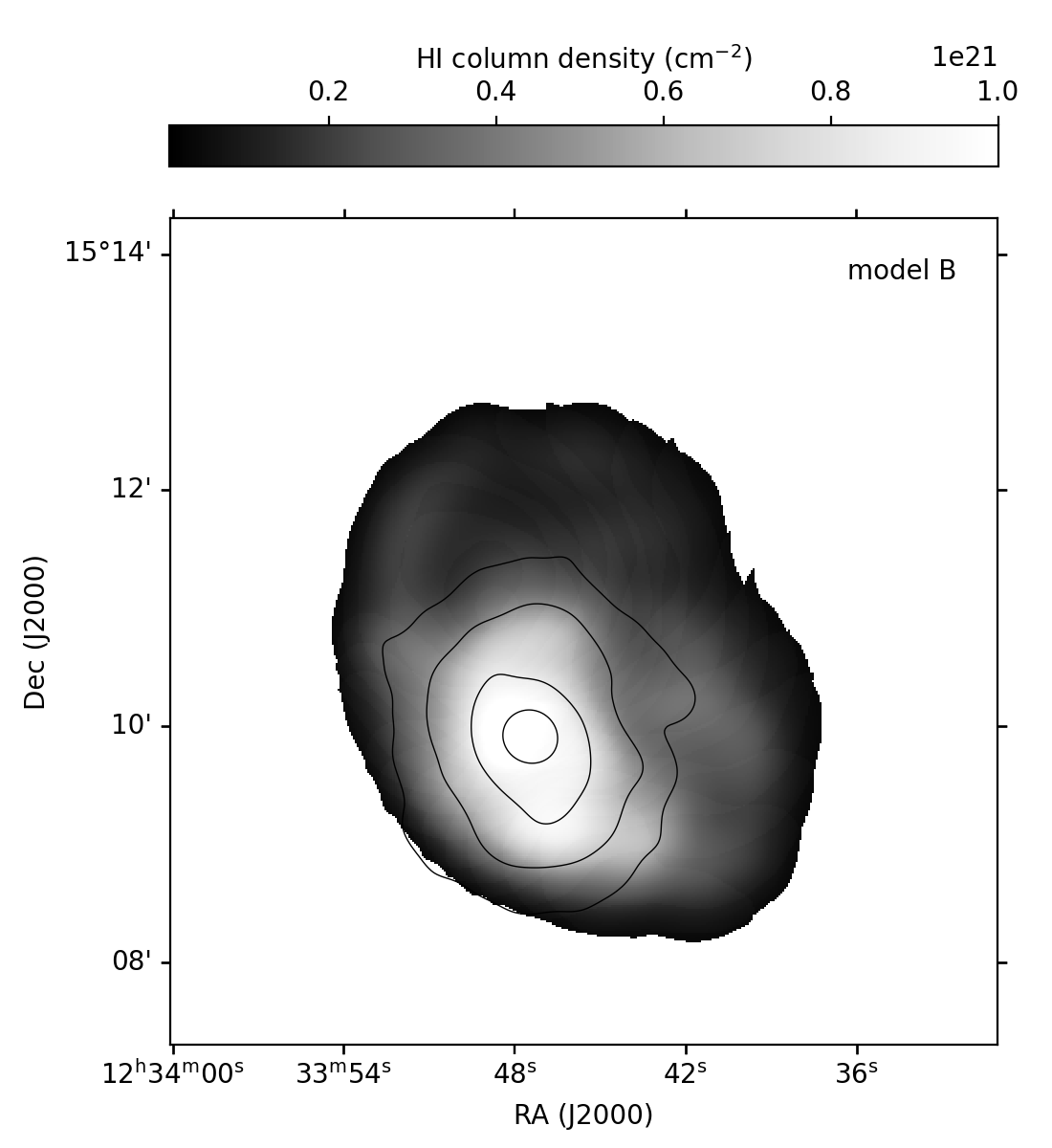}
\includegraphics[width=0.49\textwidth]{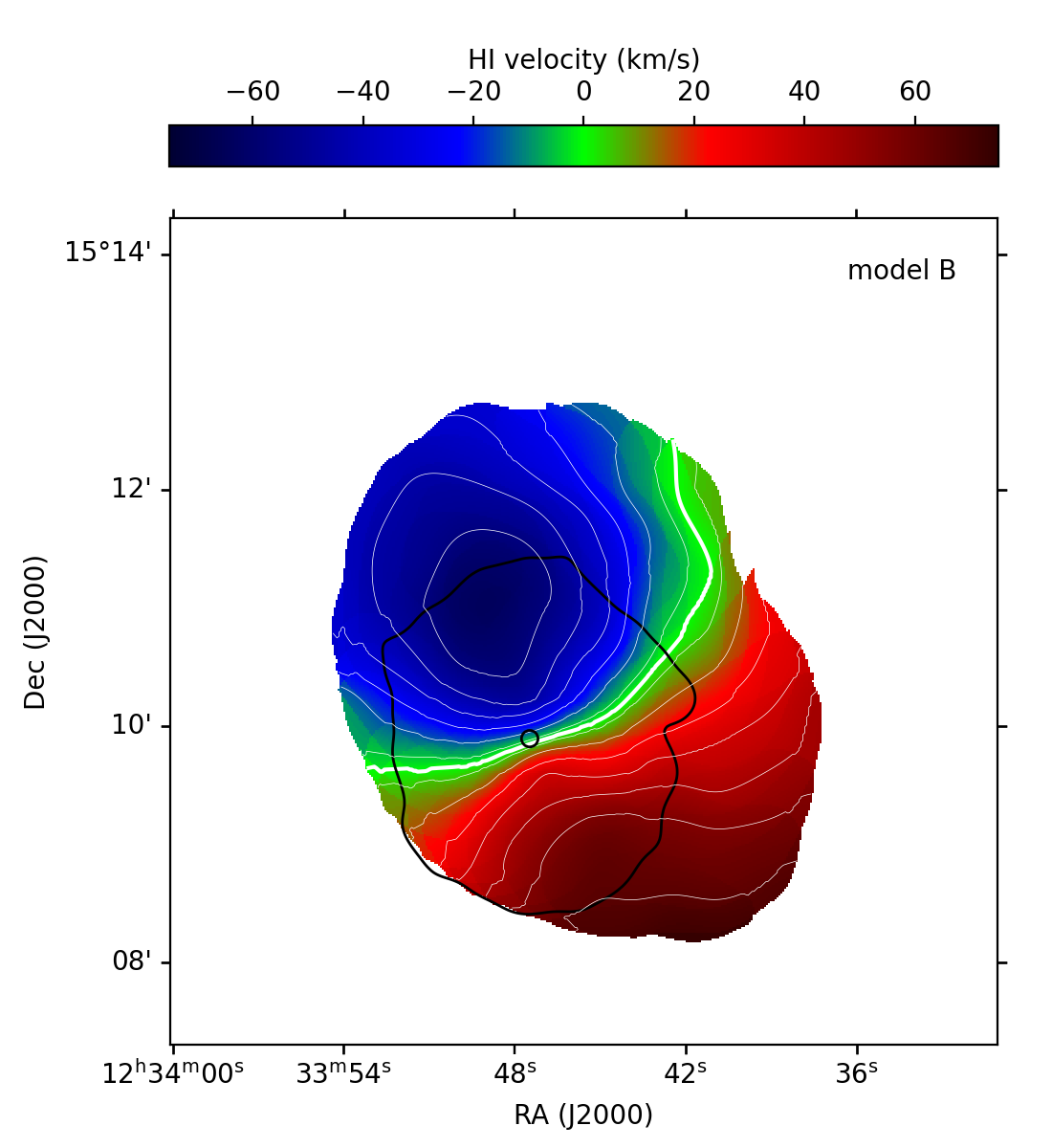}\\
\includegraphics[width=0.49\textwidth]{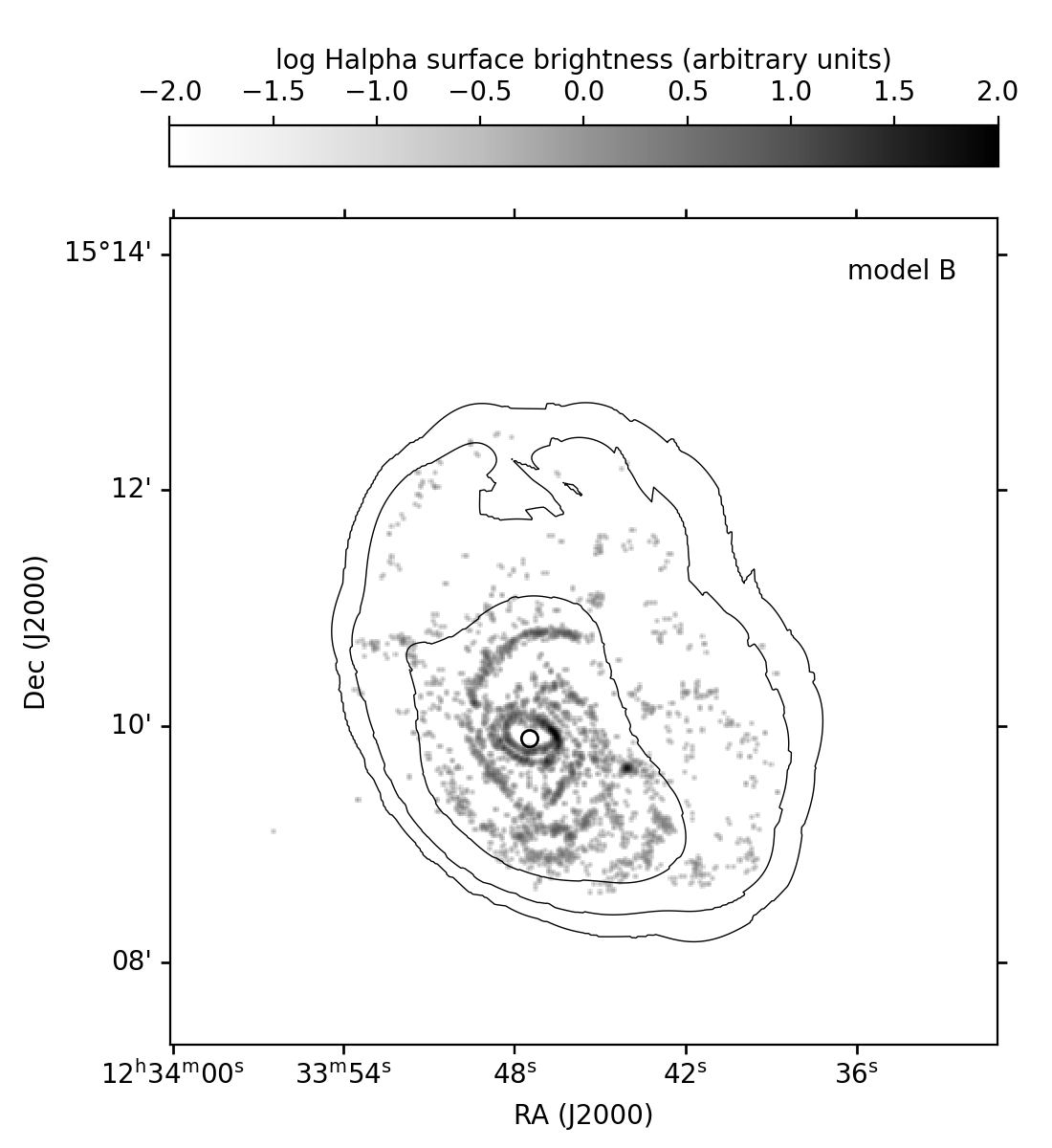}
\includegraphics[width=0.49\textwidth]{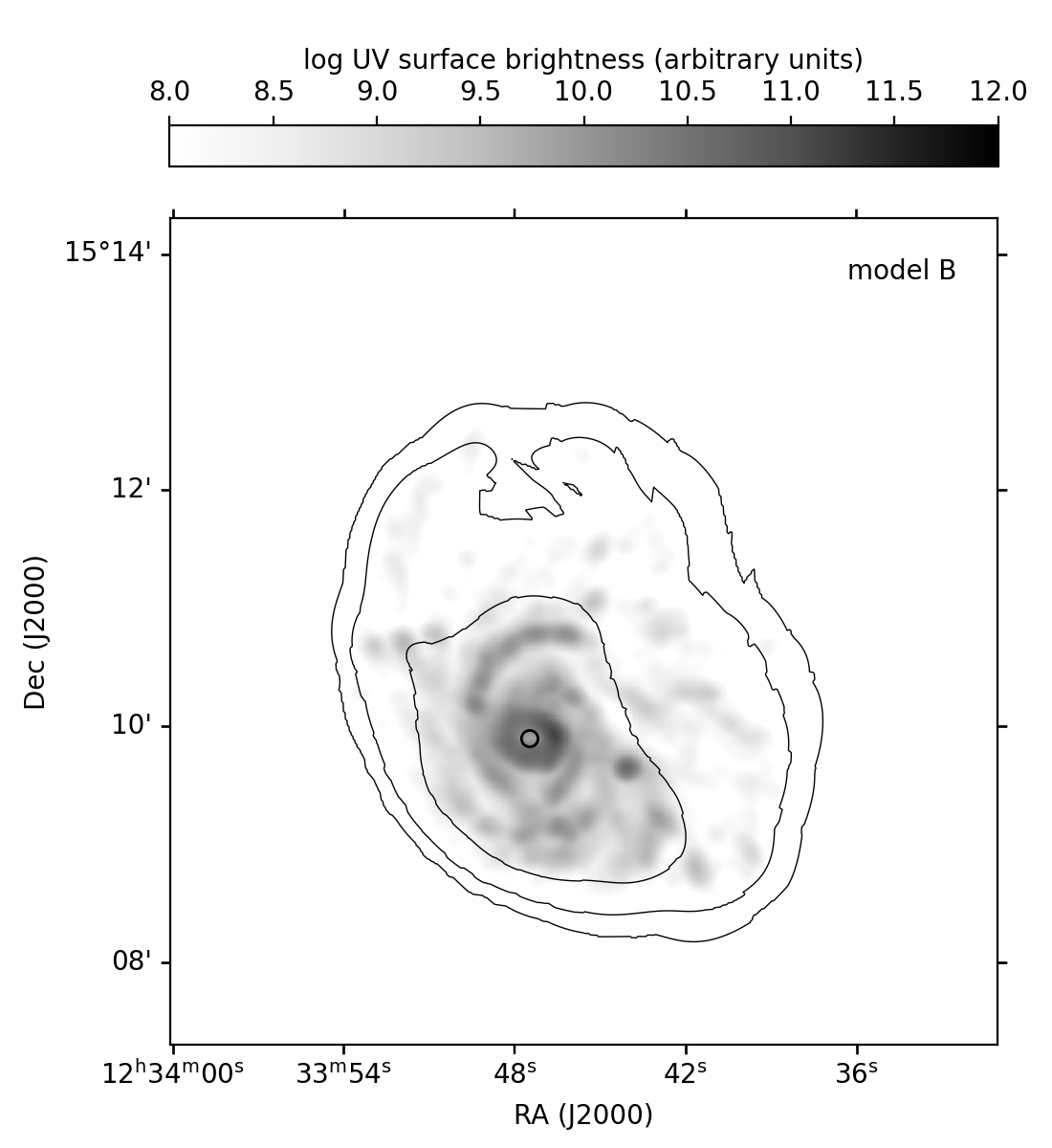}\\
\caption{Upper panels: model~B; left: model H{\sc i} moment~0 map, the contour levels show the model stellar distribution at 
10 $\times$ 10$^{0.5 \times n}$ ($n$=0,1,2,..) M$_{\odot}$ pc$^{-2}$;
right: model H{\sc i} moment~1 map, the white contour levels are from $-70$ to $70$~km\,s$^{-1}$ in steps of $10$~km\,s$^{-1}$, the black contour the model stellar distribution at 10 M$_{\odot}$ pc$^{-2}$. The empty dot shows the kinematic centre.
Lower panels: left: model H{\sc i} contours on model H$\alpha$ distribution; right:
model \hi\ contours on model FUV distribution. The \hi\ contour levels on the H$\alpha$ and the FUV images are 
$N$(\hi) = $3.9 \times 10^{19} \times 10^{0.5 n}$ cm$^{-2}$ ($n$=0,1,2...).}
\label{modelB}
\end{figure*}

\begin{figure*}
\centering
\includegraphics[width=0.49\textwidth]{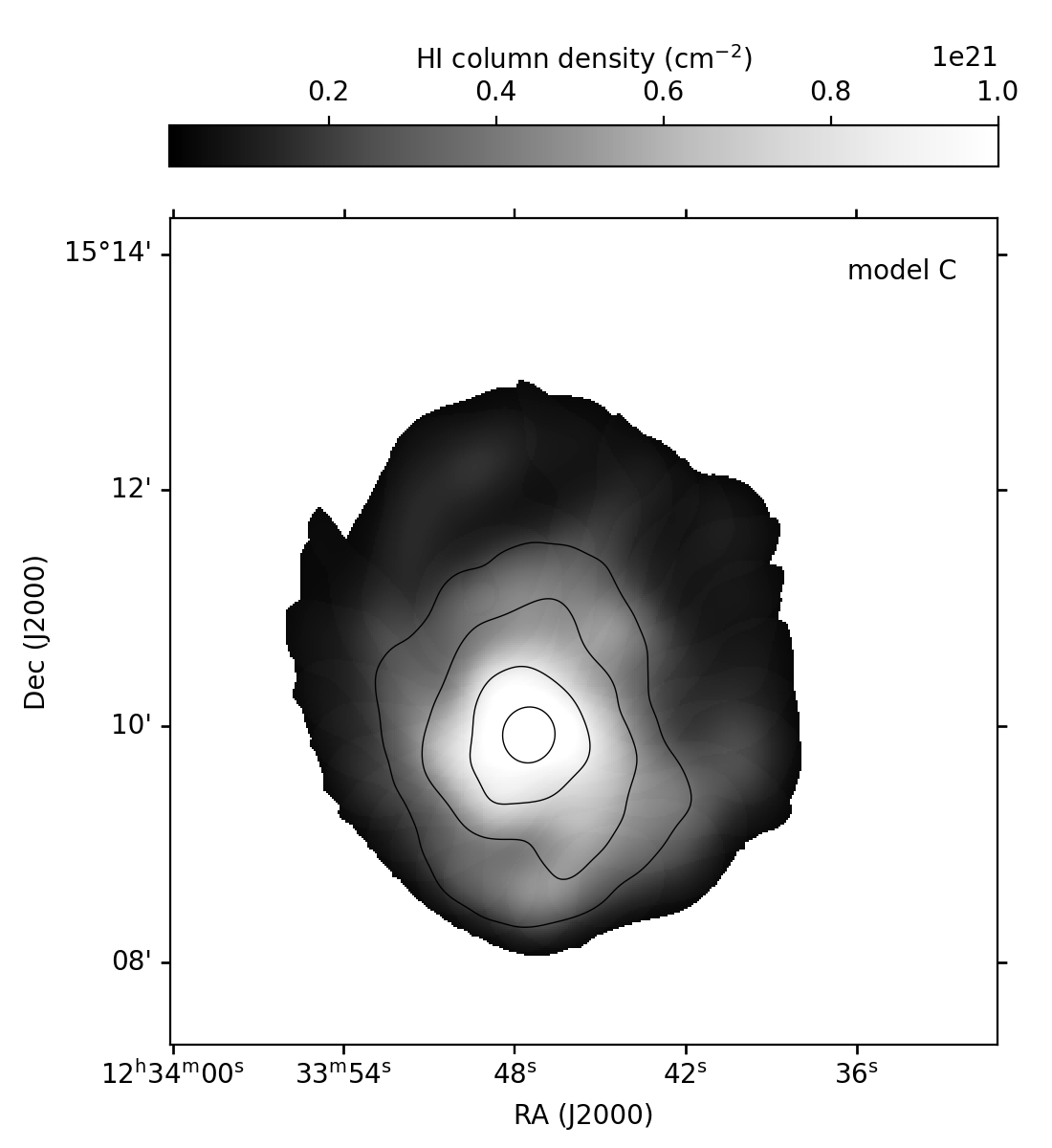}
\includegraphics[width=0.49\textwidth]{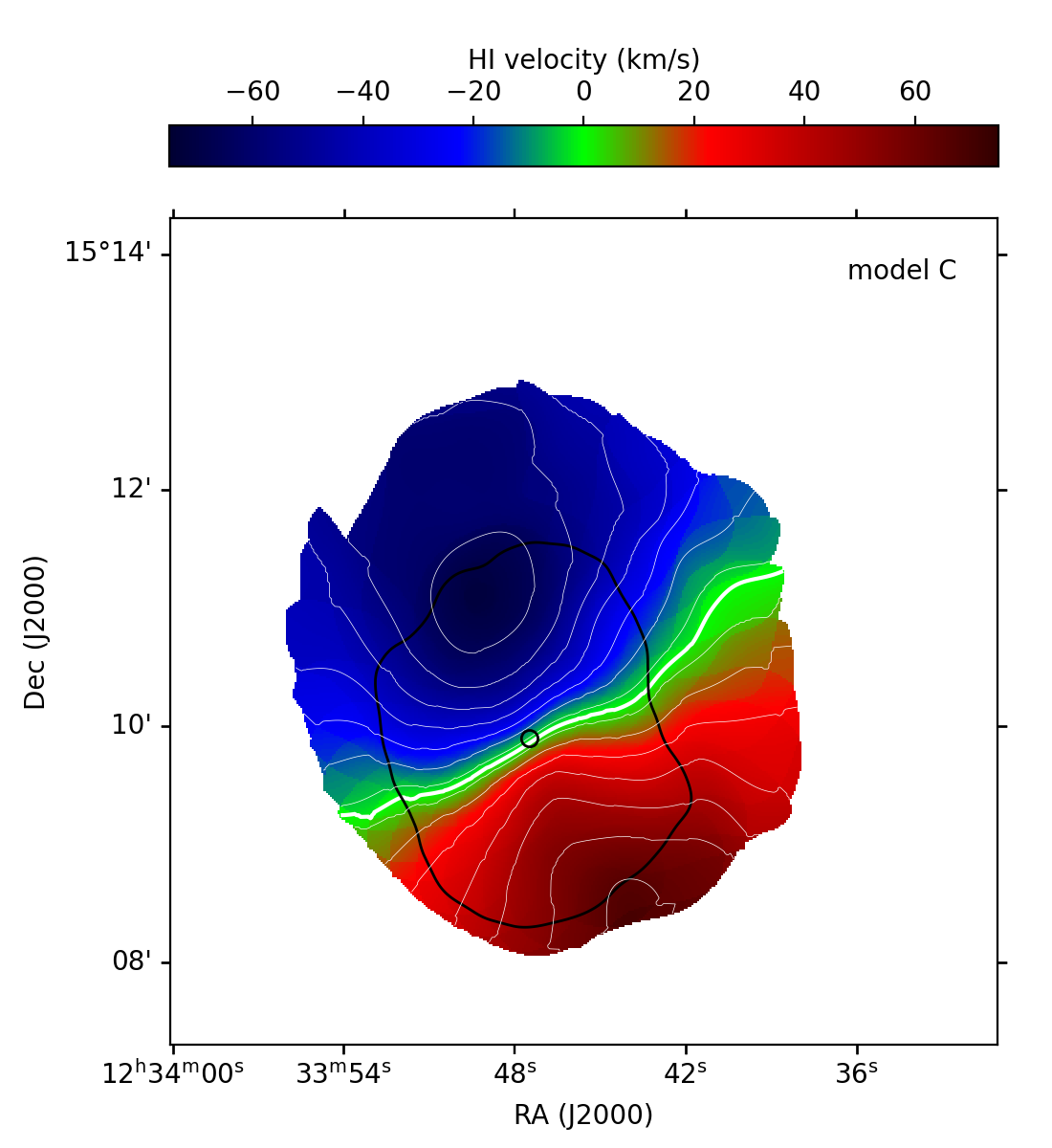}\\
\includegraphics[width=0.49\textwidth]{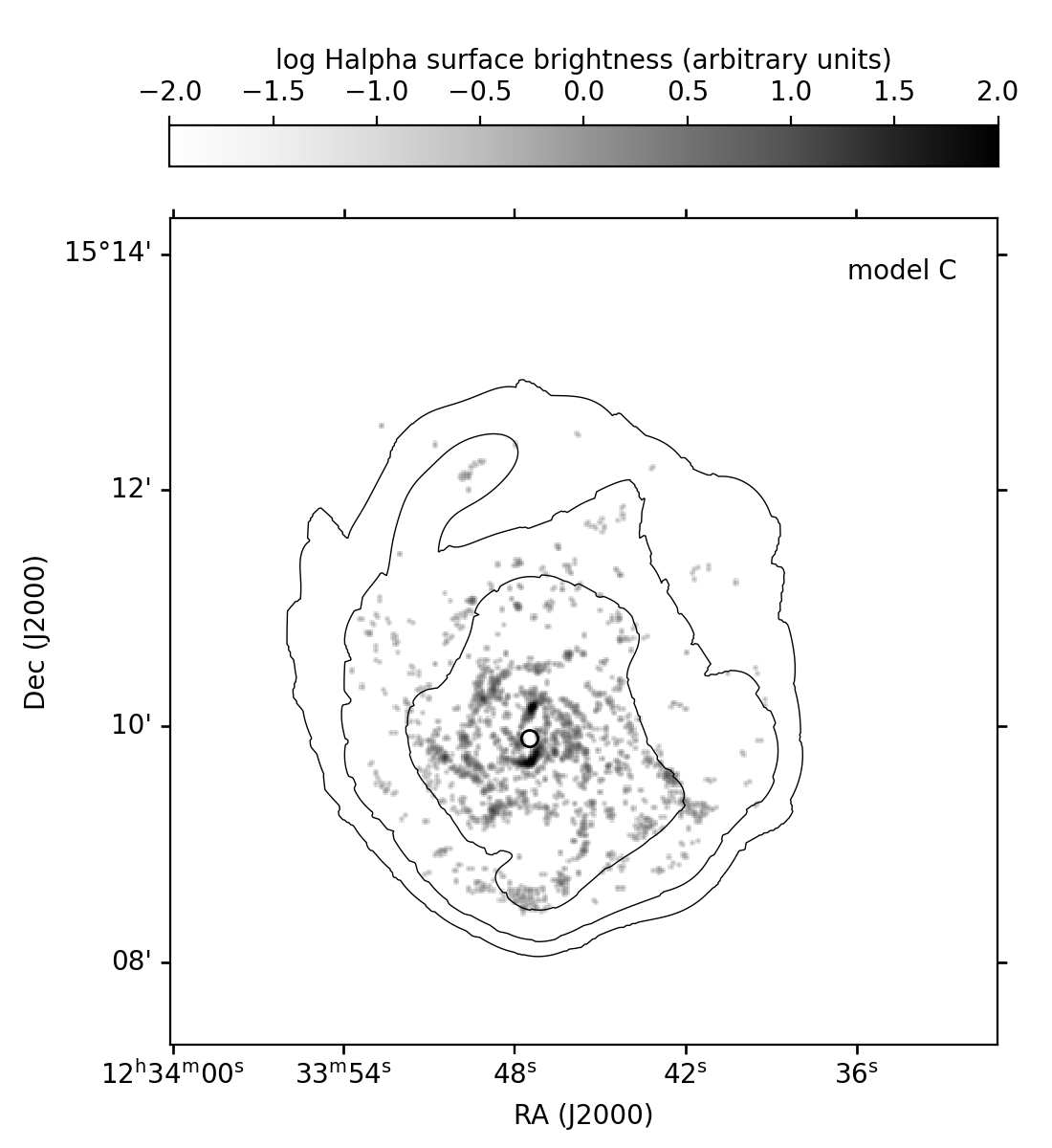}
\includegraphics[width=0.49\textwidth]{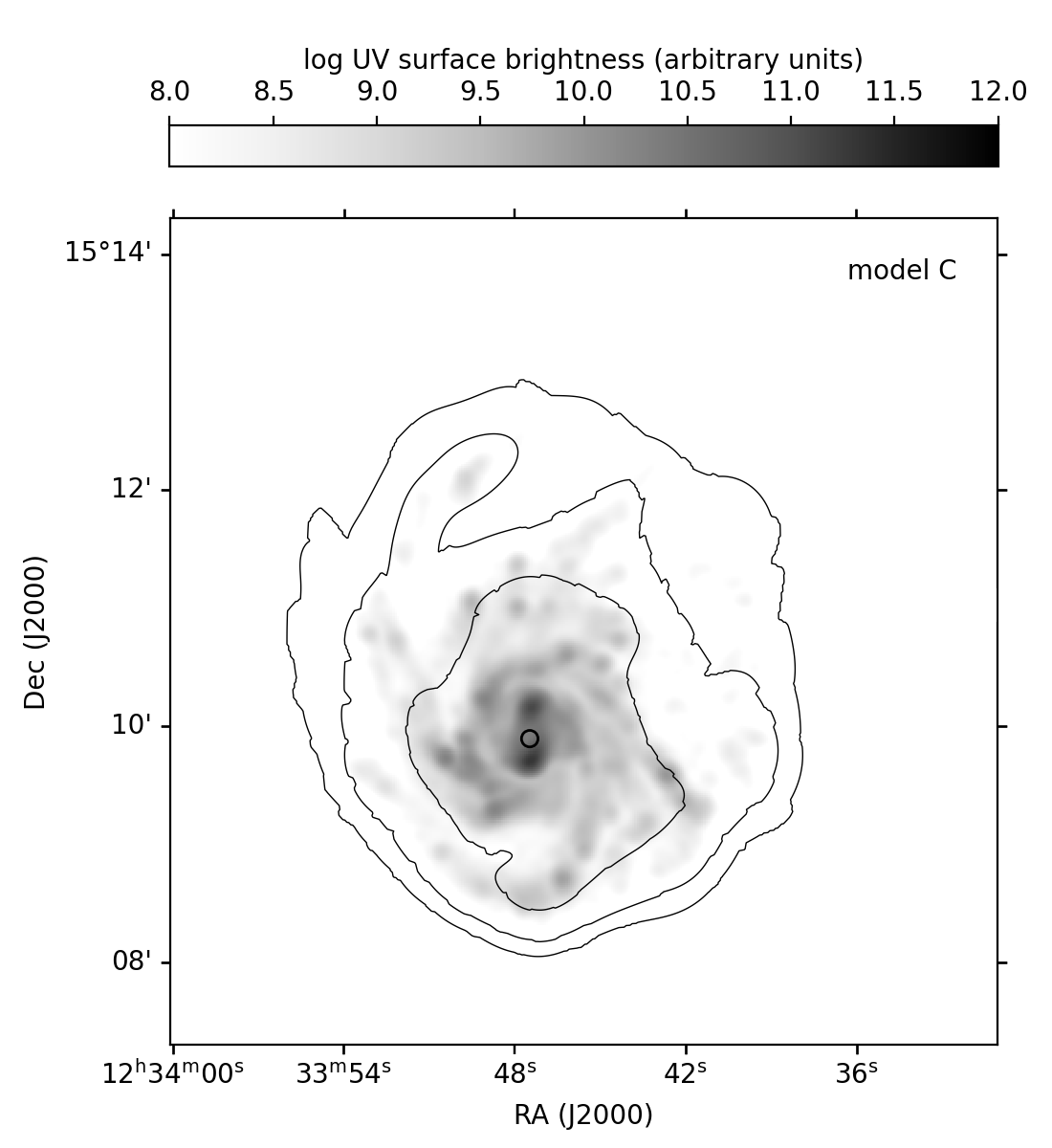}\\
\caption{Upper panels: model~C; left: model H{\sc i} moment~0 map, the contour levels show the model stellar distribution at 
10 $\times$ 10$^{0.5 \times n}$ ($n$=0,1,2,..) M$_{\odot}$ pc$^2$;
right: model H{\sc i} moment~1 map, the white contour levels are from $-70$ to $70$~km\,s$^{-1}$ in steps of $10$~km\,s$^{-1}$, the black contour the model stellar distribution at 10 M$_{\odot}$ pc$^{-2}$. The empty dot shows the kinematic centre.
Lower panels: left: model H{\sc i} contours on model H$\alpha$ distribution; right:pc$^{-2}$
model H{\sc i} contours on model FUV distribution. The H{\sc i} contour levels on the H$\alpha$ and the FUV images are $N(HI)$ = 3.9$\times$10$^{19}$ $\times$ 10$^{0.5\times n}$ ($n$=0,1,2...) cm$^{-2}$. }
\label{modelC}
\end{figure*}

\begin{figure*}
\centering
\includegraphics[width=0.49\textwidth]{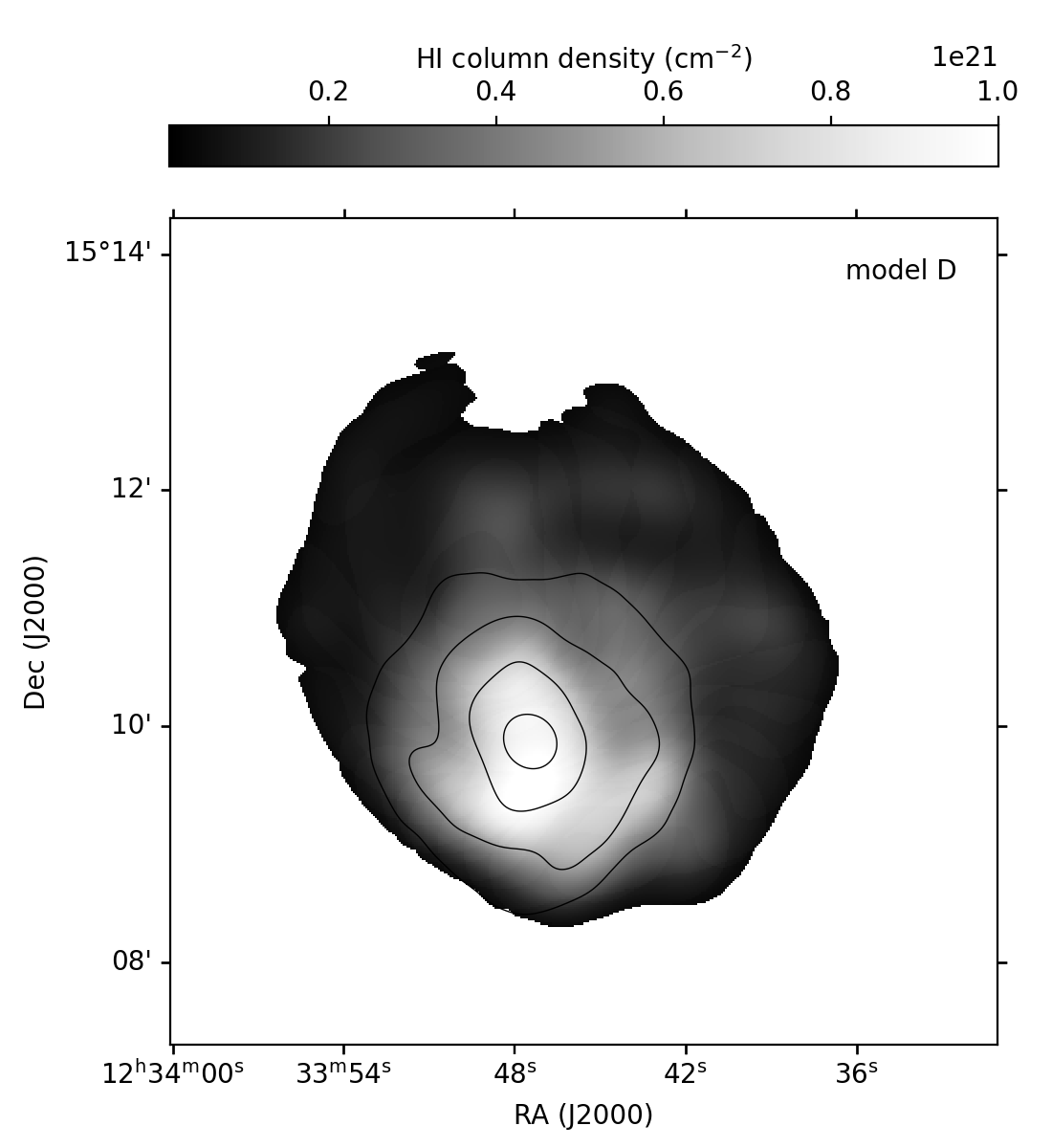}
\includegraphics[width=0.49\textwidth]{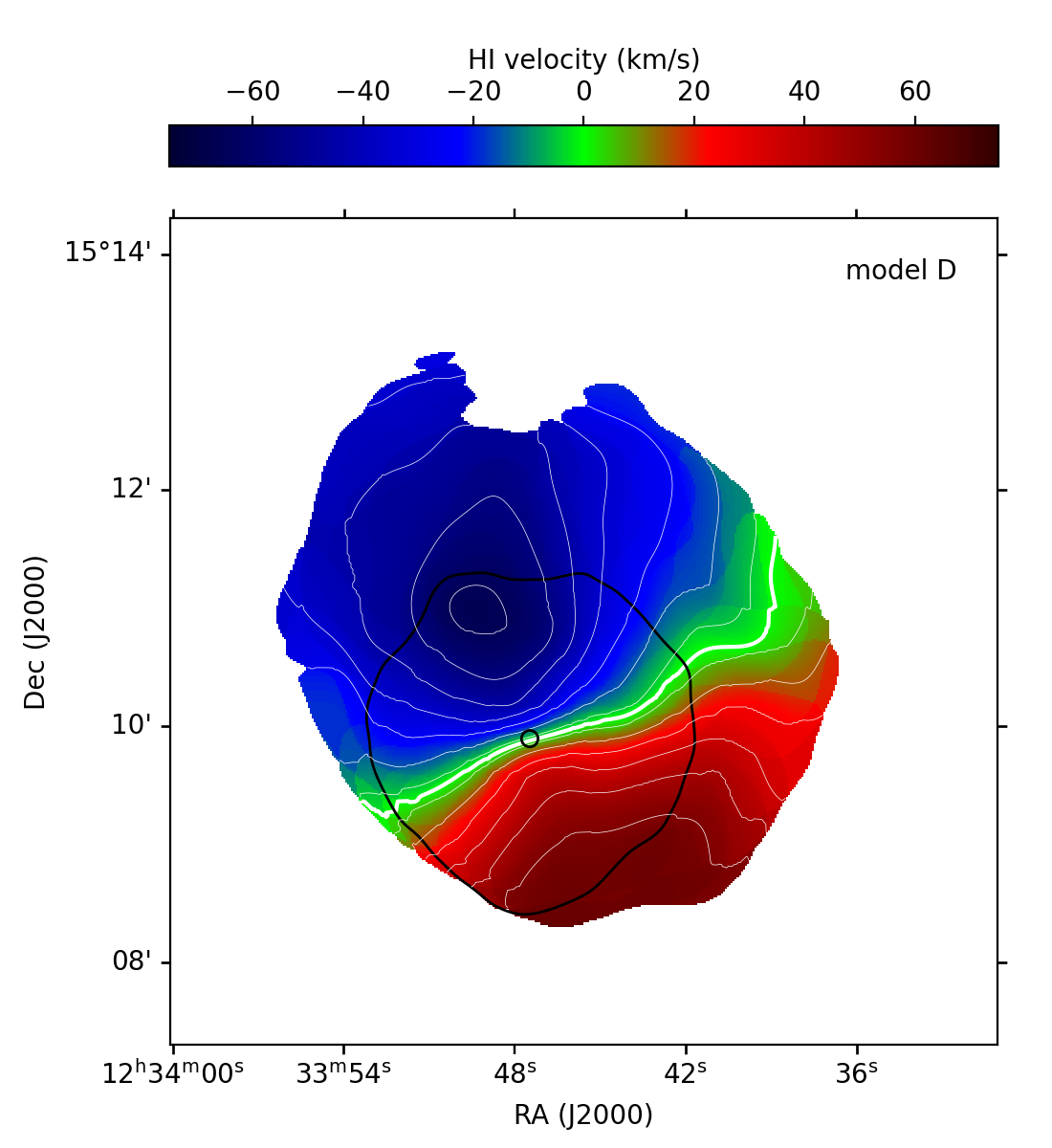}\\
\includegraphics[width=0.49\textwidth]{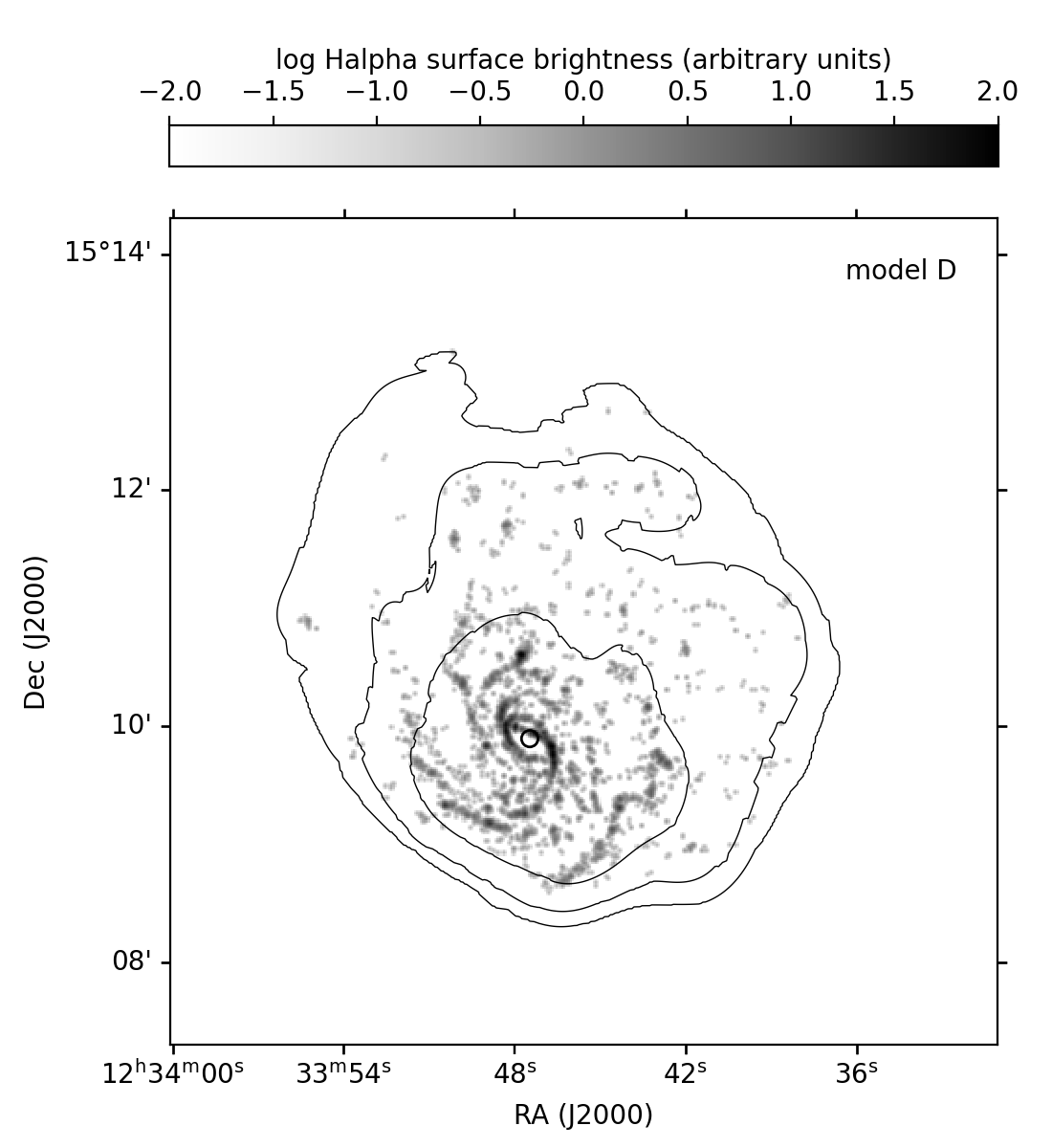}
\includegraphics[width=0.49\textwidth]{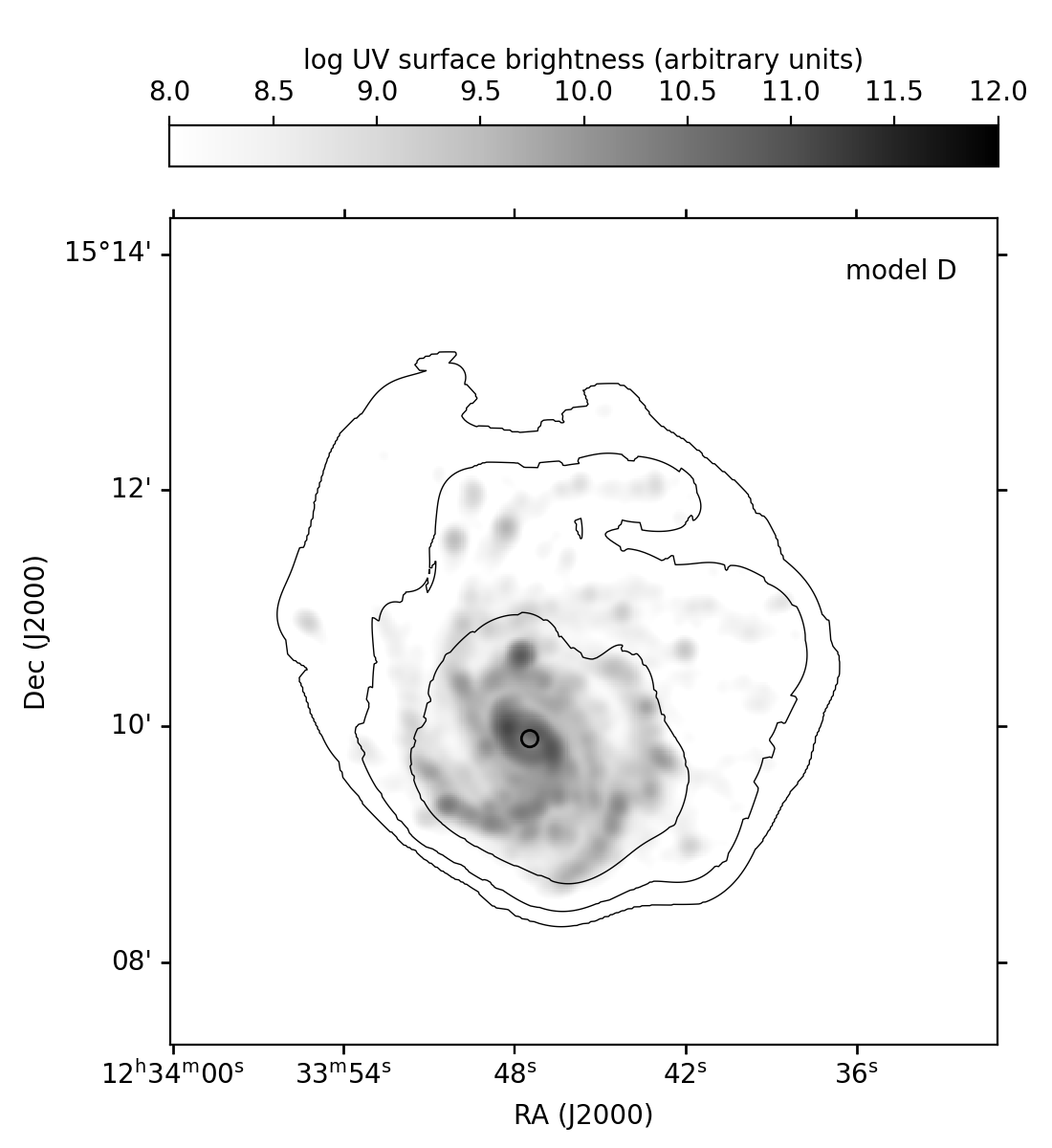}\\
\caption{Upper panels: model~D; left: model H{\sc i} moment~0 map, the contour levels show the model stellar distribution at 10 $\times$ 10$^{0.5 \times n}$ ($n$=0,1,2,..) M$_{\odot}$ pc$^{-2}$;
right: model H{\sc i} moment~1 map, the white contour levels are from $-70$ to $70$~km\,s$^{-1}$ in steps of $10$~km\,s$^{-1}$, the black contour the model stellar distribution at 10 M$_{\odot}$ pc$^{-2}$. The empty dot shows the kinematic centre.
Lower panels: left: model H{\sc i} contours on model H$\alpha$ distribution; right:
model H{\sc i} contours on model FUV distribution. The H{\sc i} contour levels on the H$\alpha$ and the FUV images are $N(HI)$ = 3.9$\times$10$^{19}$ $\times$ 10$^{0.5\times n}$ ($n$=0,1,2...) cm$^{-2}$. }
\label{modelD}
\end{figure*}

\begin{figure*}
\centering
\includegraphics[width=0.49\textwidth]{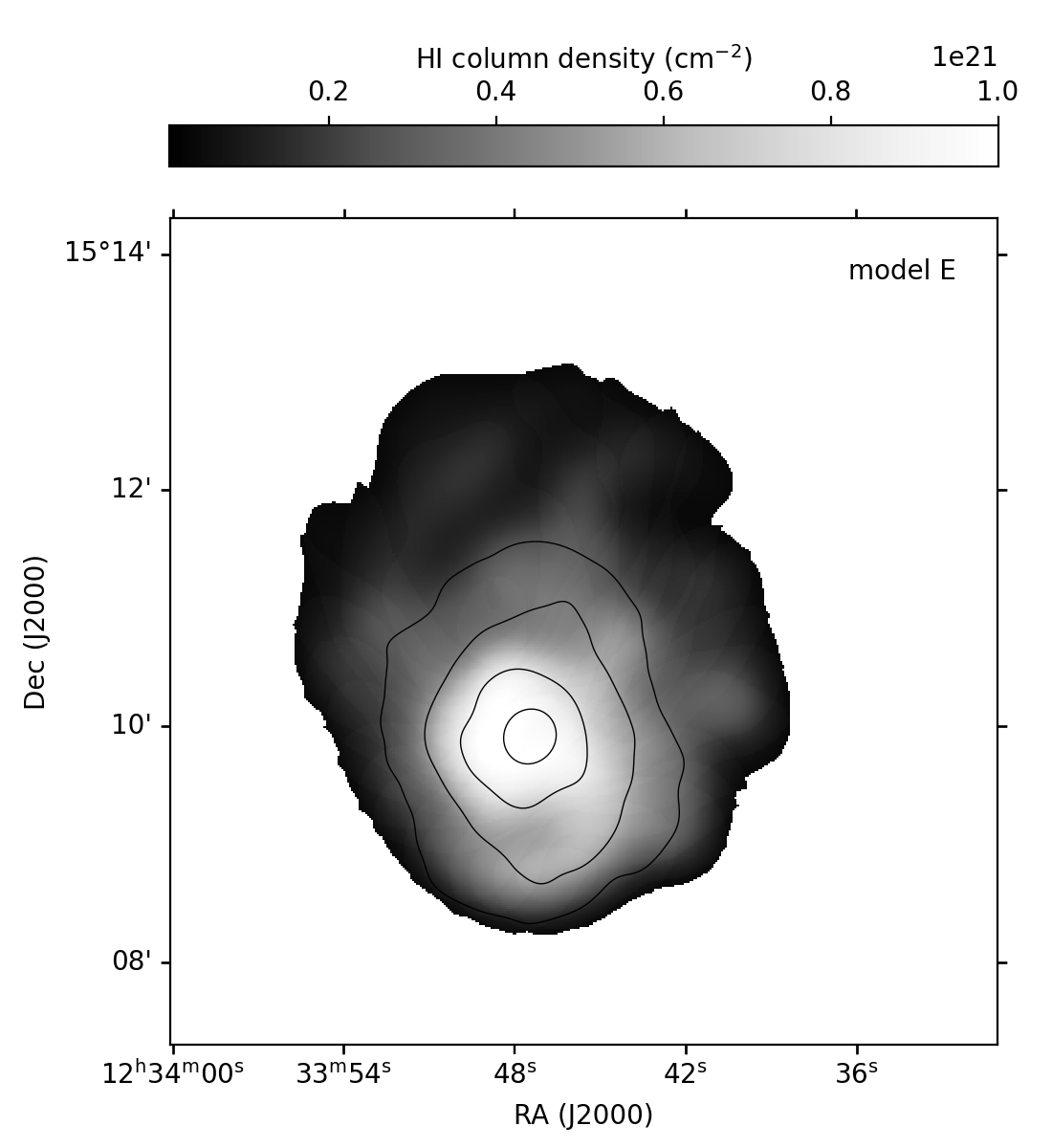}
\includegraphics[width=0.49\textwidth]{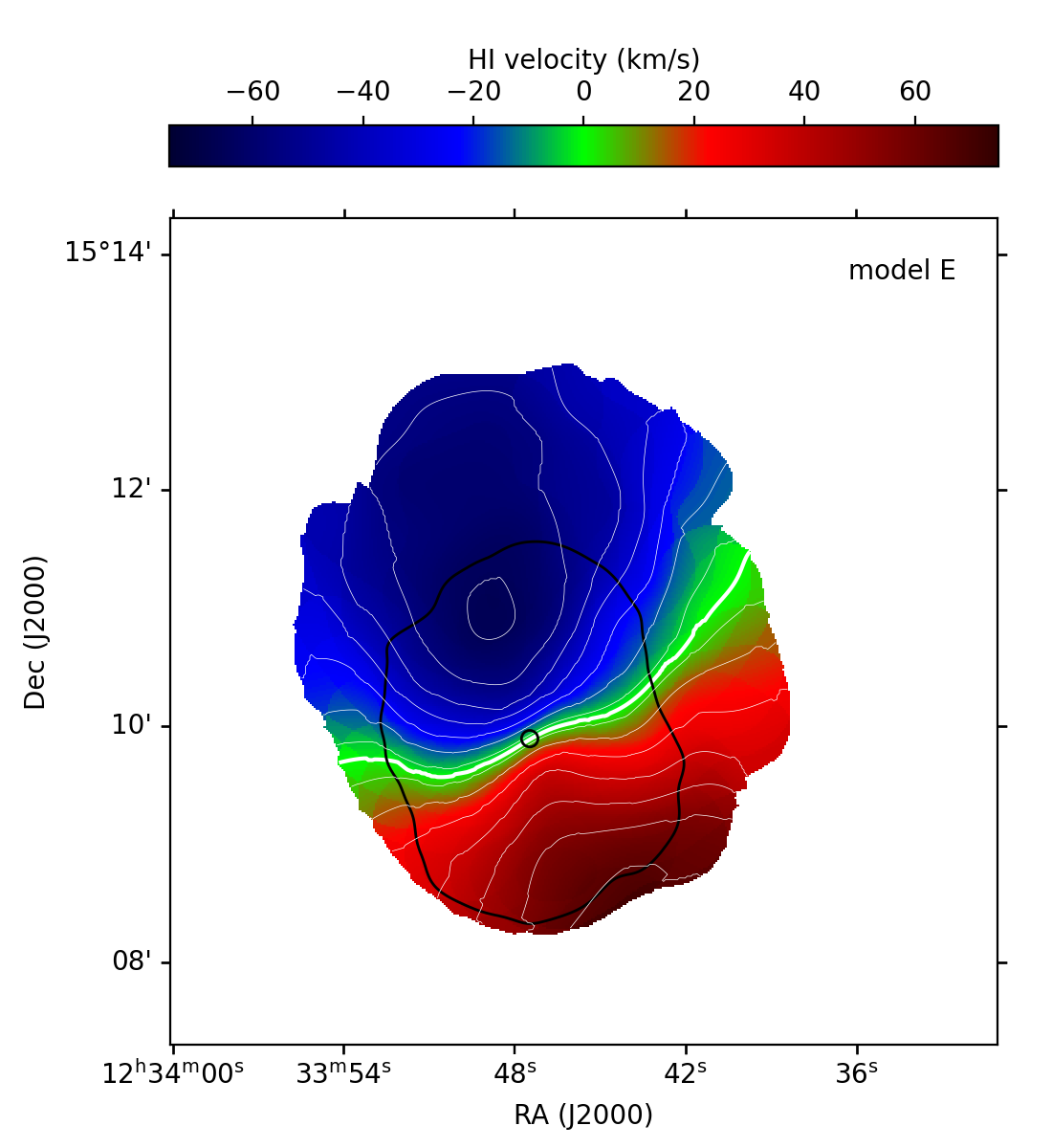}\\
\includegraphics[width=0.49\textwidth]{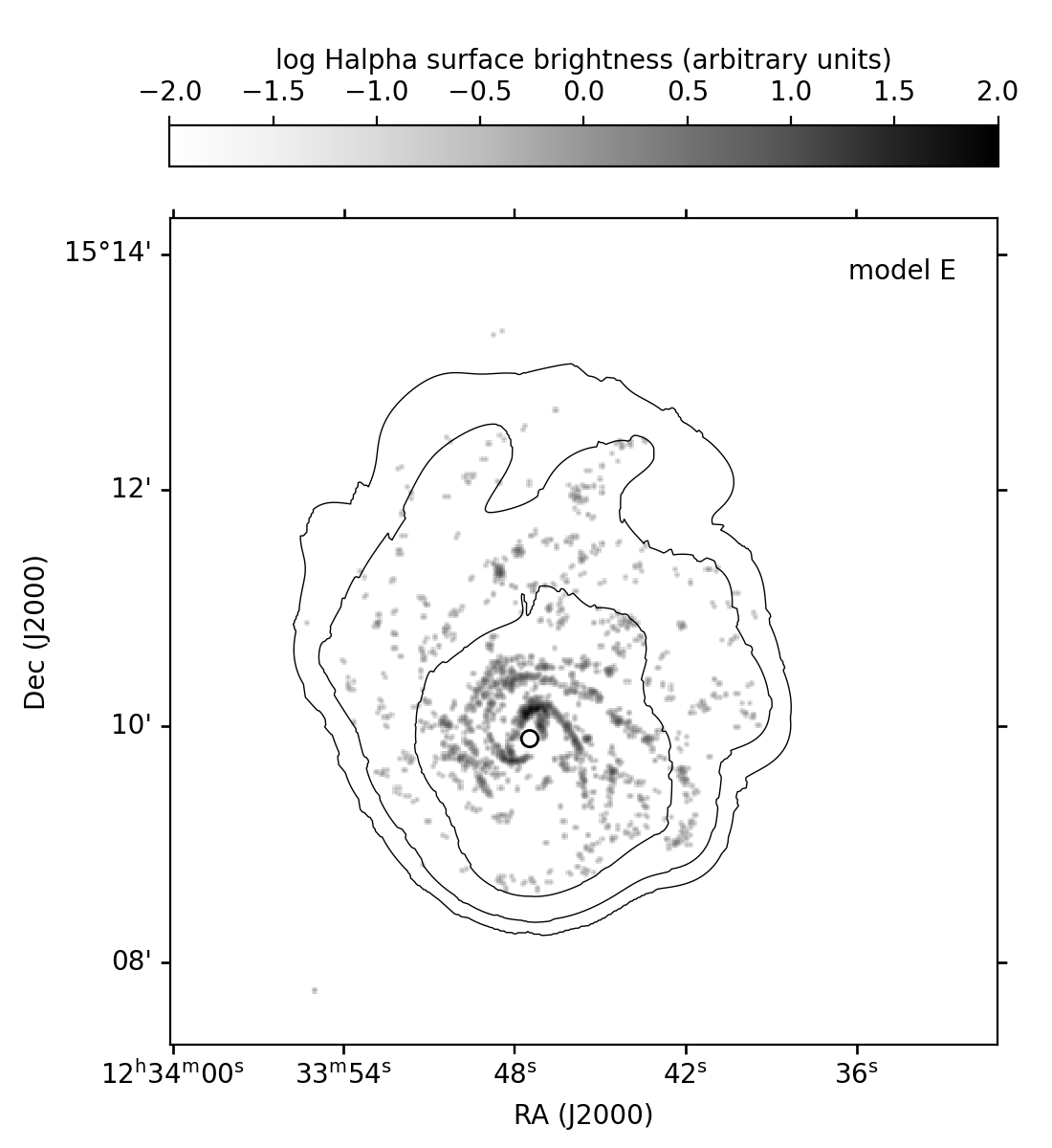}
\includegraphics[width=0.49\textwidth]{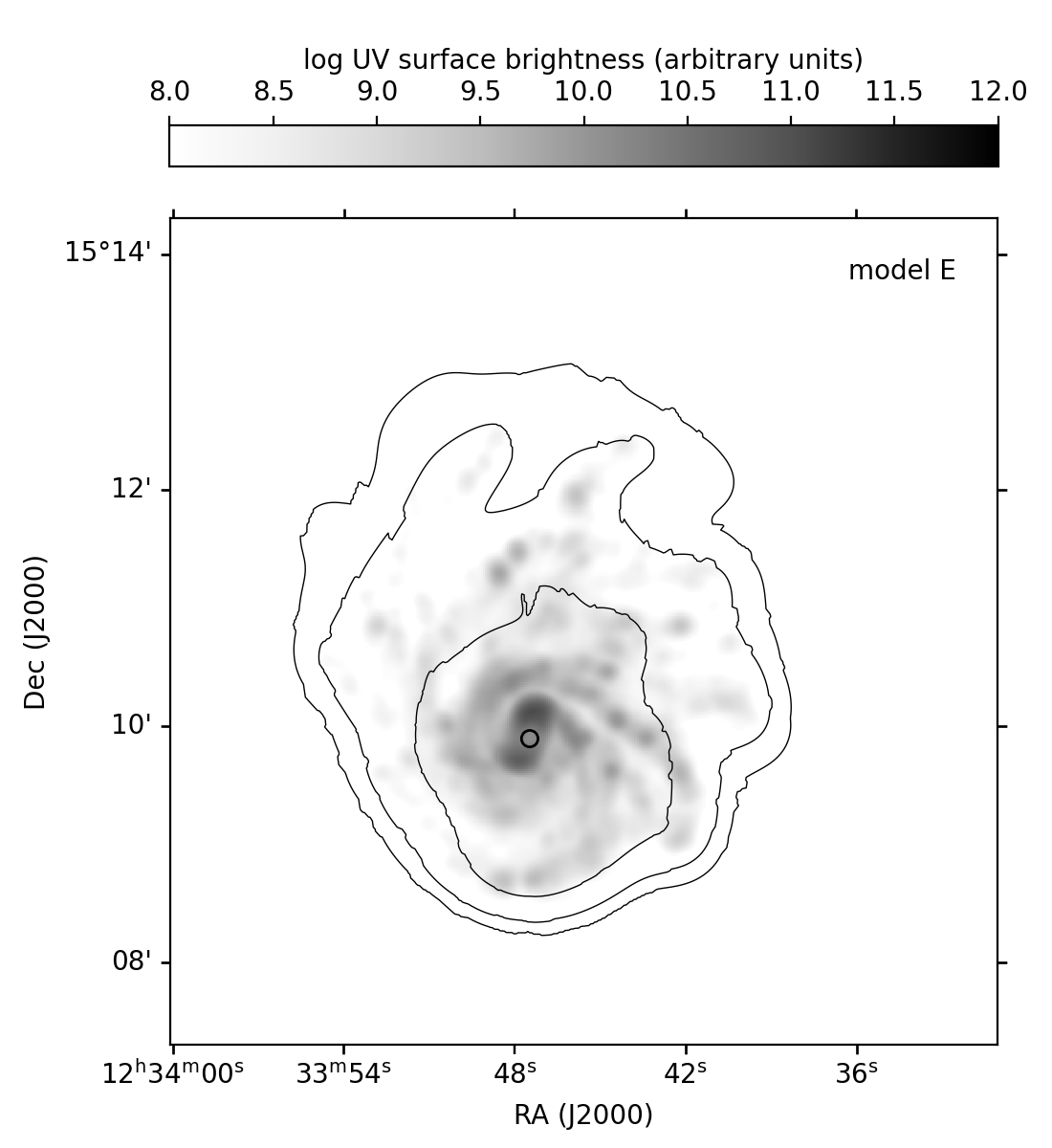}\\
\caption{Upper panels: model~E; left: model H{\sc i} moment~0 map, the contour levels show the model stellar distribution at 10 $\times$ 10$^{0.5 \times n}$ ($n$=0,1,2,..) M$_{\odot}$ pc$^{-2}$;
right: model H{\sc i} moment~1 map, the white contour levels are from $-70$ to $70$~km\,s$^{-1}$ in steps of $10$~km\,s$^{-1}$, the black contour the model stellar distribution at 10 M$_{\odot}$ pc$^{-2}$. The empty dot shows the kinematic centre.
Lower panels: left: model H{\sc i} contours on model H$\alpha$ distribution; right:
model H{\sc i} contours on model FUV distribution. The H{\sc i} contour levels on the H$\alpha$ and the FUV images are $N(HI)$ = 3.9$\times$10$^{19}$ $\times$ 10$^{0.5\times n}$ ($n$=0,1,2...) cm$^{-2}$. }
\label{modelE}
\end{figure*}

\begin{figure*}
\centering
\includegraphics[width=0.49\textwidth]{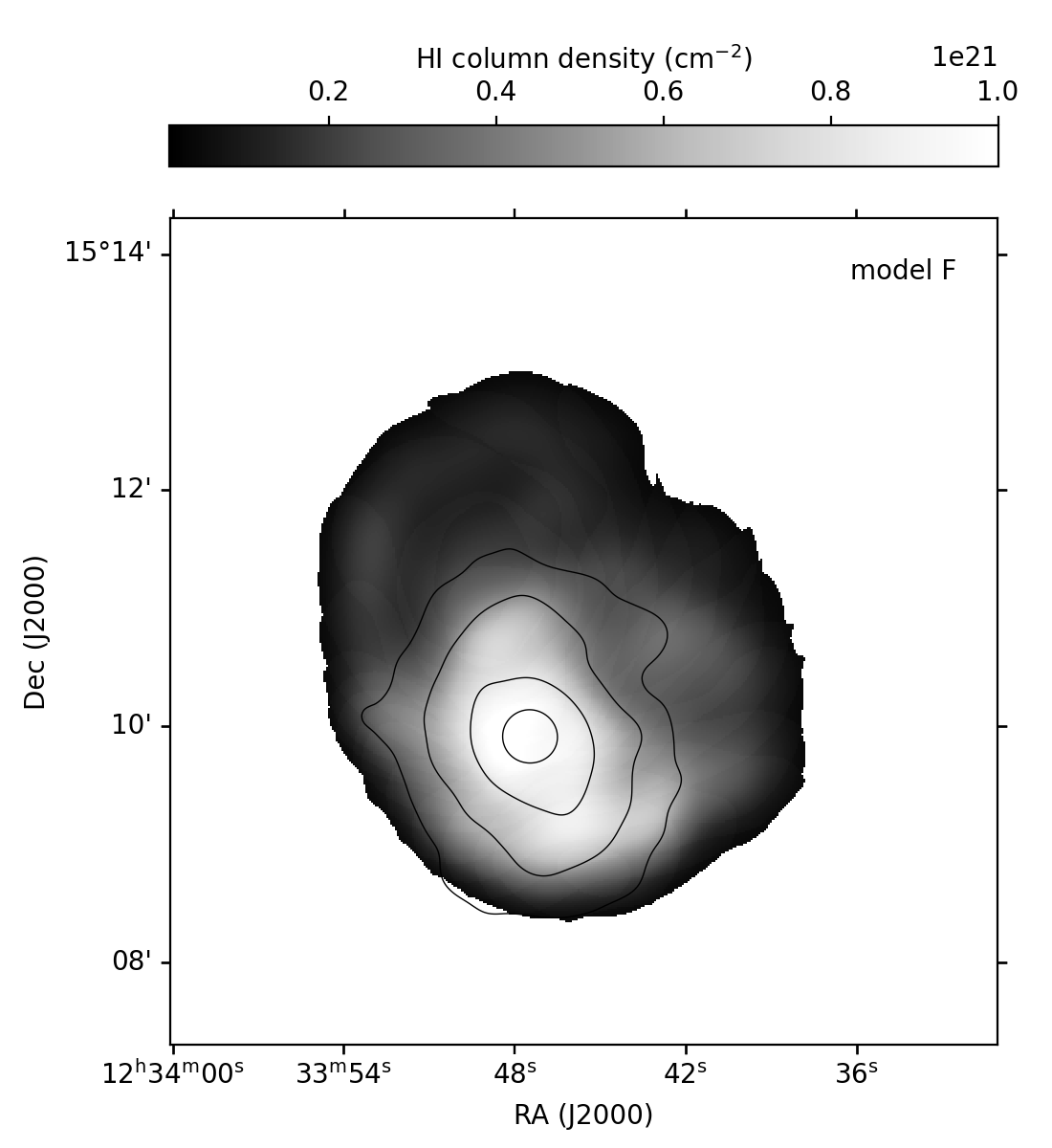}
\includegraphics[width=0.49\textwidth]{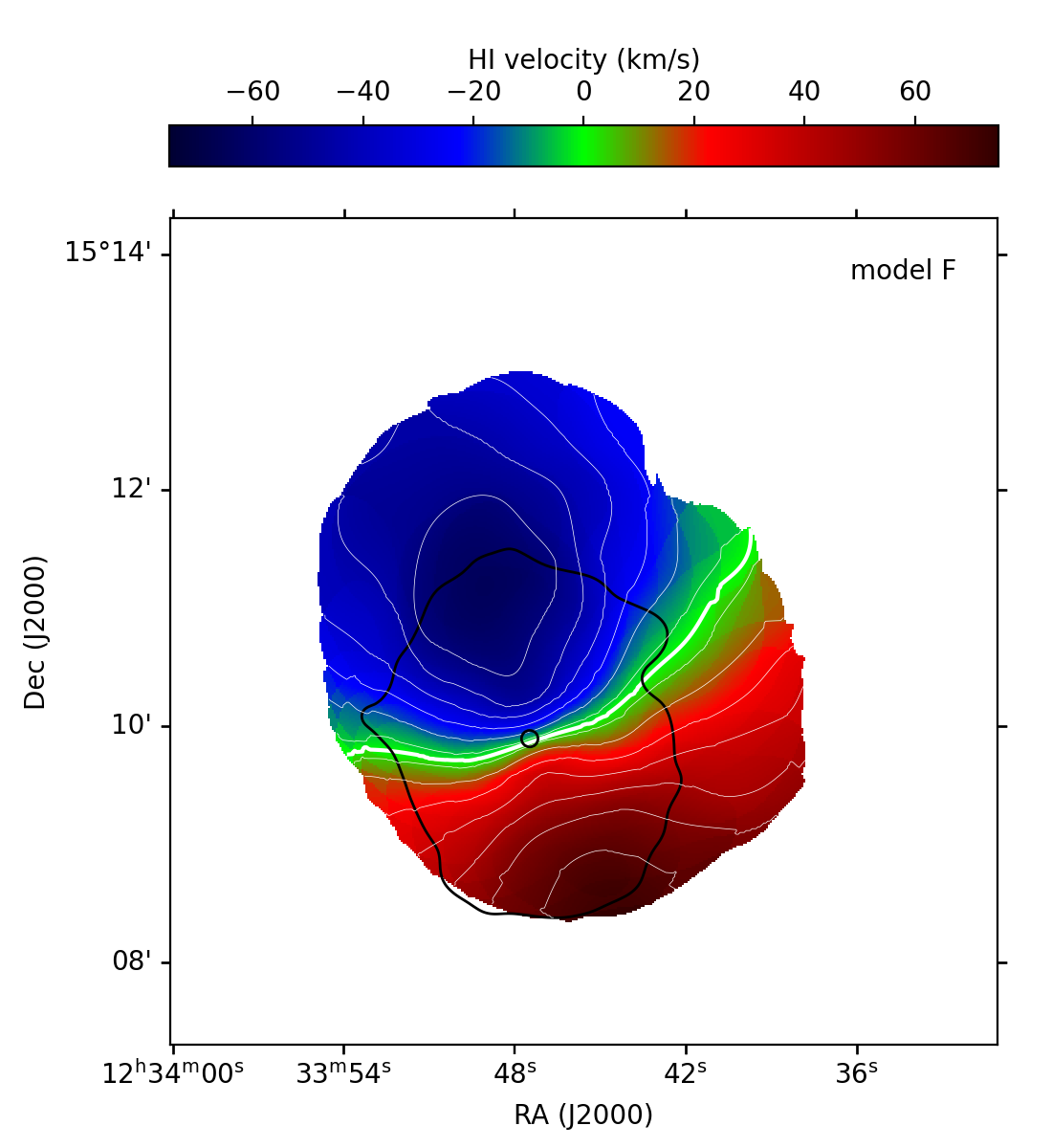}\\
\includegraphics[width=0.49\textwidth]{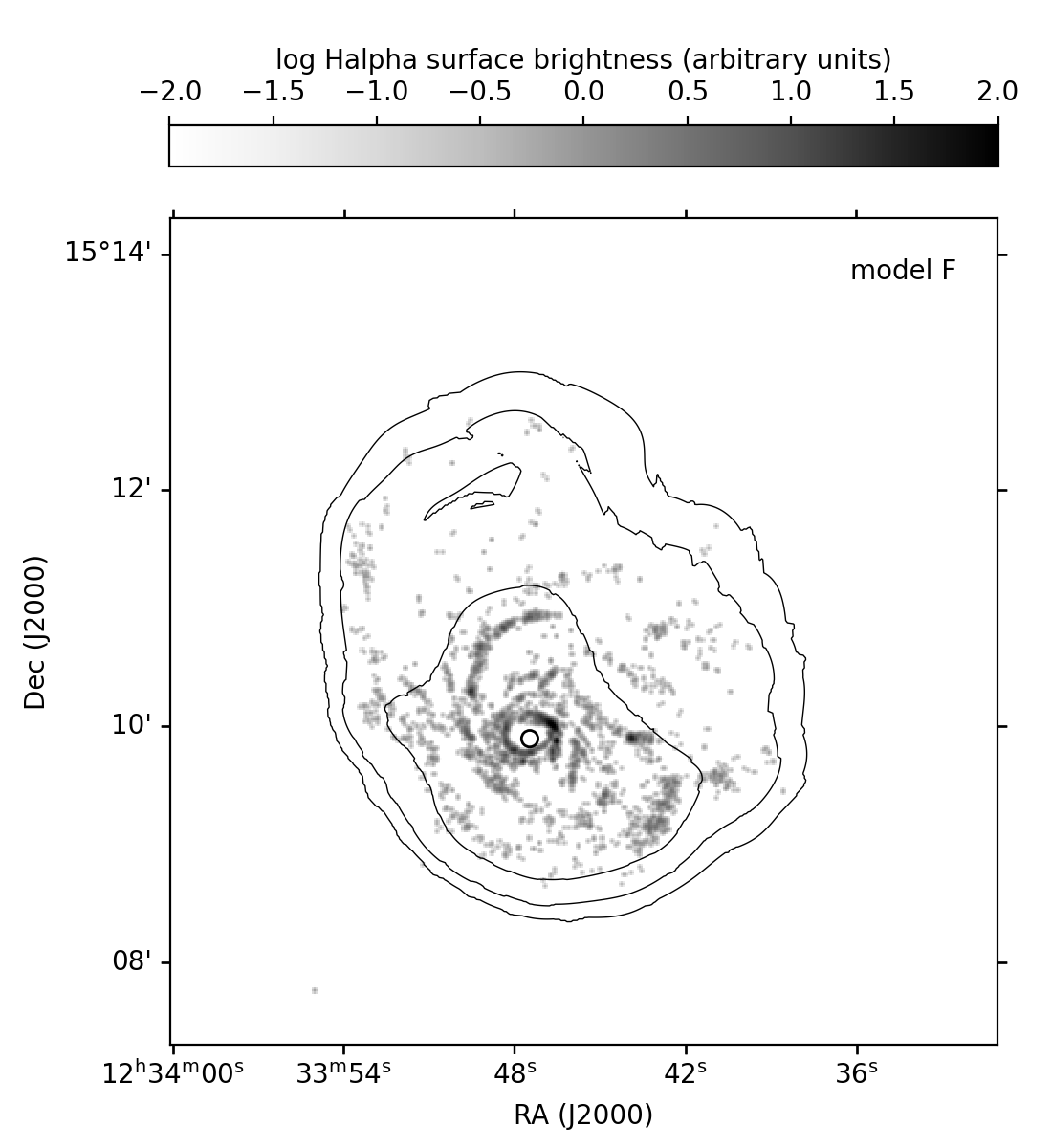}
\includegraphics[width=0.49\textwidth]{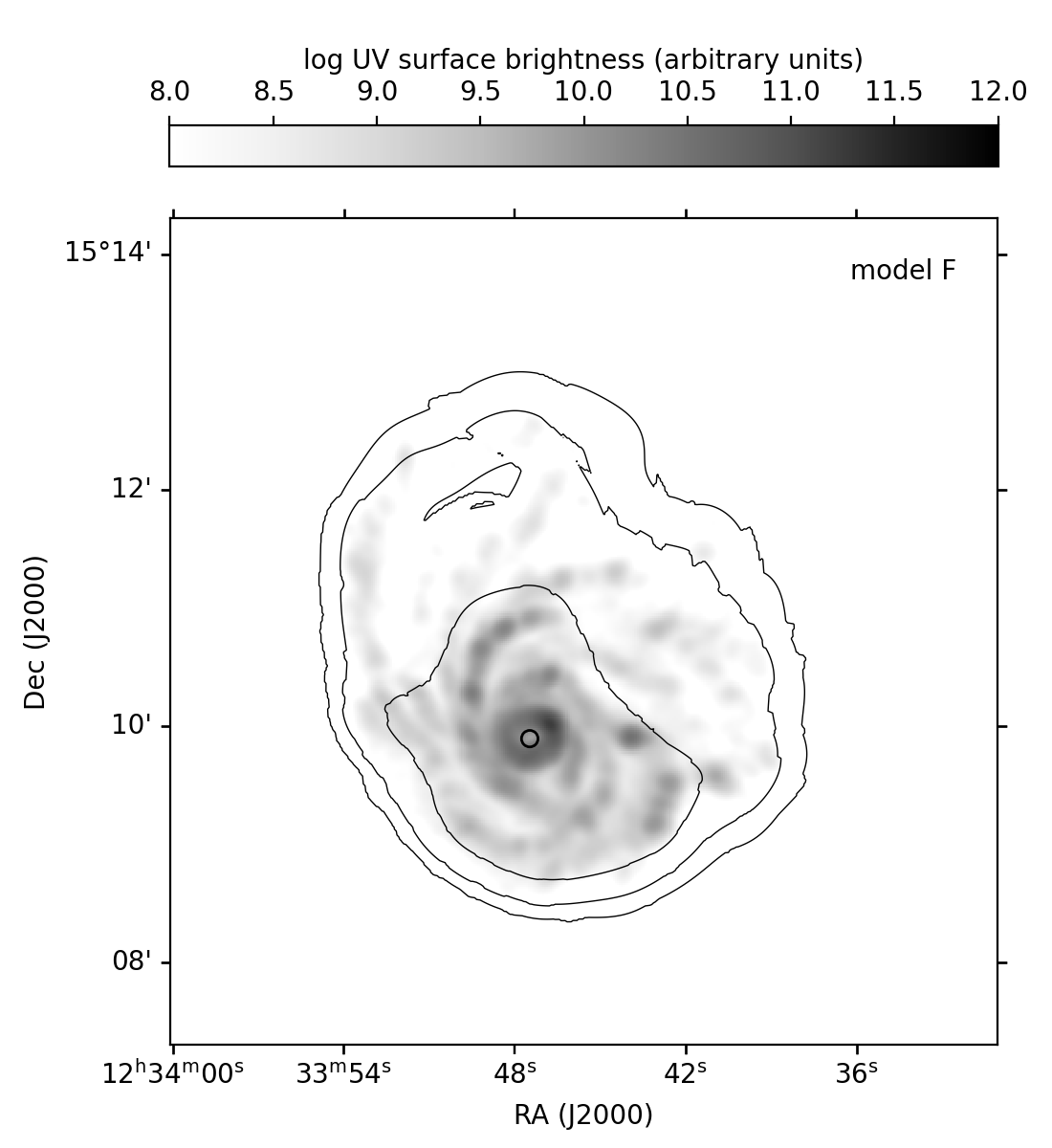}\\
\caption{Upper panels: model~F; left: model H{\sc i} moment~0 map, the contour levels show the model stellar distribution at 10 $\times$ 10$^{0.5 \times n}$ ($n$=0,1,2,..) M$_{\odot}$ pc$^{-2}$;
right: model H{\sc i} moment~1 map, the white contour levels are from $-70$ to $70$~km\,s$^{-1}$ in steps of $10$~km\,s$^{-1}$, the black contour the model stellar distribution at 10 M$_{\odot}$ pc$^{-2}$. The empty dot shows the kinematic centre.
Lower panels: left: model H{\sc i} contours on model H$\alpha$ distribution; right:
model H{\sc i} contours on model FUV distribution. The H{\sc i} contour levels on the H$\alpha$ and the FUV images are $N(HI)$ = 3.9$\times$10$^{19}$ $\times$ 10$^{0.5\times n}$ ($n$=0,1,2...) cm$^{-2}$. }
\label{modelF}
\end{figure*}

Figure \ref{modelAHR} shows the outputs of the high resolution simulations for model A. The outputs of the simulations for models B, C, D, E, and F are shown in Figs. \ref{modelB}, \ref{modelC}, \ref{modelD}, \ref{modelE}, and \ref{modelF} (see Sec. 6).

\section{Orbital parameters for model A}

\begin{figure*}
\centering
\includegraphics[width=1.0\textwidth]{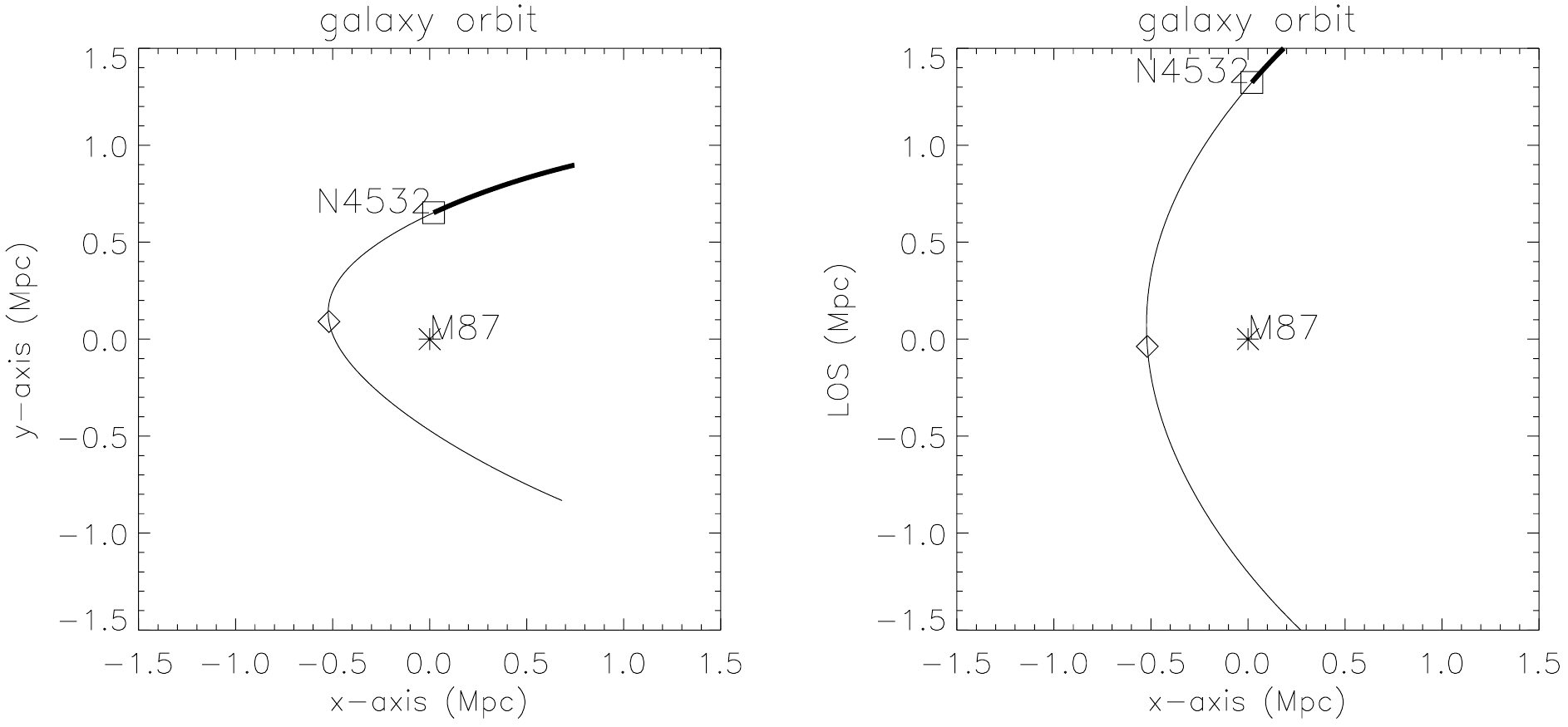}
\caption{Orbit of NGC 4523 within the Virgo cluster on the plane of the sky (left panel) and on the LoS (right panel) as derived for model A. The square box gives the position of the galaxy, the star the position of the centre of the cluster (M87), the diamond the expected position at maximum ram pressure (pericentre), which will occur in 1.2 Gyr. The thick line shows the orbit in the last 1.2 Gyr. The parameters of this orbit are given in Table \ref{parorbit}.}
\label{orbit}
\end{figure*}

The orbital parameters derived for model A are given in Table \ref{parorbit}, and the associated orbit is shown in Fig. \ref{orbit}.

\begin{table}
\caption{Orbital parameters for model A (Fig. \ref{orbit})}
\label{parorbit}
{
\[
\begin{tabular}{cc}
\hline
\noalign{\smallskip}
\hline
Orbital~parameter                        & Value                                                 \\      
\hline
present projected distance from M87     & 0.65 Mpc  \\
present position with respect to M87    & (0.02,0.65,1.32) Mpc  \\
present velocity vector                 & (-0.62,-0.29, -0.72) \\
total present velocity                  & 970 km s$^{-1}$   \\
present ICM density                     & 6 $\times$    10$^{-5}$ cm$^{-3}$ \\
ratio of maximum to present ram pressure& 9 \\
minimum distance from M87 at maximum ram pressure   & 0.53 Mpc  \\
maximum total velocity at maximum ram pressure & 1440 km s$^{-1}$  \\
maximum ICM density at maximum ram pressure    & 2.4 $\times$ 10$^{-4}$ cm$^{-3}$ \\
\noalign{\smallskip}
\hline
\end{tabular}
\]
}
\end{table}

\end{appendix}
\end{document}